\PassOptionsToPackage{table}{xcolor}
\documentclass[nonacm,sigplan]{acmart}

\usepackage[]{hyperref}

\AtBeginDocument{%
  }

\usepackage{xspace}
\usepackage{multirow}
\usepackage{comment}
\usepackage{xfrac}
\usepackage{fancybox, fancyvrb, calc}
\usepackage[center, tight]{subfigure}
\usepackage{amsmath}
\usepackage{amsthm} 
\usepackage{amsfonts} 
\usepackage[font=small]{caption}
\usepackage{xurl}
\usepackage{hyperref}
\usepackage{graphicx}
\usepackage{pbox}
\usepackage{algorithm, algorithmicx}
\usepackage[noend]{algpseudocode}
\usepackage{balance}
\usepackage{wrapfig}
\usepackage{tikz, pgfplots}
\usetikzlibrary{patterns,matrix,positioning,shadows,backgrounds,arrows,calc,fit,automata,shapes.geometric,arrows,decorations.pathreplacing,decorations.markings,shadings,shapes.symbols,arrows.meta}
\usepackage[capitalise]{cleveref}
\usepackage{tabularx}
\usepackage[T1]{fontenc}
\usepackage[utf8]{inputenc}
\usepackage[normalem]{ulem}
\usepackage{wasysym}
\usepackage{enumitem}
\usepackage{sidecap}
\usepackage{caption}
\usepackage{pifont}
\usepackage{setspace}
\usepackage{listings}
\usepackage{footnote}
\usepackage{makecell}
\makesavenoteenv{tabular}
\makesavenoteenv{table}
\usepackage{soul}

\usepackage{hyperref}
\hypersetup{
  colorlinks=true,      
  linkcolor=blue,       
  citecolor=magenta,    
  filecolor=cyan,       
  urlcolor=red          
}
\definecolor{listinggray}{gray}{0.9}
\definecolor{lbcolor}{rgb}{0.9,0.9,0.9}
\definecolor{dkgreen}{rgb}{0,0.6,0}
\definecolor{gray}{rgb}{0.5,0.5,0.5}
\definecolor{mauve}{rgb}{0.58,0,0.82}

\DeclareCaptionStyle{centered_lstlisting}{justification=centering, labelfont=bf, font=small, skip=5pt}
\graphicspath{{fig/}}

\usepackage{etoolbox}

\usepackage{tabularx} 
\usepackage{booktabs} 

\usepackage{bytefield}
\usepackage[table]{xcolor}

\theoremstyle{plain}

\definecolor{boxclr}{gray}{0.9}

\pgfdeclarelayer{bg}    
\pgfsetlayers{bg,main}  

\makeatletter
\newcommand{\thickhline}{%
    \noalign {\ifnum 0=`}\fi \hrule height 0.8pt
    \futurelet \reserved@a \@xhline
}
\newcolumntype{"}{@{\vrule width 0.8pt}}
\newcolumntype{[}{@{\vrule width 0.8pt\hskip\tabcolsep}}
\newcolumntype{]}{@{\hskip\tabcolsep\vrule width 0.8pt}}
\newcolumntype{!}{@{\hskip\tabcolsep\vrule width 0.8pt\hskip\tabcolsep}}
\makeatother

\newcommand{\cppsnippet}[1]{%
  \begin{lstlisting}[gobble=4]
    #1
  \end{lstlisting}
}

\tikzset{
    position/.style args={#1:#2 from #3}{
        at=(#3.#1), anchor=#1+180, shift=(#1:#2)
    }
}

\tikzset{
  half fill/.style 2 args={fill=#2, path picture={
    \fill[#1, sharp corners] (path picture bounding box.west) --
                         (path picture bounding box.east) --
                         (path picture bounding box.south east) --
                         (path picture bounding box.south west) -- cycle;}},
}

\tikzset{
  nil fill/.style 2 args={fill=#2, path picture={
    \fill[#1, sharp corners] (path picture bounding box.155) --
                         (path picture bounding box.25) --
                         (path picture bounding box.south east) --
                         (path picture bounding box.south west) -- cycle;}},
}

\tikzset{
  almost fill/.style 2 args={fill=#2, path picture={
    \fill[#1, sharp corners] (path picture bounding box.205) --
                         (path picture bounding box.335) --
                         (path picture bounding box.south east) --
                         (path picture bounding box.south west) -- cycle;}},
}

\definecolor{oldcolor}{HTML}{c66541}

\theoremstyle{nonumberplain}

\algdef{SE}[DOWHILE]{Do}{doWhile}{\algorithmicdo}[1]{\algorithmicwhile\ #1}%

\newcommand{\cut}[1]{}

\newcommand{\paragraphb}[1]{\vspace{0.05in}\noindent{\bf #1.}}
\newcommand{\paragrapha}[1]{\vspace{0.05in}\noindent{\bf #1}}
\newcommand{\paragraphc}[1]{\vspace{0.05in}\noindent{\em #1}}

\colorlet{soulgreen}{green!30}

\colorlet{soulblue}{blue!20}
\colorlet{soulred}{red!20}

\colorlet{soulpink}{red!20}

\tikzset{ 
table/.style={
  matrix of nodes,
  row sep=-\pgflinewidth,
  column sep=-\pgflinewidth,
  nodes={rectangle,thick,draw=black,text width={},align=center,font=\small},
  text depth=0.25ex,
  text height=1.25ex,
  nodes in empty cells
},
map/.style={
  matrix of nodes,
  row sep=-\pgflinewidth,
  column sep=-\pgflinewidth,
  nodes={rectangle,draw=black,text width=5em,align=center,font=\small},
  text depth=0.25ex,
  text height=1.25ex,
  nodes in empty cells
},
bigmap/.style={
  matrix of nodes,
  row sep=-\pgflinewidth,
  column sep=-\pgflinewidth,
  nodes={rectangle,draw=black,text width=26em,align=center,font=\small},
  text depth=0.25ex,
  text height=1.25ex,
  nodes in empty cells
},
memcell/.style={
  draw, 
  very thick, 
  text width=0.25em, 
  text height=0.25em
},
}

\tikzstyle{startstop} = [rectangle, rounded corners, minimum width=3em, minimum height=1em,text centered, draw=black, fill=red!30]
\tikzstyle{io} = [trapezium, trapezium left angle=70, trapezium right angle=120, minimum width=2.5em, minimum height=1em, text centered, draw=black, fill=blue!30]
\tikzstyle{process} = [rectangle, minimum width=1.5em, minimum height=1em, align=center, draw=black, fill=gray!30]
\tikzstyle{decision} = [diamond, minimum width=3em, minimum height=1em, align=center, draw=black, fill=SkyBlue!30]
\tikzstyle{arrow} = [thick,->,>=stealth]
\tikzstyle{monolog} = [fill=SkyBlue!30]

\tikzset{
  on each segment/.style={
    decorate,
    decoration={
      show path construction,
      moveto code={},
      lineto code={
        \path [#1]
        (\tikzinputsegmentfirst) -- (\tikzinputsegmentlast);
      },
      curveto code={
        \path [#1] (\tikzinputsegmentfirst)
        .. controls
        (\tikzinputsegmentsupporta) and (\tikzinputsegmentsupportb)
        ..
        (\tikzinputsegmentlast);
      },
      closepath code={
        \path [#1]
        (\tikzinputsegmentfirst) -- (\tikzinputsegmentlast);
      },
    },
  },
  mid arrow/.style={postaction={decorate,decoration={
        markings,
        mark=at position .5 with {\arrow[#1]{stealth}}
      }}},
}

\newcommand{\specialcell}[2][l]{%
  \begin{tabular}[#1]{@{}l@{}}#2\end{tabular}}

\newcolumntype{L}[1]{>{\RaggedRight\hspace{0pt}}p{#1}}
\newcolumntype{R}[1]{>{\RaggedLeft\hspace{0pt}}p{#1}}

\algnewcommand{\IIf}[1]{\State\algorithmicif\ #1\ \algorithmicthen}%
\algnewcommand{\EndIIf}{\unskip\ }%
\algdef{SE}[DOWHILE]{Do}{doWhile}{\algorithmicdo}[1]{\algorithmicwhile\ #1}%
\algnewcommand\algorithmicforeach{\textbf{for each}}%
\algdef{S}[FOR]{ForEach}[1]{\algorithmicforeach\ #1\ \algorithmicdo}%

\newcommand{\code}[1]{{\fontsize{9}{11}\selectfont\texttt{#1}}}
\newcommand{\smallcode}[1]{{\fontsize{8}{8}\selectfont\texttt{#1}}}

\newcommand\redsout{\bgroup\markoverwith{\textcolor{red}{\rule[0.5ex]{2pt}{0.4pt}}}\ULon}

\def\ie{{i.e.}}
\def\eg{{\em e.g.}\xspace}

\def\etc{etc.}

\def\namex{\textrm{\textsc{pulse}}}
\def\name{\namex\xspace}
\def\namey{\textrm{\textsc{pulse-acc}}}
\def\nameacc{\namey\xspace}

\def\nameh{\textrm{\textsc{pulse-asic}}}
\def\nameasic{\nameh\xspace}

\begin{document}
\sloppy


\title{\name: Accelerating Distributed Pointer-Traversals on Disaggregated Memory (Extended Version)}
\date{}

\author{Yupeng Tang}
\affiliation{%
    \institution{Yale University}
    \city{New Haven}
    \country{United States}
}
\author{Seung-seob Lee}
\affiliation{%
    \institution{Yale University}
    \city{New Haven}
    \country{United States}
}
\author{Abhishek Bhattacharjee}
\affiliation{%
    \institution{Yale University}
    \city{New Haven}
    \country{United States}
}
\author{Anurag Khandelwal}
\affiliation{%
    \institution{Yale University}
    \city{New Haven}
    \country{United States}
}
\begin{abstract}
Caches at CPU nodes in disaggregated memory architectures amortize the high data access latency over the network. However, such caches are fundamentally unable to improve performance for workloads requiring pointer traversals across linked data structures. We argue for accelerating these pointer traversals closer to disaggregated memory in a manner that preserves expressiveness for supporting various linked structures, ensures energy efficiency and performance, and supports distributed execution. We design \name, a distributed pointer-traversal framework for rack-scale disaggregated memory to meet all the above requirements. Our evaluation of \name shows that it enables low-latency, high-throughput, and energy-efficient execution for a wide range of pointer traversal workloads on disaggregated memory that fare poorly with caching alone.
\end{abstract}

\maketitle
\footnotetext{This paper is an extended version of our work originally published in the Proceedings of the 30th ACM International Conference on Architectural Support for Programming Languages and Operating Systems (ASPLOS 2025).}

\section{Introduction}
\label{sec:intro}

\begin{figure}[t]
  \centering
  \includegraphics[width=0.98\columnwidth]{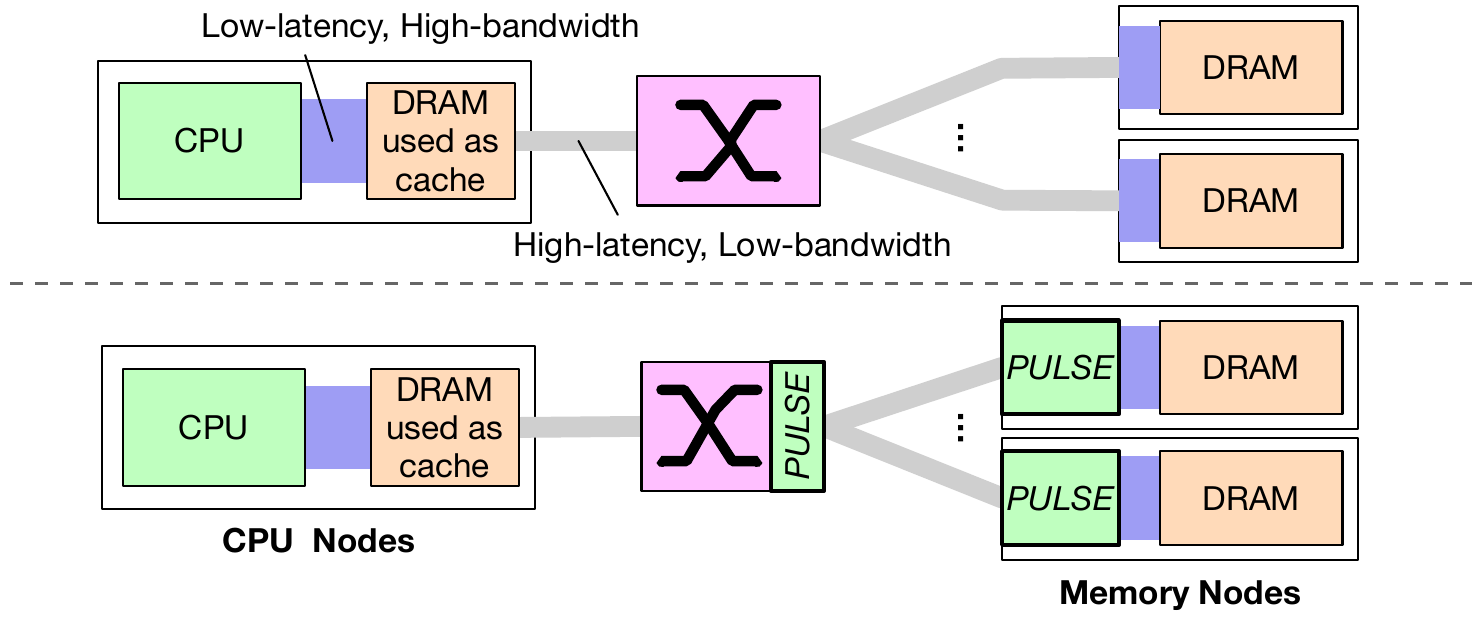}
  \vspace{-0.7em}
  \caption{\textbf{Need for accelerating pointer traversals.} \textit{(top)} The performance of pointer traversals in disaggregated architectures is bottlenecked by slow memory interconnect. \textit{(bottom)} Just as caches offer limited but fast memory near CPUs, we argue that memory needs a counterpart for traversal-heavy workloads: a lightweight but fast accelerator for cache-unfriendly pointer traversals.} 
  \label{fig:disagg}
\end{figure}

\noindent
Driven by increasing demands for memory capacity and bandwidth~\cite{scuba, cachelib, tao, memcache, flighttracker, twittercache, spark}, poor scaling~\cite{memscaling2, memscaling3, memscaling1} and resource inefficiency~\cite{infiniswap, memoverprovisioning} of DRAM, and improvements in Ethernet-based network speeds~\cite{terabitethernet, remotememory}, recent years have seen significant efforts towards memory disaggregation~\cite{fastswap, memdisagg1, infiniswap, mind, legoos}. Rather than scaling up a server's DRAM capacity and bandwidth, such proposals advocate disaggregating much of the memory over the network. The result is a set of CPU nodes equipped with a small amount of DRAM used as cache\footnote{Not to be confused with die-stacked hardware DRAM caches~\cite{jevdjic2013stacked, jevdjic2014unison, young2018accord}.}, accessing memory across a set of network-attached memory nodes with large DRAM pools (Fig.~\ref{fig:disagg}~(top)). With allocation flexibility across CPU and memory nodes, disaggregation enables high utilization and elasticity.

\begin{figure*}[ht!]
    \centering
    \subfigure[Our empirical analysis]{
        \includegraphics[width=0.49\textwidth]{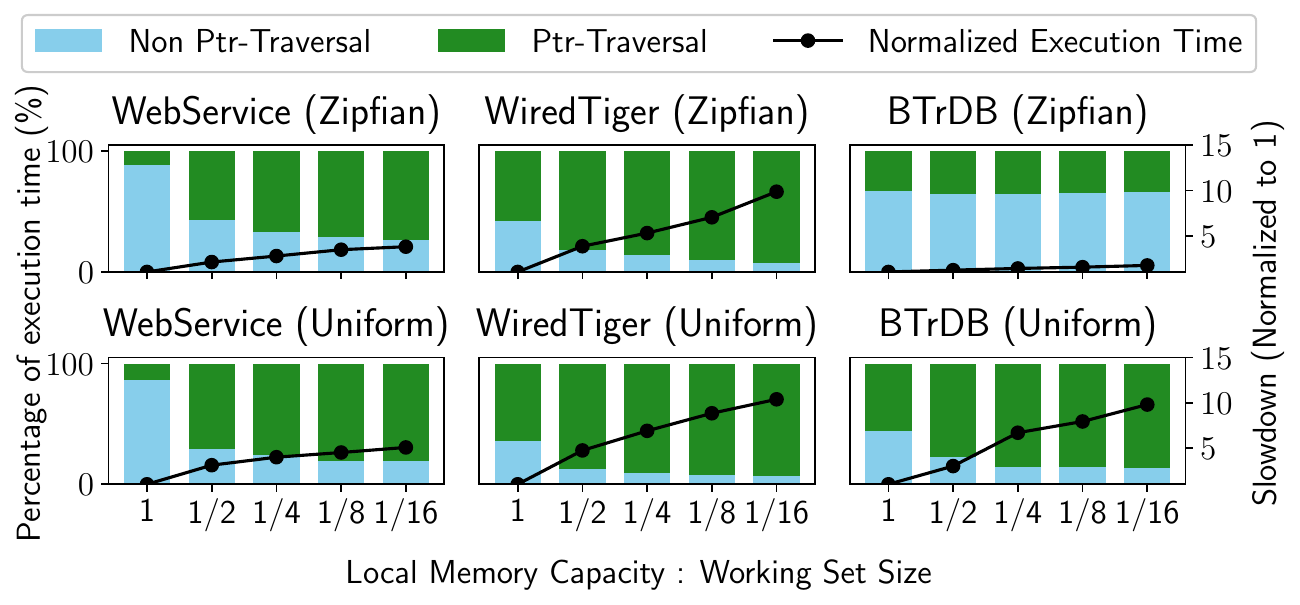}
        \label{fig:motivation_experiment}
    }
    \subfigure[\% of distributed traversals]{
        \includegraphics[width=0.21\textwidth]{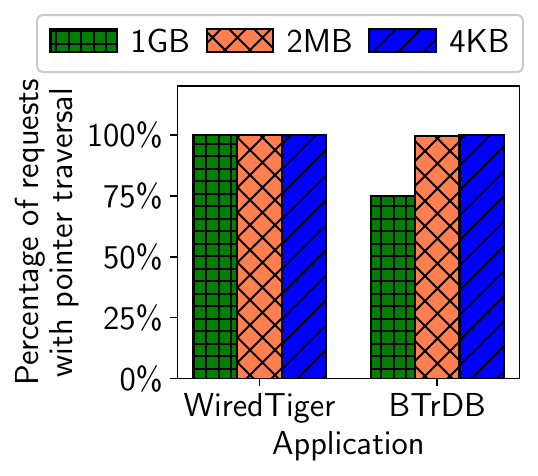}
        \label{fig:distributed_percentage}
    }
    \subfigure[CDF of distributed traversals]{
        \includegraphics[width=0.24\textwidth]{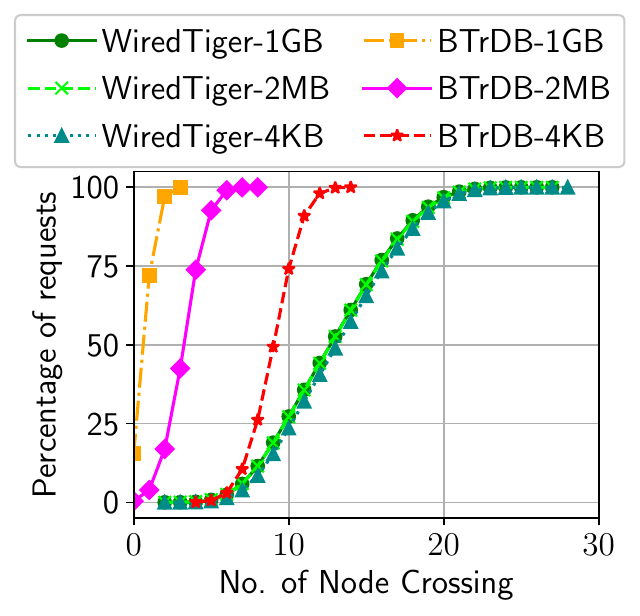}
        \label{fig:distributed_cdf_wiredtiger}
    }
    \vspace{-1em}
    \caption{\textbf{Time cloud applications spend in pointer traversals.} See \S\ref{ssec:need} for details.} 
    \label{fig:motivation}
\end{figure*}

Despite drastic improvements in recent years, the limited bandwidth and latency to network-attached memory remain a hurdle in adopting disaggregated memory, with speed-of-light constraints making it impossible to improve network latency beyond a point. Even with near-terabit links and hardware-assisted protocols like RDMA~\cite{rdmalatency}, remote memory accesses are an order of magnitude slower than local memory accesses~\cite{disagg}. Emerging CXL interconnects~\cite{cxl} share a similar trend --- around $300$ ns of CXL memory latency compared to $10$--$20$ ns of L3 cache latency~\cite{pond}. Although efficient caching strategies at the CPU node can reduce average memory access latency and volume of network traffic to remote memory, the benefit of such strategies is limited by data locality and the size of the cache on the CPU node. In many cases, remote memory accesses are unavoidable, especially for applications that rely on efficient in-memory pointer traversals on linked data structures, such as lookups on index structures~\cite{hash1, hash2, hash3, succinct, trie2, btree1, btree2, trie1, blowfish, trie3, surf} in databases and key-value stores, and traversals in graph analytics~\cite{powergraph, graphx, graphchi, pagerank} (Fig.~\ref{fig:motivation}, \S\ref{sec:overview}). 

Similar to how CPUs have small but fast memory (\ie, caches) for quick access to popular data, we argue that memory nodes should also include lightweight but fast processing units with high-bandwidth, low-latency access to memory to speed up pointer-traversals (Fig.~\ref{fig:disagg}~(bottom)). Moreover, the interconnect should facilitate efficient and scalable distributed traversals for deployments with multiple memory nodes that cater to large-scale linked data structures. Prior works have explored systems and API designs for such processing units under multiple settings, ranging from near-memory processing and processing-in-memory approaches~\cite{ahn2015scalable, asghari2016chameleon,  dai2018graphh, schuiki2018scalable, mutlu2019processing, lockerman2020livia, tu2022redcim, devic2022_PIM, wang2022_Nearstream, xie2023mpu, mutlu2022modern, oliveira2022accelerating, eckert2022eidetic, chi2016prime, seshadri2017simple, kwon2019_TensorDIMM, boroumand2019_codna, cho2020_data, ke2020_RecNMP, wang2021stream, xie2021spacea, ke2021near, singh2021fpga, olgun2022pidram, dai2022dimmining, gu2020ipim, gomez2023evaluating, walkers, impica} for single-server architectures, to the use of CPUs~\cite{storagefunctions, splinter, aifm, kayak_nsdi_21, storm_systor_19, zhang2022_teleport} or FPGAs~\cite{clio, strom} near remote/disaggregated memory, but have several key shortcomings. 

Specifically, existing approaches are limited in scale and expose a three-way tradeoff between expressiveness, energy efficiency, and performance. First, and perhaps most crucially, none of the existing approaches can accelerate pointer traversals that span \emph{multiple} network-attached memory nodes. 

This limits memory utilization and elasticity since applications must confine their data to a single memory node to accelerate pointer traversals. Their inability to support distributed pointer traversals stems from complex management of address translation state that is required to identify if a traversal can occur locally or must be re-routed to a different memory node (\S\ref{ssec:prior}). Second, existing single-node approaches use full-fledged CPUs for expressive and performant execution of pointer-traversals~\cite{storagefunctions, splinter, aifm, kayak_nsdi_21}. However, coupling large amounts of processing capacity with memory --- which has utility in reducing data movement in PIM architectures~\cite{ahn2015scalable, dai2018graphh, schuiki2018scalable, mutlu2019processing, mutlu2022modern, oliveira2022accelerating, eckert2022eidetic, xie2023mpu, tu2022redcim, lockerman2020livia, asghari2016chameleon, devic2022_PIM, wang2022_Nearstream} ---  goes against the very spirit of memory disaggregation since it leads to poor utilization of compute resources and, consequently, poor energy efficiency. 

Approaches that use wimpy processors at SmartNICs~\cite{rmc_hotnets20, redn} instead of CPUs retain expressiveness, but the limited processing speeds of wimpy nodes curtail their performance and, ultimately lead to lower energy efficiency due to their lengthened executions (\S\ref{ssec:application-study},~\cite{clio}). Lastly, FPGA-based~\cite{clio, strom, sun2023demystifying} and ASIC-based~\cite{impica, walkers} approaches achieve performance and energy efficiency by hard-wiring pointer traversal logic for specific data structures, limiting their expressiveness.  

We design \name\footnote{\textbf{P}rocessing \textbf{U}nit for \textbf{L}inked \textbf{S}tructur\textbf{E}s.}, a distributed pointer-traversal framework for rack-scale disaggregated memory, to meet all of the above needs --- namely, expressiveness, energy efficiency, performance --- via a principled redesign of near-memory processing for disaggregated memory. Central to \name's design is an expressive iterator interface that readily lends itself to a unifying abstraction across most pointer traversals in linked data structures used in key-value stores~\cite{redis, memcached}, databases~\cite{mongodb, btree1, btree2, trie1, trie3}, and big-data analytics~\cite{powergraph, graphx, graphchi, pagerank} (\S\ref{sec:interface}). \name's use of this abstraction not only makes it immediately useful in this large family of real-world traversal-heavy use cases, but also enables (i) the use of familiar compiler toolchains to support these use cases with little to no application modifications and (ii) the design of tractable hardware accelerators and efficient distributed traversal mechanisms that exploit properties unique to iterator abstractions.

In particular, \name enables transparent and efficient execution of pointer traversals for our iterator abstraction via a novel accelerator that employs a \emph{disaggregated} architecture to decouple logic and memory pipelines, exploiting the inherently sequential nature of compute and memory accesses in iterator execution (\S\ref{sec:accelerator}). This permits high utilization by provisioning more memory and fewer logic pipelines to cater to memory-centric pointer traversal workloads. A scheduler breaks pointer traversal logic from multiple concurrent workloads across the two sets of pipelines and employs a novel multiplexing strategy to maximize their utilization. While our implementation leverages an FPGA-based SmartNIC due to the high cost and complexity of ASIC fabrication, our ultimate vision is an ASIC-based realization for improved performance and energy efficiency. 

We enable distributed traversals by leveraging the insight that pointer traversal across network-attached memory nodes is equivalent to packet routing at the network switch (\S\ref{sec:distributed}). As such, \name leverages a programmable network switch to inspect the next pointer to be traversed within iterator requests and determine the next memory node to which the request should be forwarded --- both at line rate. 

We implement a real-system prototype of \name on a disaggregated rack of commodity servers, SmartNICs, and a programmable switch with full-system effects. None of \name's hardware or software changes are invasive or overly complex, ensuring deployability.  Our evaluation of end-to-end real-world workloads shows that \name outperforms disaggregated caching systems with $9$--$34\times$ lower latency and $28$--$171\times$ higher throughput. Moreover, our Xilinx XRT~\cite{xilinx_xrt} and Intel RAPL~\cite{intel_rapl}-based power analysis shows that \name consumes $4.5$--$5\times$ less energy than RPC-based schemes (\S\ref{sec:evaluation}).

\section{Motivation and \name Overview}
\label{sec:overview}

\subsection{Need for Accelerating Pointer Traversals}
\label{ssec:need}

Memory-intensive applications~\cite{scuba, cachelib, tao, memcache, flighttracker, twittercache, spark} often require traversing linked structures like lists, hash tables, trees, and graphs. 
While disaggregated architectures provide large memory pools across network-attached memory nodes, traversing pointers over the network is still slow~\cite{disagg}. Recent proposals~\cite{disagg, legoos, mind, infiniswap, fastswap} alleviate this slowdown by using the DRAM at the CPU nodes to cache ``hot'' data, but such caches often fare poorly for pointer traversals, as we show next.

\paragraphb{Pointer traversals in real-world workloads} Prior studies~\cite{graphchi, monetdb, spark, voltdb, memc3, db1000, memcached} have shown that real-world data-centric cloud applications spend anywhere from $21\%$ to $97\%$ of execution time traversing pointers. We empirically analyze the time spent in pointer traversals for three representative cloud applications --- a WebService frontend~\cite{aifm}, indexing on WiredTiger~\cite{mongodb}, and time-series analysis on BTrDB~\cite{btrdb} --- with swap-based disaggregated memory~\cite{infiniswap}\footnote{We defer the details of the data structures and workloads employed by these applications, as well as the disaggregated memory setup to \S\ref{sec:evaluation}.}. We vary the cache size at the CPU node from $6.25$\%-$100$\% of each application's working set size. Fig.~\ref{fig:motivation_experiment} shows that (i) all three applications spend a significant fraction of their execution time ($13.6$\%, $63.7$\%, and $55.8$\%, respectively) traversing pointers even when their entire working set is cached, and (ii) the time spent traversing pointers (and thus, the end-to-end execution time) increases with smaller CPU node caches. While the impact of access skew is application-dependent, pointer traversals dominate application execution times when more of the application's working set size is remote.

\paragraphb{Distributed traversals} As the number of applications and the working-set size per application grows larger, disaggregated architectures must allocate memory across multiple memory nodes to keep up. Such approaches~\cite{legoos, mind, infiniswap, fastswap} tend to strive for the smallest viable allocation granularity with reasonable metadata overheads (e.g., $1$ GB in~\cite{legoos}, $2$ MB in~\cite{mind}) since smaller allocations permit better load balancing and high memory utilization. Unfortunately, finer-grained allocations may cause an application's linked structures to get fragmented across multiple network-attached memory nodes, necessitating many \emph{distributed} traversals. 

Fig.~\ref{fig:distributed_percentage} illustrates this impact on a setup with $1$ compute and $4$ memory nodes: even with large $1$ GB allocations, WiredTiger and BTrDB require over 97\% and 75\% of their requests, respectively, to cross memory node boundaries at least once, with the volume of cross-node traffic increasing at smaller granularities. Fig.~\ref{fig:distributed_cdf_wiredtiger} shows the CDF of requests that require a certain number of memory node crossings. While the randomly ordered data in WiredTiger necessitate many cross-node traversals even for large allocations, the time-ordered data in BTrDB reduce cross-node traversals for larger allocation granularities by confining large time windows to the same memory node. However, smaller to moderate allocation granularities --- required for high memory utilization --- still require many cross-node traversals.

\subsection{Shortcomings of Prior Approaches}
\label{ssec:prior}

No prior work achieves all four properties required for pointer traversals on disaggregated memory: distributed execution, expressiveness, energy efficiency, and performance. We focus on network-attached memory, although a similar analysis extends to in-memory processing~\cite{walkers, ahn2015scalable, impica, asghari2016chameleon, chi2016prime, seshadri2017simple, dai2018graphh, schuiki2018scalable, mutlu2019processing, kwon2019_TensorDIMM, boroumand2019_codna, gu2020ipim, lockerman2020livia, cho2020_data, ke2020_RecNMP, wang2021stream, xie2021spacea, ke2021near, singh2021fpga, olgun2022pidram, mutlu2022modern, oliveira2022accelerating, eckert2022eidetic, tu2022redcim, dai2022dimmining, devic2022_PIM, wang2022_Nearstream, gomez2023evaluating, xie2023mpu}.

\paragraphb{No support for distributed execution} Distributed pointer traversals are required to ensure applications can efficiently access large pools of network-attached memory nodes. Unfortunately, to our knowledge, none of the prior works support efficient multi-node pointer traversals. Therefore, applications must confine their data to a single node for efficient traversals, exposing a tradeoff between application performance and scalability. Recent proposals~\cite{sherman, clover, fusee, rolex, marlin, sephash, ditto} explore specialized data structures that co-design partitioning and allocation policies to reduce distributed pointer traversals atop disaggregated memory. Such approaches complement our work since they still require efficient distributed traversals when their optimizations are not applicable, \eg, not many data structures benefit from such specialized co-designs. 

\begin{figure*}[ht!]
  \centering
  \includegraphics[width=0.85\textwidth]{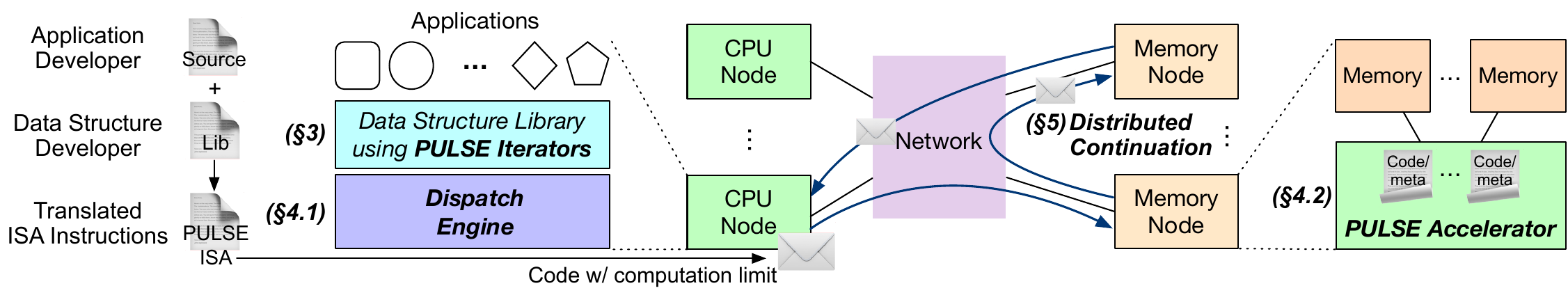}
  \vspace{-1em}
  \caption{\textbf{\name Overview.} Developers use \name's iterator interface (\S\ref{sec:interface}) to express pointer traversals, translated to \name ISA by its dispatch engine (\S\ref{ssec:compute_node}). During execution, \name accelerator ensures energy efficiency (\S\ref{ssec:architecture}) and in-network design enable distributed traversals (\S\ref{sec:distributed}).} 
  \label{fig:general}
\end{figure*}

\paragraphb{Poor performance with prefetching approaches} Cache-based designs for remote memory often employ prefetching techniques~\cite{prefetch1, prefetch2, prefetch3} that pipeline remote memory access with computations at the CPU nodes. Unfortunately, such pipelining does not improve performance for pointer traversal workloads for two main reasons. First, the remote memory access latency is typically far greater than the computation required for pointer traversals, so network round trips to prefetch the data would remain the bottleneck. Second, speculating the next unit of data to prefetch for more complex data structures like B+Trees or graphs, where each node contains pointers to many ``children'' nodes, tends to have much lower accuracy in practice. As such, prefetching can even add overheads due to unnecessary data fetches.

\paragraphb{Poor utilization/power-efficiency in CPUs} Many prior works have explored remote procedure calls (RPCs) to enable offloading computation to CPUs on memory nodes~\cite{aifm, kayak_nsdi_21, splinter, storagefunctions, storm_systor_19}. While CPUs are performant and versatile enough to support most general-purpose computations, the same versatility makes them overkill for pointer traversal workloads in disaggregated architectures --- the CPUs on memory nodes are likely to be underutilized and, consequently, waste energy (\S\ref{sec:evaluation}), since such workloads are memory-intensive and bounded by memory bandwidth rather than CPU cycles. 

Since inefficient power usage resulting from coupled compute and memory resources is the main problem disaggregation aims to resolve, leveraging CPUs at memory nodes essentially nullifies these benefits. 

\paragraphb{Limited expressiveness in FPGA/ASIC accelerators} Another approach explored in recent years uses FPGAs~\cite{clio,strom} or ASICs~\cite{impica, walkers} at memory nodes for performance and energy efficiency. FPGA approaches exploit circuit programmability to realize performant on-path data processing, albeit only for specific data structures, limiting their expressiveness. Although some FPGA approaches aim for greater expressiveness by serving RPCs~\cite{coyote}, RPC logic must be pre-compiled before it is deployed and physically consumes FPGA resources. This limits how many RPCs can be deployed on the FPGA concurrently and also elides runtime resource elasticity for different pointer traversal workloads. ASIC approaches either support a single data structure or provide limited ISA specialized for a single data structure (\eg, linked-lists~\cite{walkers}), limiting their general applicability. 

\paragraphb{Poor performance/power efficiency in wimpy SmartNICs} The emergence of programmable SmartNICs has driven work on offloading computations to the onboard network processors. Some approaches utilize wimpy processors (\eg, ARM or RISC-V processors)~\cite{rmc_hotnets20} or RDMA processing units (PUs)~\cite{redn} to support general-purpose computations near memory. While these wimpy processors can eliminate multiple network round trips in pointer traversal workloads, their processing speeds are far slower than CPU-based or FPGA-based accelerators. Often, such PUs can become a performance bottleneck, especially at high memory bandwidth ($\sim$500 Gbps)~\cite{redn, disagg}. Moreover, wimpy processors tend not to be energy-efficient since their slower execution tends to waste more static power, resulting in higher energy per pointer traversal offload --- an observation noted in prior work~\cite{clio} and confirmed in our evaluation (\S\ref{sec:evaluation}).

\subsection{\name Design Overview}
\label{ssec:overview}

\name innovates on three key design elements (Fig.~\ref{fig:general}). Central to \name's design is its iterator-based programming model (\S\ref{sec:interface}) that requires minimal effort to port real-world data structure traversals. \name supports \emph{stateful} traversals using a \emph{scratchpad} of pre-configured size, where developers can store and update arbitrary intermediate states (\eg, aggregators, arrays, lists, \etc) during the iterator's execution. Properties specific to iterator patterns enable tractable accelerator design and efficient distributed traversals in \name.  

The iterator code provided by the data structure developer is translated into \name's instruction set architecture (ISA) to be executed by \name accelerators (\S\ref{sec:accelerator}). \name achieves energy efficiency and performance through a novel accelerator that employs disaggregated logic and memory pipelines and an ISA specifically designed for the iterator pattern. Our accelerator employs a scheduler specialized for its disaggregated architecture to ensure high utilization \emph{and} performance.

\name supports scalable distributed pointer traversals by leveraging programmable network switches to reroute any requests that must cross memory node boundaries (\S\ref{sec:distributed}). \name employs hierarchical address translation \emph{in the network}, where memory node-level address translation is performed at the switch (\ie, a request is routed to the memory node based on its target address), and the memory node accelerator performs translation and protection for local accesses. During traversal, a memory node accelerator can return a request to the switch if it determines the address is not local; the switch re-routes the request to the correct memory node.

\paragraphb{Assumptions} \name does not offload synchronization to its accelerators but instead requires the application logic at the CPU node to explicitly acquire/release appropriate locks for the offloaded operation. Recent efforts enable locking primitives on NICs~\cite{sherman, clover} and programmable switches~\cite{netlock}; these are orthogonal to our work and can be incorporated into \name. 
Finally, \name does not innovate on caching and adapts the caching scheme from prior work~\cite{aifm}, which maintains a transparent cache within the data structure library. 

\section{\name Programming Model}
\label{sec:interface}
\label{ssec:iterators}
\label{ssec:iteratorexample}

We begin with \name's programming model since a carefully crafted interface is crucial to enable wide applicability for real-world traversal-heavy applications, as well as the design of tractable pointer traversal accelerators and efficient distributed traversal mechanisms. \name's interface is intended for data structure library developers to offload pointer traversals in linked data structures. Since \name code modifications are restricted to data structure libraries, existing applications utilizing their interfaces require no modifications. 

We analyzed the implementations of a wide range of popular data structures~\cite{stl, boost, javaiterator, c++iterator}  
to determine the structures common to them in pointer traversals. We found that most traversals (1) initialize a start pointer using data structure-specific logic, (2) iteratively use data structure-specific logic to determine the next pointer to look up, and (3) check a data structure-specific termination condition at the end of each iteration to determine if the traversal should end. 
This structure resembles that of the \emph{iterator} design pattern, establishing its universality as a design motif common to almost all languages~\cite{javaiterator}. This is precisely what makes it an ideal candidate for the interface between the hardware and software layers for pointer traversals. As such, \name allows developers to program their data structure traversals using the iterator interface shown in Listing~\ref{lst:iterator}. 

The interface exposes three functions that must be implemented by the user: (1) \code{init()}, which takes as input arbitrary data structure-specific state to initialize the start pointer, (2) \code{next()}, that updates the current pointer to the next pointer it must traverse to, and, (3) \code{end()}, that determines if the pointer traversal should end (either in success or failure) based on the current pointer. \name then uses the provided implementations for these functions to execute the pointer traversal iteratively, using the \code{execute()} function. We discuss two key novel aspects of our iterator abstraction that were necessary to increase and limit the expressiveness of operations on linked data structures.

\begin{figure}
\centering
\footnotesize
\begin{lstlisting}[caption={\name interface.},label={lst:iterator},escapechar=|]
class pulse_iterator {
    void init(void *) = 0; // Implemented by developer
    void *next() = 0; // Implemented by developer
    bool end() = 0; // Implemented by developer
    
    unsigned char *execute() { // Non-modifiable logic
      unsigned int num_iter = 0;
      while (!end() && num_iter++ < MAX_ITER)
        cur_ptr = next();
      return scratch_pad;|\label{line:scratch_return}|
    }
    uintptr_t cur_ptr;
    unsigned char scratch_pad[MAX_SCRATCHPAD_SIZE];
}
\end{lstlisting}
\end{figure}

\paragraphb{Stateful traversals} Pointer traversals in many data structures are stateful, and the nature of the state can vary widely. For instance, in hash table lookups, the state is the search key that must be compared against a linked list of keys in a hash bucket. In contrast, summing up values across a range of keys in a B-Tree requires maintaining a running variable for storing the sum and updating it for each value encountered in the range. To facilitate this, \name iterators maintain a \code{scratch\_pad} that the developer can use to store an arbitrary state. The \code{scratch\_pad} acts as a continuation~\cite{continuation} in the programming language sense, allowing the state to persist across iterations. It is initialized in \code{init()}, updated in \code{next()}, and finalized in \code{end()}. Since \code{execute()} in \name's iterator interface returns the contents of \code{scratch\_pad} (Line~\ref{line:scratch_return}), developers can place the state they want to retrieve in it.

\paragraphb{Bounded computations} \name accelerators support only lightweight processing in memory-intensive operations for high memory bandwidth utilization. While \code{init()} is executed on the CPU node, \code{next()} and \code{end()} are offloaded to \name accelerators; hence, \name limits what memory accesses and computations can be performed in them in two ways. Within each iteration, \name disallows nondeterministic executions, such as unbounded loops, \ie, loops that cannot be unrolled to a fixed number of instructions. Across iterations, \code{execute()} in Listing~\ref{lst:iterator} limits the maximum number of iterations that a single request is allowed to perform. This ensures that a particularly long traversal does not block other requests for a long time. 
If a request exceeds the maximum iteration count, \name terminates the traversal and returns the \code{scratch\_pad} value to the CPU node, which can issue a new request to continue the traversal from that point. 

\begin{figure}[t]
\centering
\begin{lstlisting}[caption={C++ STL realization for \code{unordered\_map::find()}.},label={lst:stl}]
struct node {
  key_type key;
  value_type value;
  struct node *next;
};

value_type find(key_type key) {
  for (struct node *cur_ptr = bucket_ptr(hash(key)); ; cur_ptr = cur_ptr->next) {
    if (key == cur_ptr->key) // Key found
      return cur_ptr->value;
    if (cur_ptr->next == nullptr) // Key not found
      break;
  }
  return KEY_NOT_FOUND;
}
\end{lstlisting}
\begin{lstlisting}[caption={\name realization for \code{unordered\_map::find()}.},label={lst:stl_mod}]
class unordered_map_find : pulse_iterator {
  init(void *key) {
    memcpy(scratch_pad, key, sizeof(key_type));
    cur_ptr = bucket_ptr(hash((key_type)*key));
  }
  
  void* next() { return cur_ptr->next; }
  
  bool end() {
    key_type key = *((key_type *)scratch_pad);
    if (key == cur_ptr->key) { // Key found
      *((value_type *)scratch_pad) = cur_ptr->value;
      return true;
    }
    if (cur_ptr->next == nullptr) { // Key not found
      *((unsigned int *)scratch_pad) = KEY_NOT_FOUND;  
      return true;
    }
    return false;
  }
}
\end{lstlisting}
\end{figure}

\paragraphb{An illustrative example} We demonstrate how the \code{find()} operation on C++ STL \code{unordered\_map} can be ported to \name. Listing~\ref{lst:stl} shows a simplified version of its implementation in STL --- the pointer traversal begins by computing a hash function and determining a pointer to the hash bucket corresponding to the hash. It then iterates through a linked list corresponding to the hash bucket, terminating if the key is found or the linked list ends without it being found.

Listing~\ref{lst:stl_mod} shows the corresponding iterator implementation in \name. Much of the implementation is unchanged, with minor restructuring for \code{init()}, \code{next()}, and \code{end()} functions. The main changes are --- how the state (the search key) is exchanged across the three functions and how the data is returned back to the user via the \code{scratch\_pad} (an error message if the key is not found, or its value if it is).   

\paragraphb{Ported Data Structures} While the \name programming model applies to various programming languages that support iterator interfaces, we have restricted our focus to C++ data structure libraries due to their widespread use. We have applied the \name programming model to 13 commonly-used data structures found in popular libraries such as STL~\cite{stl}, Boost~\cite{boost}, and Google Btree~\cite{google-btree} (Table \ref{table:extra-apps-2}); we defer a comprehensive description of their details to Appendix.

\section{Accelerating Pointer Traversals on a Node}
\label{sec:accelerator}

\subsection{\name Dispatch Engine}\label{ssec:compute_node}
The dispatch engine is a software framework running at the CPU node for two purposes.  First, it translates the iterator realization for pointer traversal provided by a data structure library developer (\S\ref{sec:interface}) into \name's ISA. Second, it determines if the accelerator can support the computations performed during the traversal, and if so, ships a request to the accelerator at the memory node. If not, the execution proceeds at the CPU node with regular remote memory accesses.

\paragraphb{Translating iterator code to \name ISA} To be readily implementable, \name plugs into existing compiler toolchains. The dispatch engine generates \name ISA instructions using widely known compiler techniques~\cite{llvm}. 
\name's ISA is a stripped-down RISC ISA, only containing operations necessary for basic processing and memory accesses to enable a simple and energy-efficient accelerator design (Table~\ref{tab:isa}). There are, however, a few notable aspects to our adapted ISA and the translation of iterator code to it. First, as noted in \S\ref{ssec:iterators}, \name does not support unbounded loops within a single iteration, \ie, the ISA only supports conditional jumps to points ahead in code. This is similar to eBPF programs~\cite{ebpfjump}, where only forward jumps are supported to prevent the program from running infinitely within the kernel. A backward jump can only occur when the next iteration starts; \name employs a special \code{NEXT\_ITER} instruction to explicitly mark this point so that the accelerator can begin scheduling the memory pipeline (\S\ref{ssec:architecture}). Second, again as noted in \S\ref{ssec:iterators}, developers can maintain state and return values using a \code{scratch\_pad} of pre-configured size; our ISA supports register operations directly on the \code{scratch\_pad} and provides special \code{RETURN} instruction that simply terminates the iterator execution and yields the contents of the \code{scratch\_pad} as the return value. Lastly, if the code cannot be compiled to the PULSE ISA --- \eg, if it involves compute-heavy or non-memory-centric tasks --- it will not be offloaded to the accelerator. Instead, such code will run on the CPU, potentially accessing memory remotely over the network (\eg, via RDMA or CXL). This design ensures that only tasks that benefit from near-memory execution are offloaded, adhering to our design philosophy of only offloading memory-bound operations.

Finally, we found that the iterator traversal pattern typically can be broken down into two types of computation --- fetching data\footnote{While the rest of the section focuses only on describing data fetches from memory, we note that writing data to memory proceeds similarly.} pointed to by \code{cur\_ptr} from memory, and processing the fetched data to determine what the next pointer should be, or if the iterator execution should terminate. If the translation from the iterator code to \name's ISA is done naively, it can result in multiple unnecessary loads within the vicinity of the memory location pointed to by \code{cur\_ptr}. For instance, the \code{unordered\_map::find()} realization shown in Listing~\ref{lst:stl_mod} makes references to \code{cur\_ptr->key}, \code{cur\_ptr->value}, and \code{cur\_ptr->next} at various points, and if each incurs a separate load, it will slow down execution and waste memory bandwidth. Consequently, \name's dispatch engine \emph{infers} the range of memory locations accessed relative to \code{cur\_ptr} in the \code{next()} and \code{end()} functions via static analysis and aggregates these accesses into a single large \code{LOAD} (of up to 256 B) at the beginning of each iteration. 

\begin{table}[!t]
  \centering
  \footnotesize
  \begin{tabular}{l|l}
    \hline
    {\bf Library} & {\bf Data Structures} \\\hline
    \hline
    \multirow{2}{*}{STL} & List~\cite{stdlist}, Forward list~\cite{stdforwardlist}, Map~\cite{stdmap}, Multimap~\cite{stdmultimap}, \\
    & Set~\cite{stdset}, Multiset~\cite{stdmultiset} \\\cline{1-2}
    \multirow{2}{*}{Boost} & Bimap~\cite{boostbimap}, Unordered map~\cite{boostunorderedmap}, Unordered set~\cite{boostunorderedset} \\ 
    & AVL tree~\cite{boostavltree}, Splay tree~\cite{boostsplaytree}, Scapegoat tree~\cite{boostscapegoattree},\\\cline{1-2}
    Google & Btree~\cite{google-btree} \\
    \hline
  \end{tabular}
  \captionof{table}{\textbf{Data structures implemented in \name (\S\ref{sec:interface}).}}
  \label{table:extra-apps-2}
\end{table}

\begin{table}[btp!]
    \centering
    \footnotesize  
    \def\arraystretch{0.98}%
    \begin{tabular}{l|l|l}
      \hline
      \textbf{Class}  & \textbf{Instructions} & \textbf{Description}\\\hline\hline
      Memory  & \smallcode{LOAD}, \smallcode{STORE} & \specialcell{Load/store data\\ from/to address.} \\  \hline
      ALU & \specialcell{\smallcode{ADD}, \smallcode{SUB}, \smallcode{MUL}, \smallcode{DIV},\\ \smallcode{AND}, \smallcode{OR}, \smallcode{NOT}} & Standard ALU operations. \\ \hline
      Register & \smallcode{MOVE} & Move data b/w registers.\\ \hline
      Branch  & \specialcell{\smallcode{COMPARE} and\\ \smallcode{JUMP\_}\{\smallcode{EQ}, \smallcode{NEQ}, \smallcode{LT}, ...\}} & \specialcell{Compare values \& jump\\ ahead based on condition\\ (\eg, equal, less than, \etc).}\\ \hline
      Terminal & \smallcode{RETURN}, \smallcode{NEXT\_ITER} & \specialcell{End traversal \& return,\\ or start next iteration.} \\
     \hline\hline
    \end{tabular}
    \caption{\textbf{\name adapts a restricted subset of RISC-V ISA} (\S\ref{ssec:compute_node}).}
    \label{tab:isa}
\end{table}

\paragraphb{Bounding complexity of offloaded code} While \name's interface and ISA already limit the \emph{types} of computation than can be performed per iteration, \name also needs to limit the \emph{amount} of computation per iteration to ensure the operations offloaded to \name accelerators remain memory-centric. To this end, \name's dispatch engine analyzes the generated ISA for the iterator to determine the time required to execute computational logic ($t_c$) and the time required to perform the single data load at the beginning of the iteration ($t_d$).
\name exploits the known execution time of its accelerators in terms of time per compute instruction, $t_i$, to determine $t_c = t_i \cdot N$, where $N$ is the number of instructions per iteration. The CPU node offloads the iterator execution only if $t_c \leq \eta \cdot t_d$, where $\eta$ is a predefined accelerator-specific threshold. Note that since we only want to offload memory-centric operations, $\eta \leq 1$. As we will show in \S\ref{ssec:architecture}, the choice of $\eta$ allows \name to maximize the memory bandwidth utilization and ensure processing never becomes a bottleneck for pointer traversals.

\paragraphb{Issuing network requests to accelerator} Once the dispatch engine decides to offload an iterator execution, it encapsulates the ISA instructions (\code{code}) along with the initial value of \code{cur\_ptr} and \code{scratch\_pad} (initialized by \code{init()}) into a network request. It issues the request, leaving the network to determine which memory node it should be forwarded to (\S\ref{sec:distributed}). To recover from packet drops, the dispatch engine embeds a request ID with the CPU node ID and a local request counter in the request packets, maintains a timer per request, and transparently retransmits requests on timeout. 

\paragraphb{Practical deployability} Our software stack is readily deployable due to its use of real-world toolchains. Our user library adapts implementations of common data structures used in key-value stores~\cite{redis, memcached}, databases~\cite{mongodb, btree1, btree2, trie1, trie3}, and big-data analytics~\cite{powergraph, graphx, graphchi, pagerank} to \name's iterator interface (\S\ref{sec:interface}). \name's dispatch engine is implemented on Intel DPDK-based~\cite{dpdk} low-latency, high-throughput UDP stack. \name compiler adapts the Sparc backend of LLVM~\cite{llvmsparc} since its ISA is close to \name's ISA. Our LLVM frontend applies a set of analysis and optimization passes~\cite{llvmpass} to enforce \name constraints and semantics: the analysis pass identifies code snippets that require offloading, while the optimization pass translates pointer traversal code to \name ISA.

\subsection{\name Accelerator Design}
\label{ssec:architecture}
\label{ssec:traversalexample}

\begin{figure}[t]
  \centering
  \includegraphics[width=0.9\columnwidth]{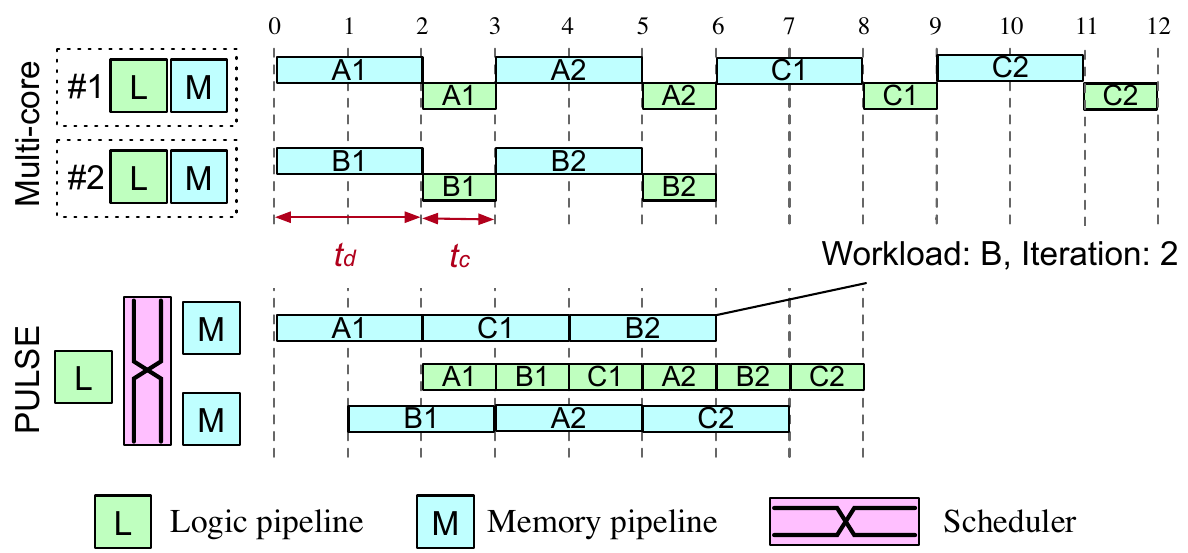}
  \vspace{-0.5em}
  \caption{\textbf{\name accelerator architecture.} (top) Traditional multi-core architectures with tightly coupled logic and memory pipelines result in low utilization and longer execution times. (bottom) \name accelerator's \emph{disaggregated} design with an unequal number of logic and memory pipelines efficiently multiplexes concurrent iterator executions across them for near-optimal utilization and performance.}
  \label{fig:architecture_overview}
\end{figure}

The accelerator is at the heart of \name design and is key to ensuring high performance for iterator executions with high resource and energy efficiency. Our motivation for a new accelerator design stems from two unique properties of iterator executions on linked structures: 

\begin{itemize}[leftmargin=*, itemsep=0pt]
  \item \textbf{Property 1:} Each iteration involves two clearly separated but sequentially dependent steps: (i) fetching data from memory via a pointer (\eg, a list or tree node), followed by (ii) executing logic on the fetched data to identify the next pointer. The logic cannot be executed concurrently with or before the data fetch, and the next data fetch cannot be performed until the logic execution yields the next pointer.
  \item \textbf{Property 2:} Iterators that benefit from offload spend more time in data fetch ($t_d$) than logic execution ($t_c$), \ie, $t_c < \eta \cdot t_d$, where $\eta \leq 1$, as noted in \S\ref{ssec:compute_node}.  
\end{itemize}
\noindent
Any accelerator for iterator executions must have a \emph{memory pipeline} and a \emph{logic pipeline} to support the execution steps (i) and (ii) above. 
The strict dependency between the steps (Property 1) renders many optimizations of traditional multi-core processors, such as out-of-order execution, ineffective. Moreover, since each core in such architectures has tightly coupled logic and memory pipelines, the memory-intensive nature of iterators (Property 2) results in the logic pipeline remaining idle most of the time. These two factors combined result in poor utilization and energy efficiency for such architectures. Fig.~\ref{fig:architecture_overview}~(top) captures this through the execution of 3 iterators (A, B, C), each with $2$ iterations (\eg, A1, A2, etc.), on a multi-core architecture. Since each iteration comprises a data fetch followed by a dependent logic execution, one of the pipelines remains idle while the other is busy. While thread-level parallelism permits iterator requests to be spread across multiple cores for increased overall throughput, per-core under-utilization of logic and memory pipelines persists, resulting in suboptimal resource and energy usage.

\paragraphb{Disaggregated accelerator design} Motivated by the unique properties of iterators, we propose a novel accelerator architecture that \emph{disaggregates memory and logic pipelines}, using a scheduler to multiplex corresponding components of iterators across them. First, such a decoupling permits an asymmetric number of logic and memory pipelines to maximize the utilization of either pipeline, in stark contrast to the tight coupling in multi-core architectures. In our design, if there are $m$ logic and $n$ memory pipelines, then the accelerator-specific threshold $\eta < 1$ we alluded to in  \S\ref{ssec:compute_node} is $\frac{m}{n}$, \ie, there are fewer logic pipelines than memory pipelines in keeping with Property 2. Fig.~\ref{fig:architecture_overview}~(bottom) shows an example of our disaggregated accelerator design with one logic pipeline and two memory pipelines (\ie, $m=1, n=2$). 

Even though data fetch and logic execution within each iterator must be sequential, the disaggregated design permits efficient multiplexing of data fetch and logic execution from different iterators across the disaggregated logic and memory pipelines to maximize utilization. To see how, recall that the logic execution time $t_c$ for each offloaded iterator execution in \name is $\leq\eta\cdot t_d$, where $t_d$ is its data fetch time (\S\ref{ssec:compute_node}). Consider the extreme case where $t_c=\eta \cdot t_d$ for all offloaded iterator executions --- in this case, it is always possible to multiplex $m+n$ concurrent iterator executions to fully utilize all $m$ logic and $n$ memory pipelines. While we omit a theoretical proof for brevity, Fig.~\ref{fig:architecture_overview}~(bottom) illustrates the multiplexed execution --- orchestrated by a scheduler in our accelerator --- for $t_c=\frac{1}{2}\cdot t_d$ with $3$ iterators. This is the ideal case --- similar multiplexing is still possible if $t_c\leq\eta\cdot t_d$ with complete utilization of memory pipelines, albeit with lower utilization of logic pipelines (since they will be idle for $\frac{t_c - \eta\cdot t_d}{t_c}$ fraction of time). As such, we provision $\eta=\frac{m}{n}$ to be as close to the expected $\frac{t_c}{t_d}$ for the workload to maximize the utilization of logic pipelines. It is possible to improve the logic pipelines' energy efficiency by dynamically down-scaling frequency~\cite{daepowerscaling}; we leave such optimizations to future work.

While the memory pipeline is stateless, the logic pipeline must maintain the state for the iterator it executes. To multiplex several iterator executions, logic pipelines need efficient mechanisms for efficient context switching. To this end, we maintain a dedicated \emph{workspace} corresponding to each iterator's execution. Each workspace stores three distinct pieces of state: \code{cur\_ptr} and \code{scratch\_pad} to track the iterator state described in \S\ref{ssec:iterators}, and \code{data}, which holds the data loaded from memory for \code{cur\_ptr}. A dedicated workspace per iterator allows the logic pipeline to switch to any iterator's execution without delay when triggered by the scheduler, although it requires maintaining multiple workspaces --- a maximum of $m+n$ to accommodate any possible schedule due to our bound on the number of concurrent iterators. We divide these workspaces equally across logic pipelines.

\begin{figure}[t]
\centering
  \includegraphics[width=\columnwidth]{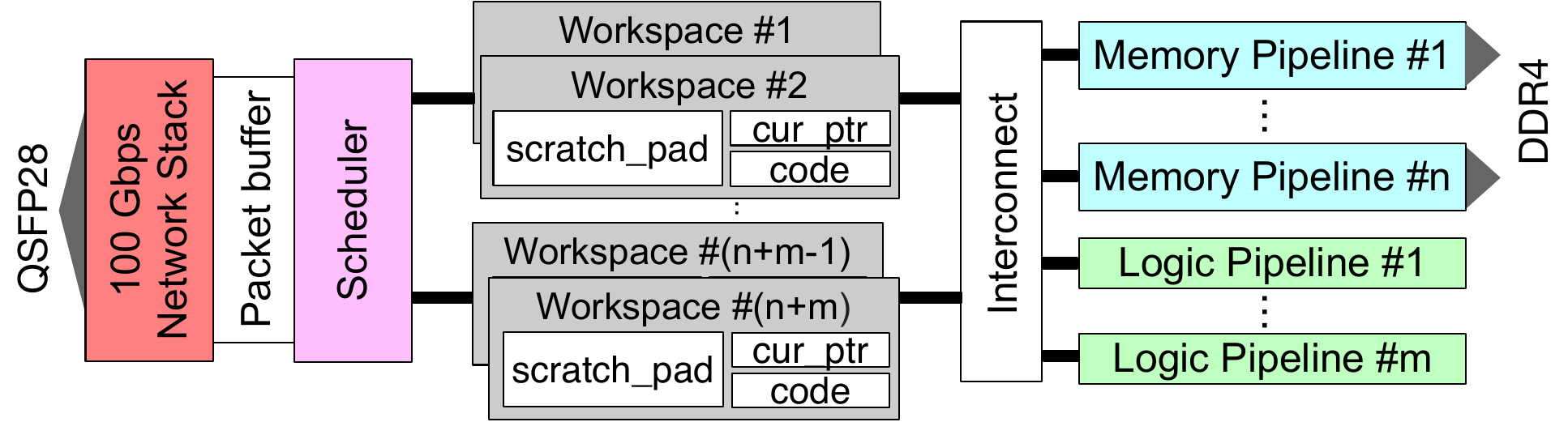}
  \vspace{-2em}
 \caption{\textbf{\name accelerator overview.} See \S\ref{ssec:architecture} for details.}
\label{fig:accelnew}
\end{figure}

\paragraphc{Memory pipeline:} Each memory pipeline loads data from the attached DRAM 
to the corresponding workspace assigned by the scheduler at the start of each iteration. This involves (i) address translation and (ii) memory protection based on page access permissions. We realize range-based address translations (simulated in prior work~\cite{range}) in our real-world implementation using TCAM to reduce on-chip storage usage. 

Once a memory access is complete, the memory pipeline signals the scheduler to continue the iterator execution or terminate it if there is a translation or protection failure.

\paragraphc{Logic pipeline:} Each logic pipeline runs \name ISA instructions other than \code{LOAD}/\code{STORE} to determine the \code{cur\_ptr} value for the next iteration or, to determine if the termination condition has been met. Our logic pipeline comprises an ALU to execute the standard arithmetic and logic instructions, as well as modules to support register manipulation, branching, and the specialized \code{RETURN} instruction execution (Table~\ref{tab:isa}). During a particular iterator's execution, the logic pipeline performs its corresponding instructions with direct reads and updates to its dedicated workspace registers. An iteration's logic can end in one of two possible ways: (i) the \code{cur\_ptr} has been updated to the next pointer, and the \code{NEXT\_ITER} instruction is reached, or (ii) the pointer traversal is complete, and the \code{RETURN} instruction is reached. In either case, the logic pipeline notifies the scheduler with the appropriate signal.

\paragraphc{Scheduler:} The scheduler handles new iterator requests received over the network and schedules each iterator's data fetch and logic execution across memory and logic pipelines: 
\begin{enumerate}[leftmargin=*, itemsep=0pt]
  \item On receiving a new request over the network, it assigns the iterator an empty workspace at a logic pipeline and signals one of the memory pipelines to execute the data fetch from memory based on the state in the workspace.\label{signal:1}
  \item On receiving a signal from the memory pipeline that a data fetch has successfully completed, it notifies the appropriate logic pipeline to continue iterator execution via the corresponding workspace.
  \item On receiving a signal from the logic pipeline that the next iteration can be started (via the \code{NEXT\_ITER} instruction), it notifies one of the memory pipelines to execute \code{LOAD} via the corresponding workspace.\label{signal:2}
  \item When it receives a signal from the memory pipeline that an address translation or memory protection failed or a signal from the logic pipeline that the iterator execution has met its terminal condition (via the \code{RETURN} instruction), it signals the network stack to prepare a response containing the iterator \code{code}, \code{cur\_ptr} and \code{scratch\_pad}.
\end{enumerate}
\noindent
While the scheduler assigns memory and logic pipelines to an iterator in steps~\ref{signal:1} and~\ref{signal:2} in a manner that maximizes utilization of all memory pipelines (\ie, Fig.~\ref{fig:architecture_overview}~(bottom)), it is possible to implement other scheduling policies.

\paragraphc{Network Stack:} The network stack receives and transmits packets; when a new request arrives, it parses/deparses the payload to extract/embed the request ID, \code{code}, and state for the offloaded iterator execution (\code{cur\_ptr}, \code{scratch\_pad}). 

The network stack uses the same format for both requests and responses, so a response can be sent back to the CPU node on traversal completion or rerouted as a request to a different memory node for continued execution (\S\ref{sec:distributed}).

\paragraphb{Implementation} We use an FPGA-based NIC (Xilinx Alveo U250) with two 100 Gbps ports, 64 GB on-board DRAM, 1,728K LUTs, and 70 Mb BRAM. Since the board has two Ethernet ports and four memory channels, we partition its resources into two \name accelerators, each with a single Ethernet port and two memory channels. Our analysis of common data structures (\S\ref{sec:evaluation}) shows their $t_c/t_d$ ratio tends to be $<0.75$. As such, we set $\eta=0.75$, \ie, there are four memory and three logic pipelines and a total of $7$ workspaces on the accelerator.
We use the Xilinx TCAM IP~\cite{tcam_ip} (for page tables), $100$ Gbps Ethernet IP, link-layer IPs~\cite{xilinx_network}, and burst data transfers~\cite{burstdatatransfer} to improve memory bandwidth. The logic and memory pipelines are clocked at 250 MHz, while the network stack operates at 322 MHz for 100 Gbps traffic. Our FPGA prototype showcases \name's potential; we believe that ASIC implementations are the next natural step. 
\section{Distributed Pointer Traversals}
\label{sec:distributed}

By restricting pointer traversals to a single memory node (\S\ref{sec:overview}), prior approaches leave applications with two undesirable options. At one extreme, they can confine their data to a single memory, but sacrifice application scalability. Conversely, they can spread their data across multiple nodes but have to return the CPU node whenever the traversal accesses a pointer on another memory node. This affords scalability but costs additional network and software processing latency at the CPU node. To avoid the cost, one may replicate the entire translation and protection state for the cluster at every memory node so they can directly forward traversal requests to other memory nodes. This comes at the cost of increased space consumption for translation, which is challenging to contain within the accelerator's translation and protection tables. Moreover, duplicating this state across memory nodes requires complex protocols for ensuring their consistency (\eg, when the state changes), which have significant performance overheads.

\begin{figure}[t]
\centering
\includegraphics[width=0.96\columnwidth]{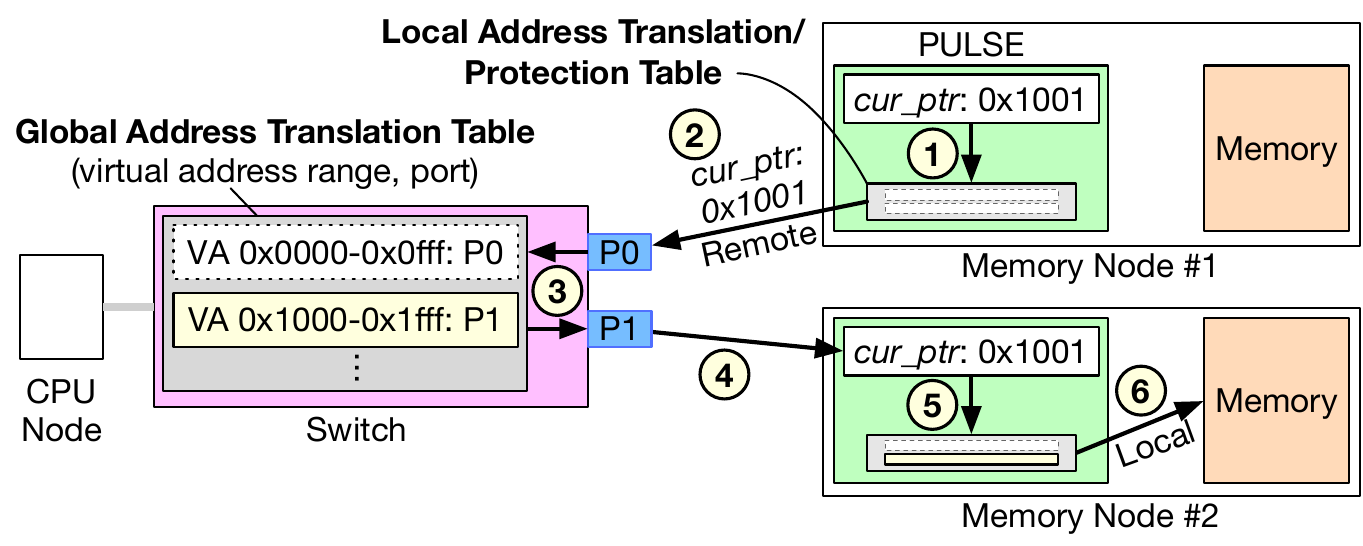}
\vspace{-1.4em}
\caption{\textbf{Hierarchical translation \& distributed traversal (\S\ref{sec:distributed}).}}
\label{fig:hierarchical}
\end{figure}

\name breaks this tradeoff between performance and scalability by leveraging a programmable network switch to support rack-scale distributed pointer traversals. In particular, if the \name accelerator on one memory node detects that the next pointer lies on a different memory node, it forwards the request to the network switch, which routes it to the appropriate memory node for continuing the traversal. This cuts the network latency by half a round trip time and avoids software overheads at the CPU node, instead performing the routing logic in switch hardware. Since continuing the traversal across memory nodes is similar to packet routing, the switch hardware is already optimized to support it.

Enabling rack-scale pointer traversals, however, requires addressing two key challenges, as we discuss next.

\paragraphb{Hierarchical translation} For the switch to forward the pointer traversal request to the appropriate memory node, it must be able to locate which memory nodes are responsible for which addresses. To minimize the logic and state maintained at the switch due to its limited resources, \name employs hierarchical address translation as shown in Fig.~\ref{fig:hierarchical}. 
In particular, the address space is range partitioned across memory nodes; \name only stores the base address to memory node mapping at the switch, while each memory node stores its own local address translation and protection metadata at the accelerator (\textcircled{1}), as outlined in \S\ref{sec:accelerator}. The routing logic at the switch inspects the \code{cur\_ptr} field in the request (\textcircled{2}) and consults its mapping to determine the target memory node (\textcircled{3}). At the memory node, the traversal proceeds until the accessed pointer is not present in the local table (as in \textcircled{1}); it then sends the request back to the switch (\S\ref{ssec:architecture}), which can re-route the request to the appropriate memory node (\textcircled{4}-\textcircled{6}), or notify the CPU node if the pointer is invalid.

\paragraphb{Continuing stateful iterator execution} One challenge of distributing iterator execution in \name lies in its stateful nature: since \name permits the storage of intermediate state in the iterator's \code{scratch\_pad}, how can such stateful iterator execution be continued on a different memory node? Fortunately, our design choices of (i) confining all of the iterator state in \code{scratch\_pad} and \code{cur\_ptr} and (ii) keeping the request and response formats identical make this straightforward. The accelerator at the memory node simply embeds the up-to-date \code{scratch\_pad} within the response before forwarding it to the switch; when the switch forwards it to the next memory node, it can simply continue execution exactly as it would have if the last memory node had the pointer.


\begin{figure*}[t]
\centering
  \includegraphics[width=0.95\textwidth]{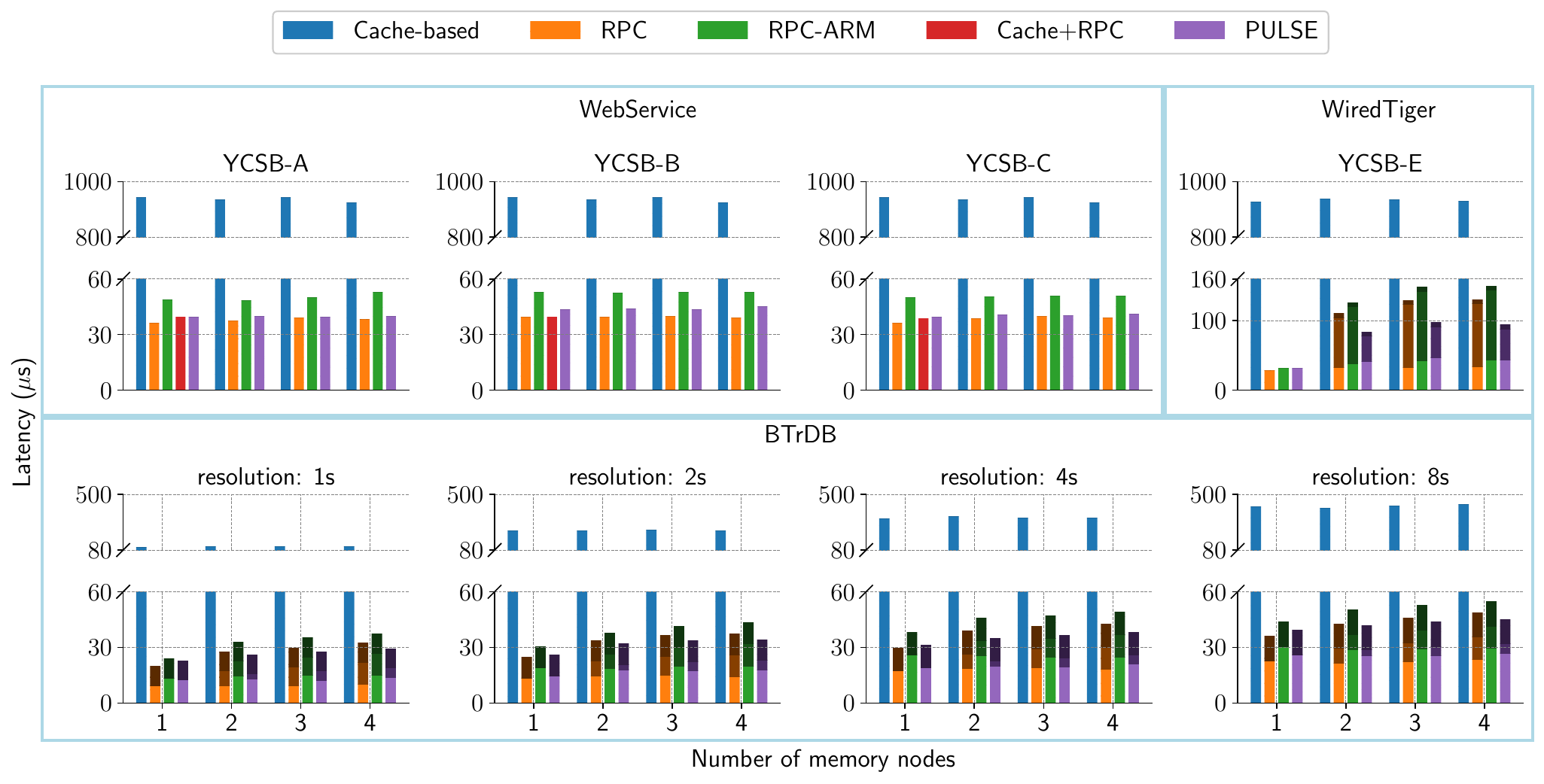}
  \\
  \includegraphics[width=0.95\textwidth]{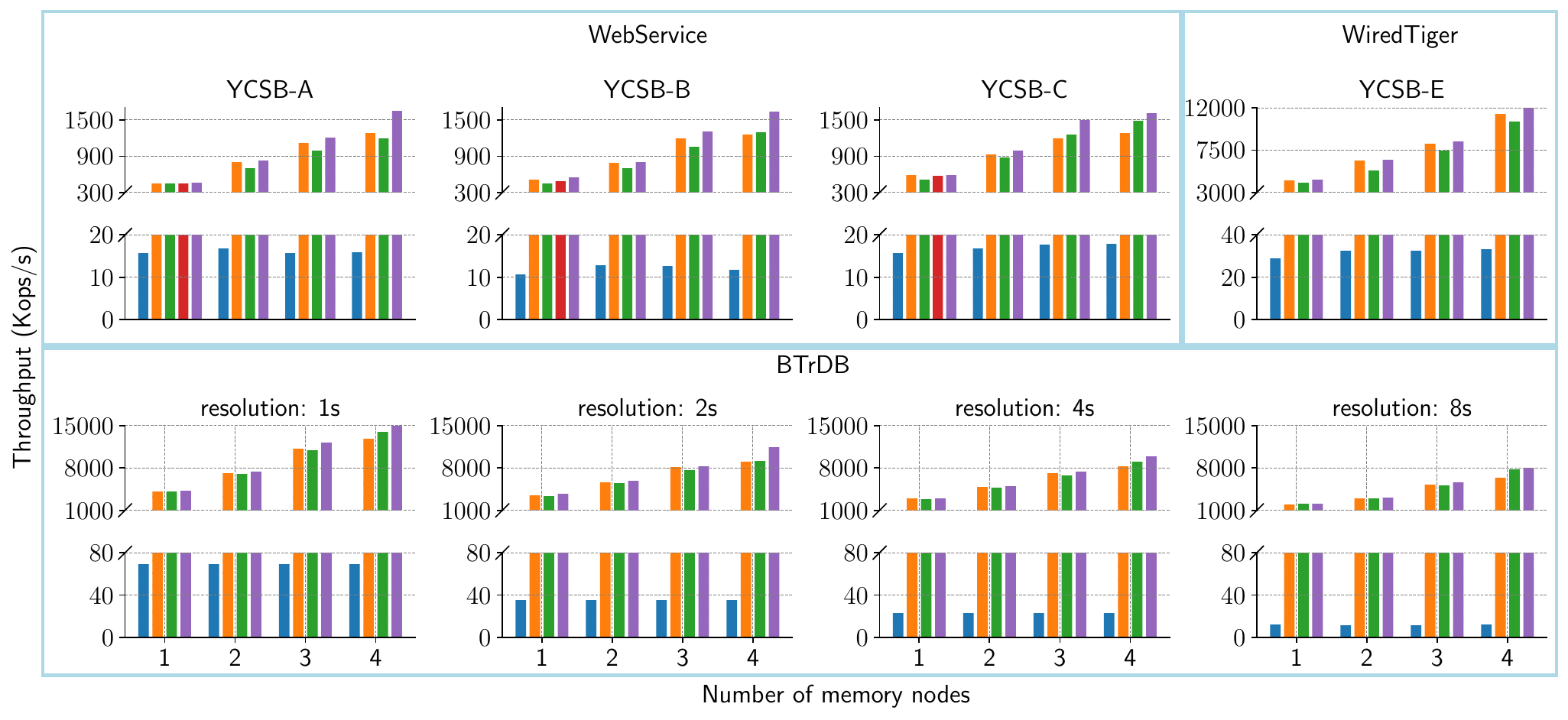}
  \vspace{-1.3em}
  \caption{\textbf{Application latency (top) \& throughput (bottom) (\S\ref{ssec:application-study}).} 
  The darker color indicates the time spent on cross-node pointer traversals, which increases with the number of memory nodes in WiredTiger and BTrDB.
  }
\label{fig:eval_perf_e2e_latency}
\label{fig:eval_perf_e2e_throughput}
\end{figure*}

\section{Evaluation}
\label{sec:evaluation}

\paragraphb{Compared systems} We compare \name against: (i) a \textbf{Cache-based} system that relies solely on caches at CPU nodes to speed up remote memory accesses; we use Fastswap~\cite{fastswap} as the representative system, (ii) an \textbf{RPC} system that offloads pointer-traversals to a CPU on memory nodes, (iii) \textbf{RPC-ARM}, an RPC system that employs a wimpy ARM processors at memory nodes, and (iv) a \textbf{Cache$+$RPC} approach that employs data structure-aware caches; we use AIFM~\cite{aifm} as the representative system. (i) and (iv) use a cache size of $2$ GB, while (ii) and (iii) use a DPDK-based RPC framework~\cite{erpc}.

\paragrapha{Our experimental setup} comprises two servers, one for the CPU node and the other for memory nodes, connected via a 32-port switch with a $6.4$ Tbps programmable Tofino ASIC. Both servers were equipped with Intel Xeon Gold 6240 Processors~\cite{intelprocessor} and $100$ Gbps Mellanox ConnectX-5 NICs. 
For a fair comparison, we limit the memory bandwidth of the memory nodes to $25$ GB/s (FPGA's peak bandwidth) using Intel Resource Director~\cite{intel_cmt_cat} and report energy consumption of the \textbf{minimum} number of CPU cores needed to saturate the bandwidth. We use Bluefield-2~\cite{bluefield} DPU as our ARM-based SmartNICs with $8$ Cortex-A72 cores and $16$ GB DRAM. For \name, we placed two memory nodes on each FPGA NIC (one per port, a total of $4$ memory nodes). Our results translate to larger setups since \name's performance or energy efficiency are independent of dataset size and cluster scale.

\begin{table}[!t]
  \centering
  \bgroup
  \small
  \def\arraystretch{0.95}%
  \begin{tabular}{l|c|c|c} 
        \hline
        \textbf{Application} & \textbf{Data Structure} & \textbf{$t_c/t_d$} & \textbf{\#Iterations} \\\hline\hline
        WebService & Hash-table & 0.06 & 48 \\\hline
        WiredTiger & \multirow{2}{*}{B+Tree} & 0.63 & 25 \\\cline{1-1}\cline{3-4}
        BTrDB ($1s$ to $8s$) & & 0.71 & $38$--$227$ \\\hline
  \end{tabular}
  \egroup
  \caption{\textbf{Workloads used in our evaluation (\S\ref{sec:evaluation}).} $t_c$ and $t_d$ correspond to compute and memory access time at the \name accelerator.} 
  \label{tab:workloads}
\end{table}

\paragraphb{Applications \& workloads} We consider $3$ applications with varying data structure complexity, compute/memory-access ratio, and iteration count per request (Table~\ref{tab:workloads}): (i) \textit{Web Service}~\cite{aifm} that processes user requests by retrieving user IDs from an in-memory hash table, using these IDs to fetch $8$ KB objects, which are then encrypted, compressed and returned to the user. Requests are generated using YCSB A (50\% read/50\% update), B (95\% read/5\% update), and C (100\% read) workloads with Zipf distribution~\cite{ycsb_workload}. (ii) \textit{WiredTiger Storage Engine} (MongoDB backend~\cite{mongodb}) uses B+Trees to index NoSQL tables. Our frontend issues range query requests over the network to WiredTiger and plots the results. Similar to prior work~\cite{aifm, xrp}, we model user queries using the YCSB E workload with Zipf distribution~\cite{ycsb_workload} on $8$ B keys and $240$ B values. (iii) \textit{BTrDB Time-series Database}~\cite{btrdb} is a database designed for visualizing patterns in time-series data. BTrDB reads the data from a B+Tree-based store for a given user query and renders the time-series data through an interactive user interface~\cite{mrplotter}. We run stateful aggregations (sum, average, min, max) for time windows of different resolutions, from $1$ s to $8$ s, on the Open $\mu$PMU Dataset~\cite{upmu} with voltage, current, and phase readings from LBNL’s power grid~\cite{btrdb}.

\subsection{Performance for Real-world Applications} 
\label{ssec:application-study}

Since AIFM~\cite{aifm} does not natively support B+-Trees or distributed execution, we restrict the Cache+RPC approach to the Web Service application on a single node.

\paragraphb{Single-node performance} Fig.~\ref{fig:eval_perf_e2e_latency} demonstrates the advantages of accelerating pointer-traversals at disaggregated memory. Compared to the Cache-based approach, \name achieves $9$--$34.4\times$ lower latency and $28$--$171\times$ higher throughput across all applications using only one network round-trip per request. RPC-based systems observe $1$--$1.4\times$ lower latency than \name due to their $9\times$ higher CPU clock rates. We believe an ASIC-based realization of \name has the potential to close or even overcome this gap. Cache$+$RPC incurs higher latency than RPC due to its TCP-based DPDK stack~\cite{ousterhout_shenango_19_nsdi, aifm} and does not outperform RPC, indicating that data structure-aware caching is not beneficial due to poor locality.

Latency depends on the number of nodes traversed during a single request and the response size. WebService experiences the highest latency due to large $8$ KB responses and long traversal length per request. In BTrDB, the latency increases (and the throughput decreases) as the window size grows due to the longer pointer traversals (see Table~\ref{tab:workloads}). Interestingly, the Cache-based approach performs significantly better for BTrDB than WebService and WiredTiger due to the better data locality in time-series analysis of chronologically ordered data. However, its throughput remains significantly lower than both \name and RPC since it is bottlenecked by the swap system performance, which could not evict pages fast enough to bring in new data. This is verified in our analysis of resource utilization (deferred to Appendix for brevity); we find that RPC, RPC-ARM, Cache$+$RPC, and \name can utilize more than 90\% of the memory bandwidth across the applications, while the Cache-based approach observes less than $1$ Gbps network bandwidth. The other systems --- \name, RPC, RPC-ARM, and Cache$+$RPC --- can also saturate available memory bandwidth (around $25$ GB/s) by offloading pointer traversals to the memory node, consuming only 0.5\%--25\% of the available network bandwidth. 

\paragraphb{Distributed pointer traversals} Fig.~\ref{fig:eval_perf_e2e_latency} shows that employing multiple memory nodes introduces two major changes in performance trends: (i) the latency increases when the pointer traversal spans multiple memory nodes, and (ii) throughput increases with the number of nodes since the systems can exploit more CPUs or accelerators. WebService is an exception to the trend: since the hash table is partitioned across memory nodes based on primary keys, the linked list for a hash bucket resides in a single memory node. 

\name observes lower latency than the compared systems due to in-network support for distributed pointer-traversals (\S\ref{sec:distributed}). The latency increases significantly from one to two memory nodes for all systems since traversing to the next pointer on a different memory node adds $5$--$10~\mu$s network latency. Also, even across two memory nodes, a request can trigger multiple inter-node pointer traversals incurring multiple network round-trips; for WiredTiger and BtrDB, $10$\%--$30$\% of pointer traversals are inter-node. However, in-network traversals allow \name to reduce latency overheads by $33$--$98$\%, with $1.1$--$1.36\times$ higher throughput than RPC.

\paragraphb{Energy consumption} We compared energy consumed per request for \name and RPC schemes at a request rate that ensured memory bandwidth was saturated for both. We measure energy consumption using Xilinx XRT~\cite{xilinx_xrt} for \name (all power rails) and Intel RAPL tools~\cite{intel_rapl} for RPC on CPUs~\cite{intelprocessor} (CPU package and DRAM only). For RPC-ARM on ARM cores, since there is no power-related performance counter~\cite{armv8registers} or open-source tool available, we adapt the measurement approach from prior work~\cite{clio}. Specifically, we calculate the CPU package's energy using application CPU cycle counts and DRAM power using Micron's estimation tool~\cite{micron}. Finally, we conservatively estimate ASIC power using our FPGA prototype: we scale down the ASIC energy only for \name accelerator using the methodology employed in prior research~\cite{asicpower} while using the unscaled FPGA energy for other components (DRAM, third-party IPs, \etc). As such, we measure an \emph{upper bound} on \name and \nameasic energy use, and a \emph{lower bound} for RPC, RPC-ARM, and Cache+RPC.

\begin{figure}[t]
\centering
\includegraphics[width=0.48\textwidth]{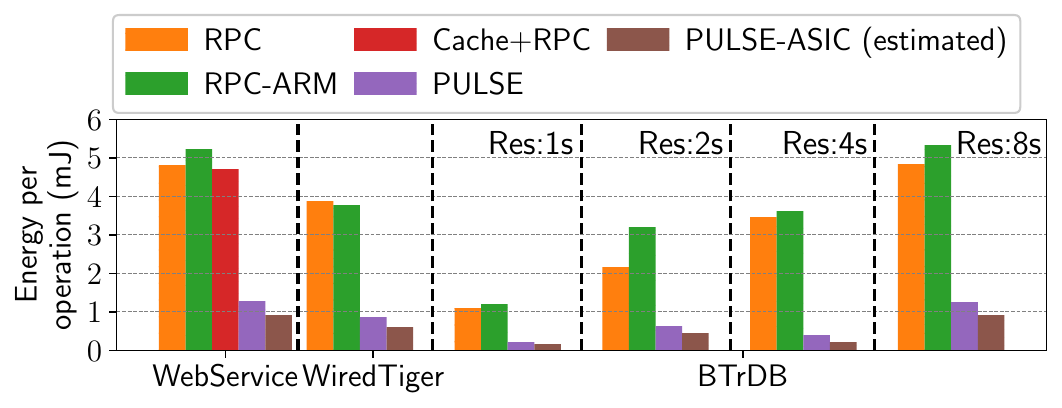}
\vspace{-2.5em}
\caption{\textbf{Application energy consumption per operation (\S\ref{ssec:application-study}).} }
\label{fig:eval_energy}
\end{figure}

Fig.~\ref{fig:eval_energy} shows that \name achieves a $4.5$--$5\times$ reduction in energy use per operation compared to RPCs on a general-purpose CPU, due to its disaggregated architecture (\S\ref{ssec:architecture}). Our estimation shows that \name's ASIC realization can conservatively reduce energy use by an additional $6.3-7\times$ factor. 
Finally, RPC-ARM's total energy consumption per request can exceed that of standard cores, as seen in the WebService workload. This observation aligns with prior studies~\cite{clio}, which attribute the increased energy use to their longer execution times, resulting in higher aggregate energy demands.

\begin{figure}[t]
\centering
\vspace{-1.5em}
\includegraphics[width=0.48\textwidth]{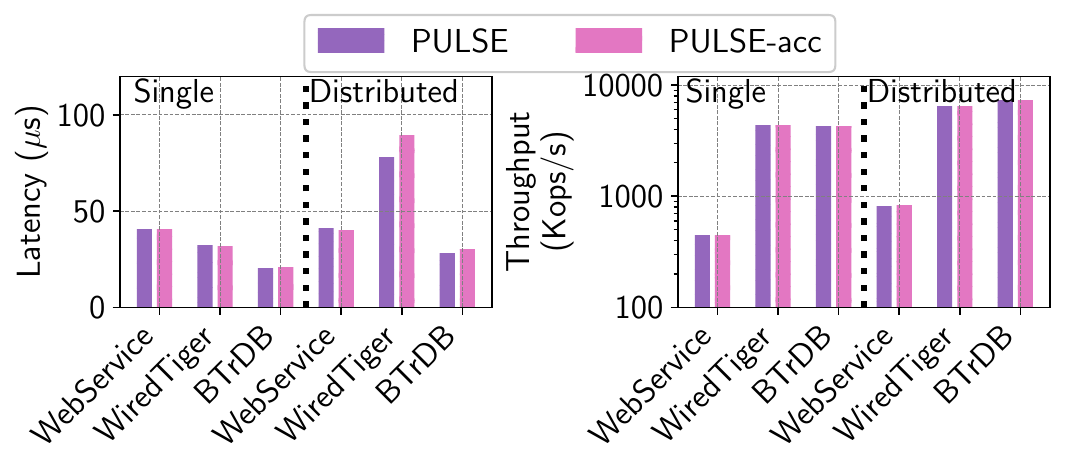}%
\vspace{-1.5em}
\caption{\textbf{Impact of distributed pointer traversals (\S\ref{ssec:breakdown}).} 
}
\label{fig:eval_breakdown}
\end{figure}

\begin{figure}[t]
  \centering	
  \vspace{-1em}
  \includegraphics[width=0.48\textwidth]{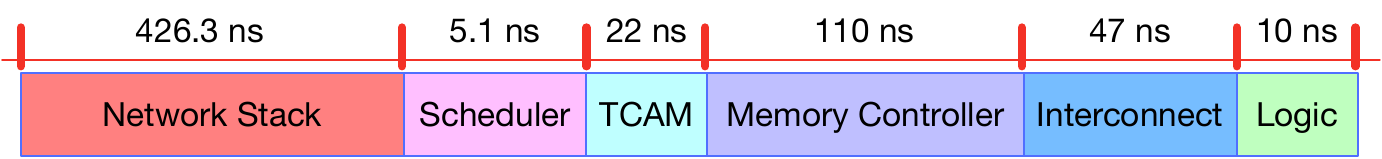}
  \vspace{-2em}
  \caption{\textbf{Latency breakdown for \name accelerator (\S\ref{ssec:breakdown}).}}
  \label{fig:eval_breakdown_latency_}
\end{figure}

\subsection{Understanding \name Performance}
\label{ssec:breakdown}

\paragraphb{Distributed pointer traversals} We evaluate the impact of distributed pointer traversals (\S\ref{sec:distributed}) by comparing \name against \nameacc, a \name variant that sends requests back to the CPU node if the next pointer is not found on the memory node. Fig.~\ref{fig:eval_breakdown} shows that while both have identical performance on a single memory node, \nameacc observes $1.02$--$1.15\times$ higher latency for two nodes. On the other hand, their throughput is the same since, under sufficient load, memory node bandwidth bottlenecks the system for both.

\paragraphb{Latency breakdown for \name accelerator} 
Fig.~\ref{fig:eval_breakdown_latency_} shows the latency contributions of various hardware components at the \name accelerator for the WebService application. The network stack first processes the pointer traversal request in about $430$ ns, after which the WebService payload is processed by the scheduler and dispatched to an idle memory access pipeline in $5.1$ ns. Then, the memory pipeline takes $\sim$$132$ ns to perform address translation, memory protection, and data fetch from DRAM. Finally, the logic pipeline takes $10$ ns to check the termination conditions and determine the next pointer to look up. This process repeats until the termination condition is met. The time to send a response back over the network stack is symmetric to the request path.

\begin{table}[t]
\centering
\scriptsize
\begin{tabularx}{\columnwidth}{@{}c|c|c|c|c|c|c@{}} 
\hline
& \specialcell{\textbf{\#Logic}\\\textbf{Pipelines}} & \specialcell{\textbf{\#Memory}\\\textbf{Pipelines}}& \textbf{LUT \%} & \textbf{BRAM \%} & \specialcell{\textbf{Throughput}\\\textbf{(Mops/s)}} & \specialcell{\textbf{Latency}\\\textbf{(us)}} \\
\hline
\multirow{4}{*}{\rotatebox[origin=c]{90}{Coupled}} & 1 & 1   & 7.37  & 7.29   & 0.41 & 33.25 \\
& 2 & 2  & 10.23 & 9.37   & 0.63 (+53\%) & 33.73 \\
& 3 & 3  & 14.33 & 15.92  & 0.87 (+112\%) & 34.66 \\
& 4 & 4   & 18.55 & 17.09  & 1.20 (+193\%) & 35.11 \\\hline
\multirow{10}{*}{\rotatebox[origin=c]{90}{\name}} &  1 & 1  & 5.88  & 8.17   & 0.51 & 37.57 \\
& 1 & 2  & 7.44  & 9.14   & 0.73 (+43\%) & 36.74 \\
& 1 & 3  & 8.32  & 11.19  & 1.01 (+98\%) & 38.46 \\
\rowcolor{yellow} & \textbf{1} & \textbf{4}  & \textbf{9.19}  & \textbf{12.92}  & \textbf{1.24 (+143\%)} & \textbf{38.37} \\
& 2 & 1  & 8.87  & 10.19  & 0.48 (-6\%) & 40.27 \\
& 2 & 2  & 10.69 & 11.19  & 0.76 (+49\%) & 39.47 \\
& 2 & 3  & 13.11 & 13.38  & 0.99 (+94\%) & 41.37 \\
& 2 & 4  & 15.07 & 15.61  & 1.19 (+133\%) & 40.37 \\
& 3 & 1  & 14.08 & 11.93  & 0.46 (-10\%) & 42.38 \\
& 3 & 2  & 15.79 & 13.78  & 0.69 (+35\%) & 43.11 \\
& 3 & 3  & 18.61 & 15.06  & 1.03 (+102\%) & 40.98 \\
& 3 & 4  & 19.20 & 17.47  & 1.17 (+129\%) & 44.02 \\
& 4 & 1  & 18.67 & 14.17  & 0.37 (-27\%) & 42.16 \\
& 4 & 2  & 20.37 & 16.02  & 0.51 (0\%) & 43.00 \\
& 4 & 3  & 22.08 & 17.86  & 1.10 (+116\%) & 43.86 \\
& 4 & 4  & 23.21 & 19.92  & 1.14 (+123\%) & 41.47 \\
\hline
\end{tabularx}
\caption{\textbf{Coupled (multi-core) vs. \name's disaggregated architecture (\S\ref{ssec:breakdown}). The highlighted configuration depicts \name's Pareto-optimal resource usage and performance.} }
\label{table:sup_architecture}
\vspace{-1em}
\end{table}

\begin{figure}[t]
  \includegraphics[width=0.75\columnwidth]{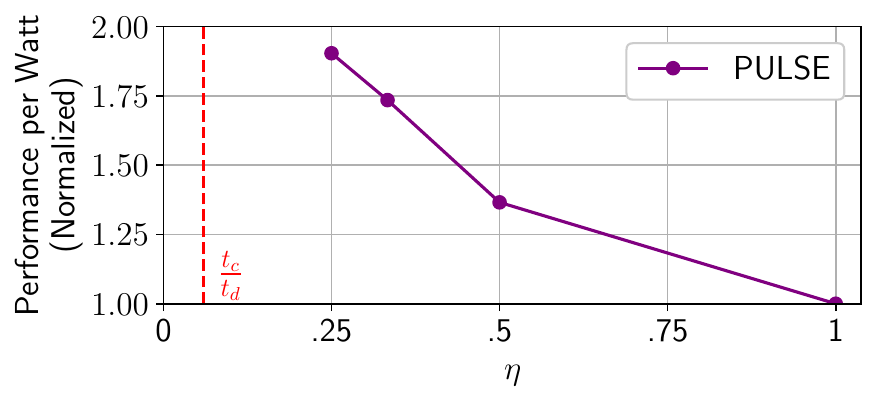}
  \vspace{-1.2em}
  \caption{\textbf{\name sensitivity to $\eta$ (\S\ref{ssec:breakdown}).} Performance-per-watt is normalized by the value at $\eta=1$.}
  \label{fig:eval_sensitivity_eta}
\end{figure}

\paragraphb{Benefits of disaggregating memory and logic pipelines} 
Table~\ref{table:sup_architecture} compares the area usage (percentage of BRAM/LUT resources used on FPGA) and performance (throughput/latency for the WebService application) of \name's disaggregated and traditional coupled (\ie, multi-core) designs, which combines logic and memory pipelines into cores. \name requires slightly more area than the coupled design when the number of logic and memory pipelines are equal to accommodate additional logic and buffers across the interconnect and the scheduler. However, due to the memory-intensive nature of pointer traversal operations (\S\ref{sec:accelerator}), \name can achieve similar performance with fewer logic pipelines and, therefore, less area. To saturate memory bandwidth (and thus maximize throughput) for the WebService application, \name only needs one logic pipeline and four memory pipelines, while a traditional core architecture must use four cores. As such, \name saves 38\% area with a marginal 7\% latency increase due to scheduling and workspace overheads.

\paragraphb{Sensitivity to $\eta$ Parameter} 
We evaluate \name’s sensitivity to $\eta$ by varying the number of memory pipelines with a single logic pipeline for the WebService application. Figure~\ref{fig:eval_sensitivity_eta} shows that as $\eta$ in \name accelerator approaches the workload's compute-to-memory ratio ($\sim 1/16$), its performance per watt improves significantly since the accelerator resources better match the workload needs. In contrast, large $\eta$ values lead to underutilization of the logic pipeline and, thus, wasted energy. For instance, decreasing the $\eta$ value from $1$ to $1/4$ increases the performance-per-watt by $1.9\times$!

\begin{figure}[t]
\centering
\vspace{-1em}
\includegraphics[width=0.48\textwidth]{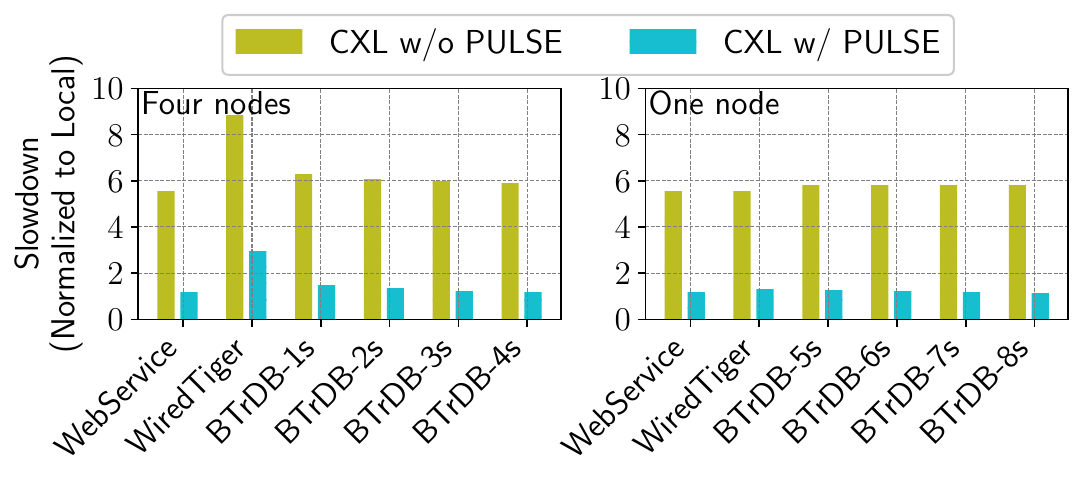}
\vspace{-2.5em}
\caption{
\textbf{Slowdown with simulated CXL interconnect (\S\ref{sec:future}).} 
}
\label{fig:eval_cxl}
\end{figure}

\section{Future Trends and Research}
\label{sec:future}

\paragraphb{Next-generation interconnects} While \name is implemented atop Ethernet, its design is interconnect-agnostic. It could be realized over emerging interconnects like CXL~\cite{cxl, cxl_azure, sun2023demystifying}. We have verified these benefits in simulation atop traces of our evaluated applications. Our simulator used $2$ GB DRAM as CPU-attached cache, while the entire working set is stored on CXL-attached memory. Following prior work~\cite{pond}, we model $10$--$20$ ns L3 cache latency, $80$ ns DRAM latency, $300$ ns CXL-attached memory latency, and $256$ B access granularity. We simulate both a four-memory-node setup, which uses a CXL switch with \name logic and a \name accelerator at each memory node, and a single-node setup with no switch. We assume a conservative overhead for \name, using our Ethernet switch and FPGA latencies.

Fig.~\ref{fig:eval_cxl} shows the slowdown of workloads on CXL memory versus local DRAM, both with and without \name. \name reduces CXL’s slowdown by $3$--$5\times$ in the four-node setup, and by $4.2$--$5.2\times$ in the single-node setup. While real hardware realization is necessary to quantify \name's benefits precisely, our simulation (with optimistic CXL latency and conservative \name overheads) highlights the potential for improving performance in such interconnects.

\paragraphb{Data encryption in memory} With increasing focus on trusted server infrastructure for secure cloud, an interesting avenue of future research is enabling near-memory processing over encrypted disaggregated memory. We identify two critical challenges. The first involves managing encryption keys securely, especially since the \name accelerator, an intermediary, could be compromised. We argue for incorporating a Trusted Execution Environment (TEE) in \name, similar to prior FPGA systems that isolate sensitive key management functions~\cite{fpgatrusted1, fpgatrusted2, fpgatrusted3, fpgatrusted4, fpgatrusted5}. The second challenge involves hiding memory access patterns as a side channel over encrypted memory~\cite{islam2012access, kellaris2016generic, leakage-attacks, kornaropoulos2019recovery}. While several recent advances in noise injection techniques permit efficient defense mechanisms against side-channel attacks~\cite{pancake, shortstack, lengthleakage}, developing performant solutions in hardware remains an open problem.

\section{Related work}
\label{sec:related}

\paragraphb{Memory disaggregation} Disaggregated memory systems span both RDMA-based~\cite{fastswap,infiniswap,legoos,mind} and CXL-based interconnects~\cite{cxl1, cxl2, cxl3, cxlperformance}. Even with gigabytes of DRAM at compute nodes, these approaches observe significant performance degradation for workloads with poor data locality when most data accesses hit slower disaggregated memory. Application-integrated disaggregated memory schemes alleviate some of the performance overheads for specific scenarios, \eg, garbage collection, key-value storage, \etc~\cite{aifm, wang2020_semeru, wang2022_memLiner,clover}, but do not generalize to other scenarios. \name aims to enable efficient execution for a large class of pointer traversal workloads by placing general but lightweight processing primitives close to the memory nodes.

\paragraphb{Near-memory processing}
Limited data bandwidth between compute and storage devices and the high cost of data movement are well-documented in the architecture community. Several works have proposed hardware architectures that move computations close to storage or memory~\cite{boroumand2019_codna,cho2020_data,devic2022_PIM,wang2022_Nearstream}, albeit with limited flexibility for expressing offloaded logic. Another related class of approaches has targeted graph-processing accelerators for machine learning workloads~\cite{ke2020_RecNMP,kwon2019_TensorDIMM} --- these still suffer from limited expressiveness for general data structures. Recent efforts in the industry have also explored placing accelerators at or close to memory devices~\cite{intel_dsa, samsung_pim}. Unfortunately, most approaches require micro-architectural modifications to enable near- or on-memory processing. \name instead focuses on leveraging programmable networks for accelerating pointer traversals on linked structures in disaggregated architectures.

\paragraphb{Prefetching} As noted in \S\ref{ssec:prior}, while prefetching~\cite{prefetch1,prefetch2,prefetch3} can be used to pipeline remote memory accesses during pointer traversals over disaggregated memory, its benefits are limited in for pointer traversals. However, prefetching and \name's approach of near-memory processing are orthogonal. Indeed, \name \emph{complements} prefetching by enhancing performance for workloads where the effectiveness of prefetching techniques is limited.

\section{Conclusion}

We have designed \name to accelerate pointer traversals across linked data structures close to disaggregated memory in a manner that preserves expressiveness, ensures energy efficiency, and supports distributed execution. \name makes a principled use of near-memory acceleration, and programmable network switches for low-latency, high-throughput pointer traversals on disaggregated memory.

\section*{Acknowledgements}
We would like to thank our shepherd Zsolt István and anonymous ASPLOS reviewers for their valuable comments and insightful feedback.
This work is supported in part by NSF Awards 2047220, 2112562, 2147946, 2118851, and a NetApp Faculty Fellowship.

\begin{sloppypar}
{
%
\bibliographystyle{ACM-Reference-Format}
\bibliography{bib/abr-short,bib/paper}


\begin{thebibliography}{163}


\ifx \showCODEN    \undefined \def \showCODEN     #1{\unskip}     \fi
\ifx \showDOI      \undefined \def \showDOI       #1{#1}\fi
\ifx \showISBNx    \undefined \def \showISBNx     #1{\unskip}     \fi
\ifx \showISBNxiii \undefined \def \showISBNxiii  #1{\unskip}     \fi
\ifx \showISSN     \undefined \def \showISSN      #1{\unskip}     \fi
\ifx \showLCCN     \undefined \def \showLCCN      #1{\unskip}     \fi
\ifx \shownote     \undefined \def \shownote      #1{#1}          \fi
\ifx \showarticletitle \undefined \def \showarticletitle #1{#1}   \fi
\ifx \showURL      \undefined \def \showURL       {\relax}        \fi
\providecommand\bibfield[2]{#2}
\providecommand\bibinfo[2]{#2}
\providecommand\natexlab[1]{#1}
\providecommand\showeprint[2][]{arXiv:#2}

\bibitem[boo(2007a)]%
        {boostavltree}
 \bibinfo{year}{2007}\natexlab{a}.
\newblock \bibinfo{title}{{Boost AVL tree}}.
\newblock
  \bibinfo{howpublished}{\url{https://www.boost.org/doc/libs/1_35_0/doc/html/intrusive/avl_set_multiset.html}}.
\newblock


\bibitem[boo(2007b)]%
        {boost}
 \bibinfo{year}{2007}\natexlab{b}.
\newblock \bibinfo{title}{{Boost library}}.
\newblock \bibinfo{howpublished}{\url{https://www.boost.org/}}.
\newblock


\bibitem[boo(2007c)]%
        {boostsplaytree}
 \bibinfo{year}{2007}\natexlab{c}.
\newblock \bibinfo{title}{{Boost splay tree}}.
\newblock
  \bibinfo{howpublished}{\url{https://www.boost.org/doc/libs/1_35_0/doc/html/intrusive/splay_set_multiset.html}}.
\newblock


\bibitem[boo(2008a)]%
        {boostscapegoattree}
 \bibinfo{year}{2008}\natexlab{a}.
\newblock \bibinfo{title}{{Boost scapegoat tree}}.
\newblock
  \bibinfo{howpublished}{\url{https://www.boost.org/doc/libs/1_38_0/doc/html/intrusive/sg_set_multiset.html}}.
\newblock


\bibitem[boo(2008b)]%
        {boostunorderedmap}
 \bibinfo{year}{2008}\natexlab{b}.
\newblock \bibinfo{title}{{Boost unordered map}}.
\newblock
  \bibinfo{howpublished}{\url{https://www.boost.org/doc/libs/1_38_0/doc/html/boost/unordered_map.html}}.
\newblock


\bibitem[boo(2008c)]%
        {boostunorderedset}
 \bibinfo{year}{2008}\natexlab{c}.
\newblock \bibinfo{title}{{Boost unordered set}}.
\newblock
  \bibinfo{howpublished}{\url{https://www.boost.org/doc/libs/1_51_0/doc/html/boost/unordered_set.html}}.
\newblock


\bibitem[goo(2011)]%
        {google-btree}
 \bibinfo{year}{2011}\natexlab{}.
\newblock \bibinfo{title}{{Google BTree}}.
\newblock
  \bibinfo{howpublished}{\url{https://code.google.com/archive/p/cpp-btree/}}.
\newblock


\bibitem[bur(2013)]%
        {burstdatatransfer}
 \bibinfo{year}{2013}\natexlab{}.
\newblock \bibinfo{title}{{AMBA AXI and ACE Protocol Specification}}.
\newblock
  \bibinfo{howpublished}{\url{https://developer.arm.com/documentation/ihi0022/e/?lang=en}}.
\newblock


\bibitem[mrp(2016)]%
        {mrplotter}
 \bibinfo{year}{2016}\natexlab{}.
\newblock \bibinfo{title}{Mr. Plotter: A Multi-Resolution Plotter compatible
  with BTrDB}.
\newblock \bibinfo{howpublished}{\url{https://github.com/BTrDB/mr-plotter}}.
\newblock


\bibitem[rdm(2017)]%
        {rdmalatency}
 \bibinfo{year}{2017}\natexlab{}.
\newblock \bibinfo{title}{{RoCE vs. iWARP Competitive Analysis}}.
\newblock
  \bibinfo{howpublished}{\url{https://www.mellanox.com/related-docs/whitepapers/WP_RoCE_vs_iWARP.pdf}}.
\newblock


\bibitem[int(2019)]%
        {intelprocessor}
 \bibinfo{year}{2019}\natexlab{}.
\newblock \bibinfo{title}{{Intel Xeon Gold 6240 Processor datasheet}}.
\newblock
  \bibinfo{howpublished}{\url{https://ark.intel.com/content/www/us/en/ark/products/192443/intel-xeon-gold-6240-processor-24-75m-cache-2-60-ghz.html}}.
\newblock


\bibitem[ter(2019)]%
        {terabitethernet}
 \bibinfo{year}{2019}\natexlab{}.
\newblock \bibinfo{title}{{Terabit Ethernet: The New Hot Trend in Data
  Centers}}.
\newblock
  \bibinfo{howpublished}{\url{https://www.lanner-america.com/blog/terabit-ethernet-new-hot-trend-data-centers/}}.
\newblock


\bibitem[blu(2020)]%
        {bluefield}
 \bibinfo{year}{2020}\natexlab{}.
\newblock \bibinfo{title}{{NIVIDIA MELLANOX BLUEFIELD-2}}.
\newblock
  \bibinfo{howpublished}{\url{https://network.nvidia.com/files/doc-2020/pb-bluefield-2-smart-nic-eth.pdf}}.
\newblock


\bibitem[boo(2022)]%
        {boostbimap}
 \bibinfo{year}{2022}\natexlab{}.
\newblock \bibinfo{title}{{Boost bimap}}.
\newblock
  \bibinfo{howpublished}{\url{https://www.boost.org/doc/libs/1_80_0/libs/bimap/doc/html/index.html}}.
\newblock


\bibitem[c++(2023)]%
        {c++iterator}
 \bibinfo{year}{2023}\natexlab{}.
\newblock \bibinfo{title}{{C++ std::iterator}}.
\newblock
  \bibinfo{howpublished}{\url{https://en.cppreference.com/w/cpp/iterator/iterator}}.
\newblock


\bibitem[stl(2023)]%
        {stl}
 \bibinfo{year}{2023}\natexlab{}.
\newblock \bibinfo{title}{{Standard containers}}.
\newblock \bibinfo{howpublished}{\url{https://cplusplus.com/reference/stl/}}.
\newblock


\bibitem[arm(2024)]%
        {armv8registers}
 \bibinfo{year}{2024}\natexlab{}.
\newblock \bibinfo{title}{{AArch64 Performance Monitors registers}}.
\newblock \bibinfo{howpublished}{\url{
  https://developer.arm.com/documentation/100095/0002/system-control/aarch64-register-summary/aarch64-performance-monitors-registers}}.
\newblock


\bibitem[std(2024a)]%
        {stdforwardlist}
 \bibinfo{year}{2024}\natexlab{a}.
\newblock \bibinfo{title}{{C++ standard forward\_list container}}.
\newblock
  \bibinfo{howpublished}{\url{https://en.cppreference.com/w/cpp/container/forward_list}}.
\newblock


\bibitem[std(2024b)]%
        {stdlist}
 \bibinfo{year}{2024}\natexlab{b}.
\newblock \bibinfo{title}{{C++ standard list container}}.
\newblock
  \bibinfo{howpublished}{\url{https://en.cppreference.com/w/cpp/container/list}}.
\newblock


\bibitem[std(2024c)]%
        {stdmap}
 \bibinfo{year}{2024}\natexlab{c}.
\newblock \bibinfo{title}{{C++ standard map container}}.
\newblock
  \bibinfo{howpublished}{\url{https://en.cppreference.com/w/cpp/container/map}}.
\newblock


\bibitem[std(2024d)]%
        {stdmultimap}
 \bibinfo{year}{2024}\natexlab{d}.
\newblock \bibinfo{title}{{C++ standard multimap container}}.
\newblock
  \bibinfo{howpublished}{\url{https://en.cppreference.com/w/cpp/container/multimap}}.
\newblock


\bibitem[std(2024e)]%
        {stdmultiset}
 \bibinfo{year}{2024}\natexlab{e}.
\newblock \bibinfo{title}{{C++ standard multiset container}}.
\newblock
  \bibinfo{howpublished}{\url{https://en.cppreference.com/w/cpp/container/multiset}}.
\newblock


\bibitem[std(2024f)]%
        {stdset}
 \bibinfo{year}{2024}\natexlab{f}.
\newblock \bibinfo{title}{{C++ standard set container}}.
\newblock
  \bibinfo{howpublished}{\url{https://en.cppreference.com/w/cpp/container/set}}.
\newblock


\bibitem[cxl(2024)]%
        {cxl}
 \bibinfo{year}{2024}\natexlab{}.
\newblock \bibinfo{title}{{Compute Express Link (CXL)}}.
\newblock \bibinfo{howpublished}{\url{https://www.computeexpresslink.org/}}.
\newblock


\bibitem[mic(2024)]%
        {micron}
 \bibinfo{year}{2024}\natexlab{}.
\newblock \bibinfo{title}{{DDR4 POWER CALC.XLSM}}.
\newblock
  \bibinfo{howpublished}{\url{https://www.micron.com/sales-support/design-tools/dram-power-calculator}}.
\newblock


\bibitem[dpd(2024)]%
        {dpdk}
 \bibinfo{year}{2024}\natexlab{}.
\newblock \bibinfo{title}{{DPDK}}.
\newblock \bibinfo{howpublished}{\url{https://www.dpdk.org/}}.
\newblock


\bibitem[int(2024)]%
        {intel_cmt_cat}
 \bibinfo{year}{2024}\natexlab{}.
\newblock \bibinfo{title}{{Intel(R) RDT Software Package}}.
\newblock \bibinfo{howpublished}{\url{https://github.com/intel/intel-cmt-cat}}.
\newblock


\bibitem[jav(2024)]%
        {javaiterator}
 \bibinfo{year}{2024}\natexlab{}.
\newblock \bibinfo{title}{{Java iterator}}.
\newblock
  \bibinfo{howpublished}{\url{https://www.w3schools.com/java/java_iterator.asp}}.
\newblock


\bibitem[llv(2024a)]%
        {llvmpass}
 \bibinfo{year}{2024}\natexlab{a}.
\newblock \bibinfo{title}{{LLVM’s Analysis and Transform Passes}}.
\newblock \bibinfo{howpublished}{\url{
  https://llvm.org/docs/Passes.html##introduction}}.
\newblock


\bibitem[mem(2024)]%
        {memcached}
 \bibinfo{year}{2024}\natexlab{}.
\newblock \bibinfo{title}{{MemCached}}.
\newblock \bibinfo{howpublished}{\url{http://www.memcached.org}}.
\newblock


\bibitem[has(2024a)]%
        {hash1}
 \bibinfo{year}{2024}\natexlab{a}.
\newblock \bibinfo{title}{{MySQL: Adaptive Hash Index}}.
\newblock
  \bibinfo{howpublished}{\url{https://dev.mysql.com/doc/refman/8.0/en/innodb-adaptive-hash.html}}.
\newblock


\bibitem[has(2024b)]%
        {hash3}
 \bibinfo{year}{2024}\natexlab{b}.
\newblock \bibinfo{title}{{Teradata: Hash Indexes}}.
\newblock
  \bibinfo{howpublished}{\url{https://docs.teradata.com/r/Enterprise_IntelliFlex_VMware/Database-Design/Join-and-Hash-Indexes/Hash-Indexes}}.
\newblock


\bibitem[llv(2024b)]%
        {llvm}
 \bibinfo{year}{2024}\natexlab{b}.
\newblock \bibinfo{title}{{The LLVM Compiler Infrastructure}}.
\newblock \bibinfo{howpublished}{\url{https://llvm.org/}}.
\newblock


\bibitem[vol(2024)]%
        {voltdb}
 \bibinfo{year}{2024}\natexlab{}.
\newblock \bibinfo{title}{{VoltDB}}.
\newblock
  \bibinfo{howpublished}{\url{http://voltdb.com/downloads/datasheets_collateral/technical_overview.pdf}}.
\newblock


\bibitem[Abraham et~al\mbox{.}(2013)]%
        {scuba}
\bibfield{author}{\bibinfo{person}{Lior Abraham}, \bibinfo{person}{John Allen},
  \bibinfo{person}{Oleksandr Barykin}, \bibinfo{person}{Vinayak Borkar},
  \bibinfo{person}{Bhuwan Chopra}, \bibinfo{person}{Ciprian Gerea},
  \bibinfo{person}{Daniel Merl}, \bibinfo{person}{Josh Metzler},
  \bibinfo{person}{David Reiss}, \bibinfo{person}{Subbu Subramanian},
  \bibinfo{person}{Janet~L. Wiener}, {and} \bibinfo{person}{Okay Zed}.}
  \bibinfo{year}{2013}\natexlab{}.
\newblock \showarticletitle{Scuba: Diving into Data at Facebook}.
\newblock \bibinfo{journal}{\emph{{{PVLDB}}}} \bibinfo{volume}{6},
  \bibinfo{number}{11} (\bibinfo{year}{2013}).
\newblock


\bibitem[Advanced Micro~Devices(2024a)]%
        {tcam_ip}
\bibfield{author}{\bibinfo{person}{Inc. Advanced Micro~Devices}.}
  \bibinfo{year}{2024}\natexlab{a}.
\newblock \bibinfo{title}{{Xilinx Content Addressable Memory (CAM)}}.
\newblock
  \bibinfo{howpublished}{\url{https://www.xilinx.com/products/intellectual-property/ef-di-cam.html}}.
\newblock


\bibitem[Advanced Micro~Devices(2024b)]%
        {xilinx_xrt}
\bibfield{author}{\bibinfo{person}{Inc. Advanced Micro~Devices}.}
  \bibinfo{year}{2024}\natexlab{b}.
\newblock \bibinfo{title}{{Xilinx Runtime Library (XRT)}}.
\newblock
  \bibinfo{howpublished}{\url{https://www.xilinx.com/products/design-tools/vitis/xrt.html}}.
\newblock


\bibitem[Agarwal et~al\mbox{.}(2015)]%
        {succinct}
\bibfield{author}{\bibinfo{person}{Rachit Agarwal}, \bibinfo{person}{Anurag
  Khandelwal}, {and} \bibinfo{person}{Ion Stoica}.}
  \bibinfo{year}{2015}\natexlab{}.
\newblock \showarticletitle{{Succinct: Enabling Queries on Compressed Data}}.
  In \bibinfo{booktitle}{\emph{USENIX NSDI}}.
\newblock


\bibitem[Aguilera et~al\mbox{.}(2017)]%
        {remotememory}
\bibfield{author}{\bibinfo{person}{Marcos~K. Aguilera}, \bibinfo{person}{Nadav
  Amit}, \bibinfo{person}{Irina Calciu}, \bibinfo{person}{Xavier Deguillard},
  \bibinfo{person}{Jayneel Gandhi}, \bibinfo{person}{Pratap Subrahmanyam},
  \bibinfo{person}{Lalith Suresh}, \bibinfo{person}{Kiran Tati},
  \bibinfo{person}{Rajesh Venkatasubramanian}, {and} \bibinfo{person}{Michael
  Wei}.} \bibinfo{year}{2017}\natexlab{}.
\newblock \showarticletitle{{Remote Memory in the Age of Fast Networks}}. In
  \bibinfo{booktitle}{\emph{SoCC}}.
\newblock


\bibitem[Ahn et~al\mbox{.}(2015)]%
        {ahn2015scalable}
\bibfield{author}{\bibinfo{person}{Junwhan Ahn}, \bibinfo{person}{Sungpack
  Hong}, \bibinfo{person}{Sungjoo Yoo}, \bibinfo{person}{Onur Mutlu}, {and}
  \bibinfo{person}{Kiyoung Choi}.} \bibinfo{year}{2015}\natexlab{}.
\newblock \showarticletitle{A scalable processing-in-memory accelerator for
  parallel graph processing}. In \bibinfo{booktitle}{\emph{ISCA}}.
  \bibinfo{pages}{105--117}.
\newblock


\bibitem[Al~Maruf and Chowdhury(2020)]%
        {prefetch1}
\bibfield{author}{\bibinfo{person}{Hasan Al~Maruf} {and}
  \bibinfo{person}{Mosharaf Chowdhury}.} \bibinfo{year}{2020}\natexlab{}.
\newblock \showarticletitle{Effectively prefetching remote memory with leap}.
  In \bibinfo{booktitle}{\emph{USENIX ATC}}. \bibinfo{pages}{843--857}.
\newblock


\bibitem[Amaro et~al\mbox{.}(2020a)]%
        {fastswap}
\bibfield{author}{\bibinfo{person}{Emmanuel Amaro},
  \bibinfo{person}{Christopher Branner-Augmon}, \bibinfo{person}{Zhihong Luo},
  \bibinfo{person}{Amy Ousterhout}, \bibinfo{person}{Marcos~K. Aguilera},
  \bibinfo{person}{Aurojit Panda}, \bibinfo{person}{Sylvia Ratnasamy}, {and}
  \bibinfo{person}{Scott Shenker}.} \bibinfo{year}{2020}\natexlab{a}.
\newblock \showarticletitle{{Can Far Memory Improve Job Throughput?}}. In
  \bibinfo{booktitle}{\emph{EuroSys}}.
\newblock


\bibitem[Amaro et~al\mbox{.}(2020b)]%
        {rmc_hotnets20}
\bibfield{author}{\bibinfo{person}{Emmanuel Amaro}, \bibinfo{person}{Zhihong
  Luo}, \bibinfo{person}{Amy Ousterhout}, \bibinfo{person}{Arvind
  Krishnamurthy}, \bibinfo{person}{Aurojit Panda}, \bibinfo{person}{Sylvia
  Ratnasamy}, {and} \bibinfo{person}{Scott Shenker}.}
  \bibinfo{year}{2020}\natexlab{b}.
\newblock \showarticletitle{Remote Memory Calls}. In
  \bibinfo{booktitle}{\emph{Proceedings of the 19th {{ACM}} Workshop on Hot
  Topics in Networks}}. \bibinfo{pages}{38--44}.
\newblock


\bibitem[An et~al\mbox{.}(2023)]%
        {marlin}
\bibfield{author}{\bibinfo{person}{Hang An}, \bibinfo{person}{Fang Wang},
  \bibinfo{person}{Dan Feng}, \bibinfo{person}{Xiaomin Zou},
  \bibinfo{person}{Zefeng Liu}, {and} \bibinfo{person}{Jianshun Zhang}.}
  \bibinfo{year}{2023}\natexlab{}.
\newblock \showarticletitle{Marlin: A Concurrent and Write-Optimized B+-tree
  Index on Disaggregated Memory}. In \bibinfo{booktitle}{\emph{ACM ICPP}}.
\newblock


\bibitem[Andersen and Culler(2016)]%
        {btrdb}
\bibfield{author}{\bibinfo{person}{Michael~P Andersen} {and}
  \bibinfo{person}{David~E. Culler}.} \bibinfo{year}{2016}\natexlab{}.
\newblock \showarticletitle{{BTrDB}: Optimizing Storage System Design for
  Timeseries Processing}. In \bibinfo{booktitle}{\emph{14th USENIX Conference
  on File and Storage Technologies (FAST 16)}}. \bibinfo{publisher}{USENIX
  Association}, \bibinfo{address}{Santa Clara, CA}, \bibinfo{pages}{39--52}.
\newblock
\showISBNx{978-1-931971-28-7}
\urldef\tempurl%
\url{https://www.usenix.org/conference/fast16/technical-sessions/presentation/andersen}
\showURL{%
\tempurl}


\bibitem[Asanovi{\'c}(2014)]%
        {memdisagg1}
\bibfield{author}{\bibinfo{person}{Krste Asanovi{\'c}}.}
  \bibinfo{year}{2014}\natexlab{}.
\newblock \showarticletitle{{FireBox: A Hardware Building Block for 2020
  Warehouse-Scale Computers}}.
\newblock


\bibitem[Asghari-Moghaddam et~al\mbox{.}(2016)]%
        {asghari2016chameleon}
\bibfield{author}{\bibinfo{person}{Hadi Asghari-Moghaddam},
  \bibinfo{person}{Young~Hoon Son}, \bibinfo{person}{Jung~Ho Ahn}, {and}
  \bibinfo{person}{Nam~Sung Kim}.} \bibinfo{year}{2016}\natexlab{}.
\newblock \showarticletitle{Chameleon: Versatile and practical near-DRAM
  acceleration architecture for large memory systems}. In
  \bibinfo{booktitle}{\emph{IEEE/ACM MICRO}}.
\newblock


\bibitem[Askitis and Sinha(2007)]%
        {trie2}
\bibfield{author}{\bibinfo{person}{Nikolas Askitis} {and}
  \bibinfo{person}{Ranjan Sinha}.} \bibinfo{year}{2007}\natexlab{}.
\newblock \showarticletitle{{HAT-trie: A Cache-conscious Trie-based Data
  Structure for Strings}}. In \bibinfo{booktitle}{\emph{ACSC}}.
\newblock


\bibitem[Bayer and McCreight(1970)]%
        {btree1}
\bibfield{author}{\bibinfo{person}{R. Bayer} {and} \bibinfo{person}{E.
  McCreight}.} \bibinfo{year}{1970}\natexlab{}.
\newblock \showarticletitle{{Organization and Maintenance of Large Ordered
  Indices}}. In \bibinfo{booktitle}{\emph{ACM-SIGMOD Workshop on Data
  Description, Access and Control}}.
\newblock


\bibitem[Berg et~al\mbox{.}(2020)]%
        {cachelib}
\bibfield{author}{\bibinfo{person}{Benjamin Berg}, \bibinfo{person}{Daniel~S.
  Berger}, \bibinfo{person}{Sara McAllister}, \bibinfo{person}{Isaac Grosof},
  \bibinfo{person}{Sathya Gunasekar}, \bibinfo{person}{Jimmy Lu},
  \bibinfo{person}{Michael Uhlar}, \bibinfo{person}{Jim Carrig},
  \bibinfo{person}{Nathan Beckmann}, \bibinfo{person}{Mor Harchol-Balter},
  {and} \bibinfo{person}{Gregory~R. Ganger}.} \bibinfo{year}{2020}\natexlab{}.
\newblock \showarticletitle{The {CacheLib} Caching Engine: Design and
  Experiences at Scale}. In \bibinfo{booktitle}{\emph{{{USENIX OSDI}}}}.
\newblock


\bibitem[Bhardwaj et~al\mbox{.}(2020)]%
        {storagefunctions}
\bibfield{author}{\bibinfo{person}{Ankit Bhardwaj}, \bibinfo{person}{Chinmay
  Kulkarni}, {and} \bibinfo{person}{Ryan Stutsman}.}
  \bibinfo{year}{2020}\natexlab{}.
\newblock \showarticletitle{Adaptive Placement for In-memory Storage
  Functions}. In \bibinfo{booktitle}{\emph{USENIX ATC}}.
\newblock


\bibitem[Boroumand et~al\mbox{.}(2019)]%
        {boroumand2019_codna}
\bibfield{author}{\bibinfo{person}{Amirali Boroumand}, \bibinfo{person}{Saugata
  Ghose}, \bibinfo{person}{Minesh Patel}, \bibinfo{person}{Hasan Hassan},
  \bibinfo{person}{Brandon Lucia}, \bibinfo{person}{Rachata Ausavarungnirun},
  \bibinfo{person}{Kevin Hsieh}, \bibinfo{person}{Nastaran Hajinazar},
  \bibinfo{person}{Krishna~T. Malladi}, \bibinfo{person}{Hongzhong Zheng},
  {and} \bibinfo{person}{Onur Mutlu}.} \bibinfo{year}{2019}\natexlab{}.
\newblock \showarticletitle{{{CoNDA}}: {{Efficient}} Cache Coherence Support
  for near-{{Data}} Accelerators}. In \bibinfo{booktitle}{\emph{ISCA}}.
  \bibinfo{pages}{629--642}.
\newblock


\bibitem[Braginsky and Petrank(2012)]%
        {btree2}
\bibfield{author}{\bibinfo{person}{Anastasia Braginsky} {and}
  \bibinfo{person}{Erez Petrank}.} \bibinfo{year}{2012}\natexlab{}.
\newblock \showarticletitle{A Lock-free B+Tree}. In
  \bibinfo{booktitle}{\emph{SPAA}}.
\newblock


\bibitem[Bronson et~al\mbox{.}(2013)]%
        {tao}
\bibfield{author}{\bibinfo{person}{Nathan Bronson}, \bibinfo{person}{Zach
  Amsden}, \bibinfo{person}{George Cabrera}, \bibinfo{person}{Prasad Chakka},
  \bibinfo{person}{Peter Dimov}, \bibinfo{person}{Hui Ding},
  \bibinfo{person}{Jack Ferris}, \bibinfo{person}{Anthony Giardullo},
  \bibinfo{person}{Sachin Kulkarni}, \bibinfo{person}{Harry Li},
  \bibinfo{person}{Mark Marchukov}, \bibinfo{person}{Dmitri Petrov},
  \bibinfo{person}{Lovro Puzar}, \bibinfo{person}{Yee~Jiun Song}, {and}
  \bibinfo{person}{Venkat Venkataramani}.} \bibinfo{year}{2013}\natexlab{}.
\newblock \showarticletitle{{TAO}: {Facebook{\textquoteright}s} Distributed
  Data Store for the Social Graph}. In \bibinfo{booktitle}{\emph{{{USENIX
  ATC}}}}.
\newblock


\bibitem[Cash et~al\mbox{.}(2015)]%
        {leakage-attacks}
\bibfield{author}{\bibinfo{person}{David Cash}, \bibinfo{person}{Paul Grubbs},
  \bibinfo{person}{Jason Perry}, {and} \bibinfo{person}{Thomas Ristenpart}.}
  \bibinfo{year}{2015}\natexlab{}.
\newblock \showarticletitle{Leakage-Abuse Attacks Against Searchable
  Encryption}. In \bibinfo{booktitle}{\emph{ACM CCS}}.
\newblock


\bibitem[Chi et~al\mbox{.}(2016)]%
        {chi2016prime}
\bibfield{author}{\bibinfo{person}{Ping Chi}, \bibinfo{person}{Shuangchen Li},
  \bibinfo{person}{Cong Xu}, \bibinfo{person}{Tao Zhang},
  \bibinfo{person}{Jishen Zhao}, \bibinfo{person}{Yongpan Liu},
  \bibinfo{person}{Yu Wang}, {and} \bibinfo{person}{Yuan Xie}.}
  \bibinfo{year}{2016}\natexlab{}.
\newblock \showarticletitle{Prime: A novel processing-in-memory architecture
  for neural network computation in reram-based main memory}.
\newblock \bibinfo{journal}{\emph{ACM SIGARCH Computer Architecture News}}
  \bibinfo{volume}{44}, \bibinfo{number}{3} (\bibinfo{year}{2016}),
  \bibinfo{pages}{27--39}.
\newblock


\bibitem[Cho et~al\mbox{.}(2020)]%
        {cho2020_data}
\bibfield{author}{\bibinfo{person}{Benjamin~Y. Cho}, \bibinfo{person}{Yongkee
  Kwon}, \bibinfo{person}{Sangkug Lym}, {and} \bibinfo{person}{Mattan Erez}.}
  \bibinfo{year}{2020}\natexlab{}.
\newblock \showarticletitle{Near Data Acceleration with Concurrent Host
  Access}. In \bibinfo{booktitle}{\emph{ISCA}}. \bibinfo{pages}{818--831}.
\newblock


\bibitem[Cooper et~al\mbox{.}(2010)]%
        {ycsb_workload}
\bibfield{author}{\bibinfo{person}{Brian~F. Cooper}, \bibinfo{person}{Adam
  Silberstein}, \bibinfo{person}{Erwin Tam}, \bibinfo{person}{Raghu
  Ramakrishnan}, {and} \bibinfo{person}{Russell Sears}.}
  \bibinfo{year}{2010}\natexlab{}.
\newblock \showarticletitle{{Benchmarking Cloud Serving Systems with YCSB}}. In
  \bibinfo{booktitle}{\emph{Proceedings of the 1st ACM Symposium on Cloud
  Computing}} (Indianapolis, Indiana, USA) \emph{(\bibinfo{series}{SoCC '10})}.
  \bibinfo{publisher}{Association for Computing Machinery},
  \bibinfo{address}{New York, NY, USA}, \bibinfo{pages}{143–154}.
\newblock
\showISBNx{9781450300360}
\urldef\tempurl%
\url{https://doi.org/10.1145/1807128.1807152}
\showDOI{\tempurl}


\bibitem[Dai et~al\mbox{.}(2018)]%
        {dai2018graphh}
\bibfield{author}{\bibinfo{person}{Guohao Dai}, \bibinfo{person}{Tianhao
  Huang}, \bibinfo{person}{Yuze Chi}, \bibinfo{person}{Jishen Zhao},
  \bibinfo{person}{Guangyu Sun}, \bibinfo{person}{Yongpan Liu},
  \bibinfo{person}{Yu Wang}, \bibinfo{person}{Yuan Xie}, {and}
  \bibinfo{person}{Huazhong Yang}.} \bibinfo{year}{2018}\natexlab{}.
\newblock \showarticletitle{GraphH: A processing-in-memory architecture for
  large-scale graph processing}.
\newblock \bibinfo{journal}{\emph{IEEE Transactions on Computer-Aided Design of
  Integrated Circuits and Systems}} \bibinfo{volume}{38}, \bibinfo{number}{4}
  (\bibinfo{year}{2018}), \bibinfo{pages}{640--653}.
\newblock


\bibitem[Dai et~al\mbox{.}(2022)]%
        {dai2022dimmining}
\bibfield{author}{\bibinfo{person}{Guohao Dai}, \bibinfo{person}{Zhenhua Zhu},
  \bibinfo{person}{Tianyu Fu}, \bibinfo{person}{Chiyue Wei},
  \bibinfo{person}{Bangyan Wang}, \bibinfo{person}{Xiangyu Li},
  \bibinfo{person}{Yuan Xie}, \bibinfo{person}{Huazhong Yang}, {and}
  \bibinfo{person}{Yu Wang}.} \bibinfo{year}{2022}\natexlab{}.
\newblock \showarticletitle{Dimmining: pruning-efficient and parallel graph
  mining on near-memory-computing}. In \bibinfo{booktitle}{\emph{ISCA}}.
  \bibinfo{pages}{130--145}.
\newblock


\bibitem[Devic et~al\mbox{.}(2022)]%
        {devic2022_PIM}
\bibfield{author}{\bibinfo{person}{Alexandar Devic},
  \bibinfo{person}{Siddhartha~Balakrishna Rai}, \bibinfo{person}{Anand
  Sivasubramaniam}, \bibinfo{person}{Ameen Akel}, \bibinfo{person}{Sean
  Eilert}, {and} \bibinfo{person}{Justin Eno}.}
  \bibinfo{year}{2022}\natexlab{}.
\newblock \showarticletitle{To {{PIM}} or Not for Emerging General Purpose
  Processing in {{DDR}} Memory Systems}. In \bibinfo{booktitle}{\emph{ISCA}}.
  \bibinfo{pages}{231--244}.
\newblock


\bibitem[Eckert et~al\mbox{.}(2022)]%
        {eckert2022eidetic}
\bibfield{author}{\bibinfo{person}{Charles Eckert}, \bibinfo{person}{Arun
  Subramaniyan}, \bibinfo{person}{Xiaowei Wang}, \bibinfo{person}{Charles
  Augustine}, \bibinfo{person}{Ravishankar Iyer}, {and}
  \bibinfo{person}{Reetuparna Das}.} \bibinfo{year}{2022}\natexlab{}.
\newblock \showarticletitle{Eidetic: An in-memory matrix multiplication
  accelerator for neural networks}.
\newblock \bibinfo{journal}{\emph{IEEE Trans. Comput.}} (\bibinfo{year}{2022}).
\newblock


\bibitem[Fan et~al\mbox{.}(2013)]%
        {memc3}
\bibfield{author}{\bibinfo{person}{Bin Fan}, \bibinfo{person}{David~G.
  Andersen}, {and} \bibinfo{person}{Michael Kaminsky}.}
  \bibinfo{year}{2013}\natexlab{}.
\newblock \showarticletitle{MemC3: Compact and Concurrent MemCache with Dumber
  Caching and Smarter Hashing}. In \bibinfo{booktitle}{\emph{USENIX NSDI}}.
\newblock


\bibitem[Gandhi et~al\mbox{.}(2016)]%
        {range}
\bibfield{author}{\bibinfo{person}{Jayneel Gandhi}, \bibinfo{person}{Vasileios
  Karakostas}, \bibinfo{person}{Furkan Ayar}, \bibinfo{person}{Adrián
  Cristal}, \bibinfo{person}{Mark~D. Hill}, \bibinfo{person}{Kathryn~S.
  McKinley}, \bibinfo{person}{Mario Nemirovsky}, \bibinfo{person}{Michael~M.
  Swift}, {and} \bibinfo{person}{Osman~S. Ünsal}.}
  \bibinfo{year}{2016}\natexlab{}.
\newblock \showarticletitle{Range Translations for Fast Virtual Memory}.
\newblock \bibinfo{journal}{\emph{IEEE Micro}} \bibinfo{volume}{36},
  \bibinfo{number}{3} (\bibinfo{year}{2016}), \bibinfo{pages}{118--126}.
\newblock
\urldef\tempurl%
\url{https://doi.org/10.1109/MM.2016.10}
\showDOI{\tempurl}


\bibitem[Gao et~al\mbox{.}(2016)]%
        {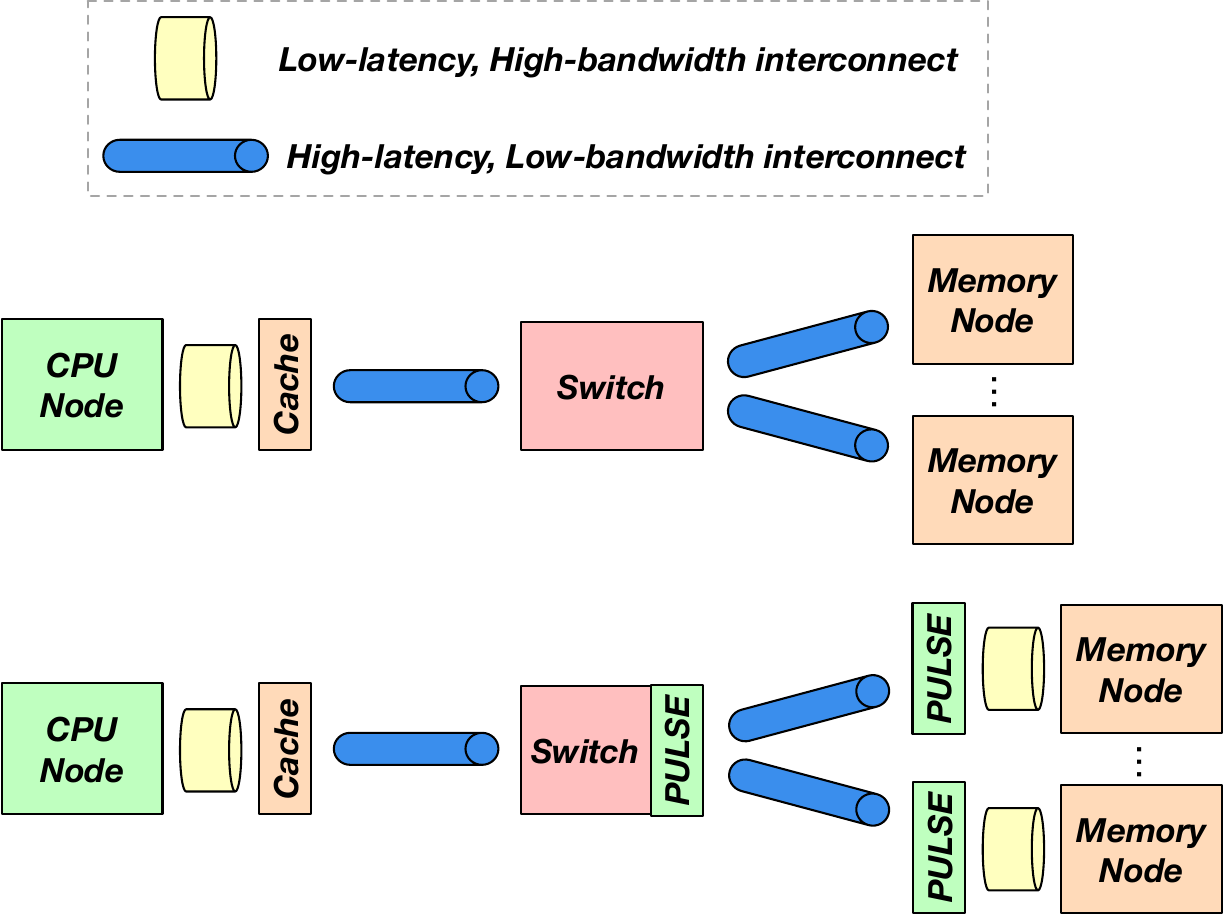}
\bibfield{author}{\bibinfo{person}{Peter~Xiang Gao}, \bibinfo{person}{Akshay
  Narayan}, \bibinfo{person}{Sagar Karandikar}, \bibinfo{person}{Joao
  Carreira}, \bibinfo{person}{Sangjin Han}, \bibinfo{person}{Rachit Agarwal},
  \bibinfo{person}{Sylvia Ratnasamy}, {and} \bibinfo{person}{Scott Shenker}.}
  \bibinfo{year}{2016}\natexlab{}.
\newblock \showarticletitle{{Network Requirements for Resource
  Disaggregation}}. In \bibinfo{booktitle}{\emph{USENIX OSDI}}.
\newblock


\bibitem[G{\'o}mez-Luna et~al\mbox{.}(2023)]%
        {gomez2023evaluating}
\bibfield{author}{\bibinfo{person}{Juan G{\'o}mez-Luna}, \bibinfo{person}{Yuxin
  Guo}, \bibinfo{person}{Sylvan Brocard}, \bibinfo{person}{Julien Legriel},
  \bibinfo{person}{Remy Cimadomo}, \bibinfo{person}{Geraldo~F Oliveira},
  \bibinfo{person}{Gagandeep Singh}, {and} \bibinfo{person}{Onur Mutlu}.}
  \bibinfo{year}{2023}\natexlab{}.
\newblock \showarticletitle{Evaluating machine learning workloads on
  memory-centric computing systems}. In \bibinfo{booktitle}{\emph{2023 IEEE
  International Symposium on Performance Analysis of Systems and Software
  (ISPASS)}}. IEEE, \bibinfo{pages}{35--49}.
\newblock


\bibitem[Gonzalez et~al\mbox{.}(2012)]%
        {powergraph}
\bibfield{author}{\bibinfo{person}{Joseph~E Gonzalez}, \bibinfo{person}{Yucheng
  Low}, \bibinfo{person}{Haijie Gu}, \bibinfo{person}{Danny Bickson}, {and}
  \bibinfo{person}{Carlos Guestrin}.} \bibinfo{year}{2012}\natexlab{}.
\newblock \showarticletitle{{PowerGraph: Distributed Graph-Parallel Computation
  on Natural Graphs}}. In \bibinfo{booktitle}{\emph{USENIX OSDI}}.
\newblock


\bibitem[Gonzalez et~al\mbox{.}(2014)]%
        {graphx}
\bibfield{author}{\bibinfo{person}{Joseph~E Gonzalez},
  \bibinfo{person}{Reynold~S Xin}, \bibinfo{person}{Ankur Dave},
  \bibinfo{person}{Daniel Crankshaw}, \bibinfo{person}{Michael~J Franklin},
  {and} \bibinfo{person}{Ion Stoica}.} \bibinfo{year}{2014}\natexlab{}.
\newblock \showarticletitle{{GraphX: Graph Processing in a Distributed Dataflow
  Framework}}. In \bibinfo{booktitle}{\emph{USENIX OSDI}}.
\newblock


\bibitem[Gouk et~al\mbox{.}(2023)]%
        {cxl3}
\bibfield{author}{\bibinfo{person}{Donghyun Gouk}, \bibinfo{person}{Miryeong
  Kwon}, \bibinfo{person}{Hanyeoreum Bae}, \bibinfo{person}{Sangwon Lee}, {and}
  \bibinfo{person}{Myoungsoo Jung}.} \bibinfo{year}{2023}\natexlab{}.
\newblock \showarticletitle{Memory pooling with cxl}.
\newblock \bibinfo{journal}{\emph{IEEE Micro}} \bibinfo{volume}{43},
  \bibinfo{number}{2} (\bibinfo{year}{2023}), \bibinfo{pages}{48--57}.
\newblock


\bibitem[Gouk et~al\mbox{.}(2022)]%
        {cxl1}
\bibfield{author}{\bibinfo{person}{Donghyun Gouk}, \bibinfo{person}{Sangwon
  Lee}, \bibinfo{person}{Miryeong Kwon}, {and} \bibinfo{person}{Myoungsoo
  Jung}.} \bibinfo{year}{2022}\natexlab{}.
\newblock \showarticletitle{Direct access, High-Performance memory
  disaggregation with DirectCXL}. In \bibinfo{booktitle}{\emph{USENIX ATC}}.
\newblock


\bibitem[Grubbs et~al\mbox{.}(2020)]%
        {pancake}
\bibfield{author}{\bibinfo{person}{Paul Grubbs}, \bibinfo{person}{Anurag
  Khandelwal}, \bibinfo{person}{Marie-Sarah Lacharit{\'e}},
  \bibinfo{person}{Lloyd Brown}, \bibinfo{person}{Lucy Li},
  \bibinfo{person}{Rachit Agarwal}, {and} \bibinfo{person}{Thomas Ristenpart}.}
  \bibinfo{year}{2020}\natexlab{}.
\newblock \showarticletitle{Pancake: Frequency smoothing for encrypted data
  stores}. In \bibinfo{booktitle}{\emph{{USENIX} Security}}.
\newblock


\bibitem[Gu et~al\mbox{.}(2017)]%
        {infiniswap}
\bibfield{author}{\bibinfo{person}{Juncheng Gu}, \bibinfo{person}{Youngmoon
  Lee}, \bibinfo{person}{Yiwen Zhang}, \bibinfo{person}{Mosharaf Chowdhury},
  {and} \bibinfo{person}{Kang~G. Shin}.} \bibinfo{year}{2017}\natexlab{}.
\newblock \showarticletitle{{Efficient Memory Disaggregation with Infiniswap}}.
  In \bibinfo{booktitle}{\emph{USENIX NSDI}}.
\newblock


\bibitem[Gu et~al\mbox{.}(2020)]%
        {gu2020ipim}
\bibfield{author}{\bibinfo{person}{Peng Gu}, \bibinfo{person}{Xinfeng Xie},
  \bibinfo{person}{Yufei Ding}, \bibinfo{person}{Guoyang Chen},
  \bibinfo{person}{Weifeng Zhang}, \bibinfo{person}{Dimin Niu}, {and}
  \bibinfo{person}{Yuan Xie}.} \bibinfo{year}{2020}\natexlab{}.
\newblock \showarticletitle{iPIM: Programmable in-memory image processing
  accelerator using near-bank architecture}. In
  \bibinfo{booktitle}{\emph{ISCA}}. IEEE, \bibinfo{pages}{804--817}.
\newblock


\bibitem[Guo et~al\mbox{.}(2022)]%
        {clio}
\bibfield{author}{\bibinfo{person}{Zhiyuan Guo}, \bibinfo{person}{Yizhou Shan},
  \bibinfo{person}{Xuhao Luo}, \bibinfo{person}{Yutong Huang}, {and}
  \bibinfo{person}{Yiying Zhang}.} \bibinfo{year}{2022}\natexlab{}.
\newblock \showarticletitle{Clio: A Hardware-Software Co-Designed Disaggregated
  Memory System}. In \bibinfo{booktitle}{\emph{{{ACM ASPLOS}}}}.
\newblock


\bibitem[Heinz et~al\mbox{.}(2002)]%
        {trie1}
\bibfield{author}{\bibinfo{person}{Steffen Heinz}, \bibinfo{person}{Justin
  Zobel}, {and} \bibinfo{person}{Hugh~E Williams}.}
  \bibinfo{year}{2002}\natexlab{}.
\newblock \showarticletitle{{Burst tries: a fast, efficient data structure for
  string keys}}.
\newblock \bibinfo{journal}{\emph{TOIS}} (\bibinfo{year}{2002}).
\newblock


\bibitem[Hsieh et~al\mbox{.}(2016)]%
        {impica}
\bibfield{author}{\bibinfo{person}{Kevin Hsieh}, \bibinfo{person}{Samira Khan},
  \bibinfo{person}{Nandita Vijaykumar}, \bibinfo{person}{Kevin~K. Chang},
  \bibinfo{person}{Amirali Boroumand}, \bibinfo{person}{Saugata Ghose}, {and}
  \bibinfo{person}{Onur Mutlu}.} \bibinfo{year}{2016}\natexlab{}.
\newblock \showarticletitle{Accelerating pointer chasing in 3D-stacked memory:
  Challenges, mechanisms, evaluation}. In
  \bibinfo{booktitle}{\emph{International Conference on Computer Design
  (ICCD)}}.
\newblock


\bibitem[Idreos et~al\mbox{.}(2012)]%
        {monetdb}
\bibfield{author}{\bibinfo{person}{Stratos Idreos}, \bibinfo{person}{F.
  Groffen}, \bibinfo{person}{Niels Nes}, \bibinfo{person}{Stefan Manegold},
  \bibinfo{person}{Sjoerd Mullender}, {and} \bibinfo{person}{Martin Kersten}.}
  \bibinfo{year}{2012}\natexlab{}.
\newblock \showarticletitle{MonetDB: Two Decades of Research in Column-oriented
  Database Architectures}.
\newblock \bibinfo{journal}{\emph{IEEE Data Eng. Bull.}}  \bibinfo{volume}{35}
  (\bibinfo{date}{01} \bibinfo{year}{2012}).
\newblock


\bibitem[{Intel Corporation}(2024)]%
        {intel_rapl}
\bibfield{author}{\bibinfo{person}{{Intel Corporation}}.}
  \bibinfo{year}{2024}\natexlab{}.
\newblock \bibinfo{booktitle}{\emph{Intel 64 and IA-32 Architectures Software
  Developer’s Manual}}.
\newblock
\newblock
\shownote{\url{https://www.intel.com/content/www/us/en/developer/articles/technical/intel-sdm.html}}.


\bibitem[Islam et~al\mbox{.}(2012)]%
        {islam2012access}
\bibfield{author}{\bibinfo{person}{Mohammad~Saiful Islam},
  \bibinfo{person}{Mehmet Kuzu}, {and} \bibinfo{person}{Murat Kantarcioglu}.}
  \bibinfo{year}{2012}\natexlab{}.
\newblock \showarticletitle{Access pattern disclosure on searchable encryption:
  Ramification, attack and mitigation}. In \bibinfo{booktitle}{\emph{{NDSS}}}.
\newblock


\bibitem[Jevdjic et~al\mbox{.}(2014)]%
        {jevdjic2014unison}
\bibfield{author}{\bibinfo{person}{Djordje Jevdjic}, \bibinfo{person}{Gabriel~H
  Loh}, \bibinfo{person}{Cansu Kaynak}, {and} \bibinfo{person}{Babak Falsafi}.}
  \bibinfo{year}{2014}\natexlab{}.
\newblock \showarticletitle{Unison cache: A scalable and effective die-stacked
  DRAM cache}. In \bibinfo{booktitle}{\emph{IEEE/ACM MICRO}}.
\newblock


\bibitem[Jevdjic et~al\mbox{.}(2013)]%
        {jevdjic2013stacked}
\bibfield{author}{\bibinfo{person}{Djordje Jevdjic}, \bibinfo{person}{Stavros
  Volos}, {and} \bibinfo{person}{Babak Falsafi}.}
  \bibinfo{year}{2013}\natexlab{}.
\newblock \showarticletitle{Die-stacked dram caches for servers: Hit ratio,
  latency, or bandwidth? have it all with footprint cache}.
\newblock \bibinfo{journal}{\emph{ACM SIGARCH Computer Architecture News}}
  \bibinfo{volume}{41}, \bibinfo{number}{3} (\bibinfo{year}{2013}),
  \bibinfo{pages}{404--415}.
\newblock


\bibitem[Jia et~al\mbox{.}(2024)]%
        {lengthleakage}
\bibfield{author}{\bibinfo{person}{Grace Jia}, \bibinfo{person}{Rachit
  Agarwal}, {and} \bibinfo{person}{Anurag Khandelwal}.}
  \bibinfo{year}{2024}\natexlab{}.
\newblock \showarticletitle{Length Leakage in Oblivious Data Access
  Mechanisms}. In \bibinfo{booktitle}{\emph{USENIX Security}}.
\newblock


\bibitem[Jiang(2019)]%
        {intel_dsa}
\bibfield{author}{\bibinfo{person}{Dave Jiang}.}
  \bibinfo{year}{2019}\natexlab{}.
\newblock \bibinfo{title}{Introducing the {{Intel}}\textregistered{} {{Data
  Streaming Accelerator}} ({{Intel}}\textregistered{} {{DSA}})}.
\newblock
  \bibinfo{howpublished}{\url{https://01.org/blogs/2019/introducing-intel-data-streaming-accelerator}}.
\newblock


\bibitem[Kalia et~al\mbox{.}(2019)]%
        {erpc}
\bibfield{author}{\bibinfo{person}{Anuj Kalia}, \bibinfo{person}{Michael
  Kaminsky}, {and} \bibinfo{person}{David Andersen}.}
  \bibinfo{year}{2019}\natexlab{}.
\newblock \showarticletitle{Datacenter {RPCs} can be General and Fast}. In
  \bibinfo{booktitle}{\emph{USENIX NSDI}}.
\newblock


\bibitem[Kang et~al\mbox{.}(2014)]%
        {memscaling2}
\bibfield{author}{\bibinfo{person}{Uksong Kang}, \bibinfo{person}{Hak-Soo Yu},
  \bibinfo{person}{Churoo Park}, \bibinfo{person}{Hongzhong Zheng},
  \bibinfo{person}{John Halbert}, \bibinfo{person}{Kuljit Bains},
  \bibinfo{person}{S Jang}, {and} \bibinfo{person}{Joo~Sun Choi}.}
  \bibinfo{year}{2014}\natexlab{}.
\newblock \showarticletitle{Co-architecting controllers and DRAM to enhance
  DRAM process scaling}. In \bibinfo{booktitle}{\emph{The memory forum}},
  Vol.~\bibinfo{volume}{14}.
\newblock


\bibitem[Ke et~al\mbox{.}(2020)]%
        {ke2020_RecNMP}
\bibfield{author}{\bibinfo{person}{Liu Ke}, \bibinfo{person}{Udit Gupta},
  \bibinfo{person}{Benjamin~Youngjae Cho}, \bibinfo{person}{David Brooks},
  \bibinfo{person}{Vikas Chandra}, \bibinfo{person}{Utku Diril},
  \bibinfo{person}{Amin Firoozshahian}, \bibinfo{person}{Kim Hazelwood},
  \bibinfo{person}{Bill Jia}, \bibinfo{person}{Hsien-Hsin~S. Lee},
  \bibinfo{person}{Meng Li}, \bibinfo{person}{Bert Maher},
  \bibinfo{person}{Dheevatsa Mudigere}, \bibinfo{person}{Maxim Naumov},
  \bibinfo{person}{Martin Schatz}, \bibinfo{person}{Mikhail Smelyanskiy},
  \bibinfo{person}{Xiaodong Wang}, \bibinfo{person}{Brandon Reagen},
  \bibinfo{person}{Carole-Jean Wu}, \bibinfo{person}{Mark Hempstead}, {and}
  \bibinfo{person}{Xuan Zhang}.} \bibinfo{year}{2020}\natexlab{}.
\newblock \showarticletitle{{{RecNMP}}: {{Accelerating}} Personalized
  Recommendation with near-Memory Processing}. In
  \bibinfo{booktitle}{\emph{ISCA}}. \bibinfo{pages}{790--803}.
\newblock


\bibitem[Ke et~al\mbox{.}(2021)]%
        {ke2021near}
\bibfield{author}{\bibinfo{person}{Liu Ke}, \bibinfo{person}{Xuan Zhang},
  \bibinfo{person}{Jinin So}, \bibinfo{person}{Jong-Geon Lee},
  \bibinfo{person}{Shin-Haeng Kang}, \bibinfo{person}{Sukhan Lee},
  \bibinfo{person}{Songyi Han}, \bibinfo{person}{YeonGon Cho},
  \bibinfo{person}{Jin~Hyun Kim}, \bibinfo{person}{Yongsuk Kwon},
  {et~al\mbox{.}}} \bibinfo{year}{2021}\natexlab{}.
\newblock \showarticletitle{Near-memory processing in action: Accelerating
  personalized recommendation with axdimm}.
\newblock \bibinfo{journal}{\emph{IEEE Micro}} \bibinfo{volume}{42},
  \bibinfo{number}{1} (\bibinfo{year}{2021}), \bibinfo{pages}{116--127}.
\newblock


\bibitem[Kellaris et~al\mbox{.}(2016)]%
        {kellaris2016generic}
\bibfield{author}{\bibinfo{person}{Georgios Kellaris}, \bibinfo{person}{George
  Kollios}, \bibinfo{person}{Kobbi Nissim}, {and} \bibinfo{person}{Adam
  O'Neill}.} \bibinfo{year}{2016}\natexlab{}.
\newblock \showarticletitle{Generic Attacks on Secure Outsourced Databases}. In
  \bibinfo{booktitle}{\emph{ACM CCS}}.
\newblock


\bibitem[Khandelwal et~al\mbox{.}(2016)]%
        {blowfish}
\bibfield{author}{\bibinfo{person}{Anurag Khandelwal}, \bibinfo{person}{Rachit
  Agarwal}, {and} \bibinfo{person}{Ion Stoica}.}
  \bibinfo{year}{2016}\natexlab{}.
\newblock \showarticletitle{BlowFish: Dynamic Storage-Performance Tradeoff in
  Data Stores.}. In \bibinfo{booktitle}{\emph{USENIX NSDI}}.
\newblock


\bibitem[Kocberber et~al\mbox{.}(2013)]%
        {walkers}
\bibfield{author}{\bibinfo{person}{Onur Kocberber}, \bibinfo{person}{Boris
  Grot}, \bibinfo{person}{Javier Picorel}, \bibinfo{person}{Babak Falsafi},
  \bibinfo{person}{Kevin Lim}, {and} \bibinfo{person}{Parthasarathy
  Ranganathan}.} \bibinfo{year}{2013}\natexlab{}.
\newblock \showarticletitle{Meet the Walkers: Accelerating Index Traversals for
  in-Memory Databases}. In \bibinfo{booktitle}{\emph{IEEE/ACM MICRO}}.
\newblock


\bibitem[Kornaropoulos et~al\mbox{.}(2019)]%
        {kornaropoulos2019recovery}
\bibfield{author}{\bibinfo{person}{Evgenios Kornaropoulos},
  \bibinfo{person}{Charalampos Papamanthou}, {and} \bibinfo{person}{Roberto
  Tamassia}.} \bibinfo{year}{2019}\natexlab{}.
\newblock \showarticletitle{Data Recovery on Encrypted Databases with k-Nearest
  Neighbor Query Leakage}. In \bibinfo{booktitle}{\emph{IEEE S\&P}}.
\newblock


\bibitem[Korolija et~al\mbox{.}(2020)]%
        {coyote}
\bibfield{author}{\bibinfo{person}{Dario Korolija}, \bibinfo{person}{Timothy
  Roscoe}, {and} \bibinfo{person}{Gustavo Alonso}.}
  \bibinfo{year}{2020}\natexlab{}.
\newblock \showarticletitle{{Do OS abstractions make sense on FPGAs?}}. In
  \bibinfo{booktitle}{\emph{USENIX OSDI}}.
\newblock


\bibitem[Koukos et~al\mbox{.}(2013)]%
        {daepowerscaling}
\bibfield{author}{\bibinfo{person}{Konstantinos Koukos}, \bibinfo{person}{David
  Black-Schaffer}, \bibinfo{person}{Vasileios Spiliopoulos}, {and}
  \bibinfo{person}{Stefanos Kaxiras}.} \bibinfo{year}{2013}\natexlab{}.
\newblock \showarticletitle{Towards More Efficient Execution: A Decoupled
  Access-Execute Approach}. In \bibinfo{booktitle}{\emph{Proceedings of the
  27th International ACM Conference on International Conference on
  Supercomputing}} (Eugene, Oregon, USA) \emph{(\bibinfo{series}{ICS '13})}.
  \bibinfo{publisher}{Association for Computing Machinery},
  \bibinfo{address}{New York, NY, USA}, \bibinfo{pages}{253–262}.
\newblock
\showISBNx{9781450321303}
\urldef\tempurl%
\url{https://doi.org/10.1145/2464996.2465012}
\showDOI{\tempurl}


\bibitem[Kulkarni et~al\mbox{.}(2018)]%
        {splinter}
\bibfield{author}{\bibinfo{person}{Chinmay Kulkarni}, \bibinfo{person}{Sara
  Moore}, \bibinfo{person}{Mazhar Naqvi}, \bibinfo{person}{Tian Zhang},
  \bibinfo{person}{Robert Ricci}, {and} \bibinfo{person}{Ryan Stutsman}.}
  \bibinfo{year}{2018}\natexlab{}.
\newblock \showarticletitle{Splinter: {Bare-Metal} Extensions for
  {Multi-Tenant} {Low-Latency} Storage}. In \bibinfo{booktitle}{\emph{{{USENIX
  OSDI}}}}.
\newblock


\bibitem[Kuon and Rose(2006)]%
        {asicpower}
\bibfield{author}{\bibinfo{person}{Ian Kuon} {and} \bibinfo{person}{Jonathan
  Rose}.} \bibinfo{year}{2006}\natexlab{}.
\newblock \showarticletitle{Measuring the Gap between FPGAs and ASICs}. In
  \bibinfo{booktitle}{\emph{Proceedings of the 2006 ACM/SIGDA 14th
  International Symposium on Field Programmable Gate Arrays}} (Monterey,
  California, USA) \emph{(\bibinfo{series}{FPGA '06})}.
  \bibinfo{publisher}{Association for Computing Machinery},
  \bibinfo{address}{New York, NY, USA}, \bibinfo{pages}{21–30}.
\newblock
\showISBNx{1595932925}
\urldef\tempurl%
\url{https://doi.org/10.1145/1117201.1117205}
\showDOI{\tempurl}


\bibitem[Kwon et~al\mbox{.}(2019)]%
        {kwon2019_TensorDIMM}
\bibfield{author}{\bibinfo{person}{Youngeun Kwon}, \bibinfo{person}{Yunjae
  Lee}, {and} \bibinfo{person}{Minsoo Rhu}.} \bibinfo{year}{2019}\natexlab{}.
\newblock \showarticletitle{{{TensorDIMM}}: {{A}} Practical near-Memory
  Processing Architecture for Embeddings and Tensor Operations in Deep
  Learning}. In \bibinfo{booktitle}{\emph{IEEE/ACM MICRO}}.
  \bibinfo{pages}{740--753}.
\newblock


\bibitem[Kyrola et~al\mbox{.}(2012)]%
        {graphchi}
\bibfield{author}{\bibinfo{person}{Aapo Kyrola}, \bibinfo{person}{Guy~E
  Blelloch}, {and} \bibinfo{person}{Carlos Guestrin}.}
  \bibinfo{year}{2012}\natexlab{}.
\newblock \showarticletitle{{GraphChi: Large-Scale Graph Computation on Just a
  PC}}. In \bibinfo{booktitle}{\emph{USENIX OSDI}}.
\newblock


\bibitem[Lattner and Adve(2004)]%
        {llvmsparc}
\bibfield{author}{\bibinfo{person}{C. Lattner} {and} \bibinfo{person}{V.
  Adve}.} \bibinfo{year}{2004}\natexlab{}.
\newblock \showarticletitle{LLVM: a compilation framework for lifelong program
  analysis \& transformation}. In \bibinfo{booktitle}{\emph{International
  Symposium on Code Generation and Optimization, 2004. CGO 2004.}}
  \bibinfo{pages}{75--86}.
\newblock
\urldef\tempurl%
\url{https://doi.org/10.1109/CGO.2004.1281665}
\showDOI{\tempurl}


\bibitem[Lee(2016)]%
        {memscaling3}
\bibfield{author}{\bibinfo{person}{Seok-Hee Lee}.}
  \bibinfo{year}{2016}\natexlab{}.
\newblock \showarticletitle{Technology scaling challenges and opportunities of
  memory devices}. In \bibinfo{booktitle}{\emph{International Electron Devices
  Meeting (IEDM)}}.
\newblock


\bibitem[Lee et~al\mbox{.}(2021)]%
        {mind}
\bibfield{author}{\bibinfo{person}{Seung-seob Lee}, \bibinfo{person}{Yanpeng
  Yu}, \bibinfo{person}{Yupeng Tang}, \bibinfo{person}{Anurag Khandelwal},
  \bibinfo{person}{Lin Zhong}, {and} \bibinfo{person}{Abhishek Bhattacharjee}.}
  \bibinfo{year}{2021}\natexlab{}.
\newblock \showarticletitle{{MIND: In-Network Memory Management for
  Disaggregated Data Centers}}. In \bibinfo{booktitle}{\emph{{{SOSP}}}}.
\newblock


\bibitem[Li et~al\mbox{.}(2023a)]%
        {pond}
\bibfield{author}{\bibinfo{person}{Huaicheng Li}, \bibinfo{person}{Daniel~S.
  Berger}, \bibinfo{person}{Lisa Hsu}, \bibinfo{person}{Daniel Ernst},
  \bibinfo{person}{Pantea Zardoshti}, \bibinfo{person}{Stanko Novakovic},
  \bibinfo{person}{Monish Shah}, \bibinfo{person}{Samir Rajadnya},
  \bibinfo{person}{Scott Lee}, \bibinfo{person}{Ishwar Agarwal},
  \bibinfo{person}{Mark~D. Hill}, \bibinfo{person}{Marcus Fontoura}, {and}
  \bibinfo{person}{Ricardo Bianchini}.} \bibinfo{year}{2023}\natexlab{a}.
\newblock \showarticletitle{Pond: CXL-Based Memory Pooling Systems for Cloud
  Platforms}. In \bibinfo{booktitle}{\emph{ACM ASPLOS}}.
\newblock


\bibitem[Li et~al\mbox{.}(2022)]%
        {cxl_azure}
\bibfield{author}{\bibinfo{person}{Huaicheng Li}, \bibinfo{person}{Daniel~S
  Berger}, \bibinfo{person}{Stanko Novakovic}, \bibinfo{person}{Lisa Hsu},
  \bibinfo{person}{Dan Ernst}, \bibinfo{person}{Pantea Zardoshti},
  \bibinfo{person}{Monish Shah}, \bibinfo{person}{Ishwar Agarwal},
  \bibinfo{person}{Mark Hill}, \bibinfo{person}{Marcus Fontoura},
  {et~al\mbox{.}}} \bibinfo{year}{2022}\natexlab{}.
\newblock \showarticletitle{First-generation Memory Disaggregation for Cloud
  Platforms}.
\newblock \bibinfo{journal}{\emph{arXiv preprint arXiv:2203.00241}}
  (\bibinfo{year}{2022}).
\newblock


\bibitem[Li et~al\mbox{.}(2023c)]%
        {prefetch2}
\bibfield{author}{\bibinfo{person}{Haifeng Li}, \bibinfo{person}{Ke Liu},
  \bibinfo{person}{Ting Liang}, \bibinfo{person}{Zuojun Li},
  \bibinfo{person}{Tianyue Lu}, \bibinfo{person}{Hui Yuan},
  \bibinfo{person}{Yinben Xia}, \bibinfo{person}{Yungang Bao},
  \bibinfo{person}{Mingyu Chen}, {and} \bibinfo{person}{Yizhou Shan}.}
  \bibinfo{year}{2023}\natexlab{c}.
\newblock \showarticletitle{HoPP: Hardware-Software Co-Designed Page
  Prefetching for Disaggregated Memory}. In \bibinfo{booktitle}{\emph{IEEE
  HPCA}}.
\newblock


\bibitem[Li et~al\mbox{.}(2023b)]%
        {rolex}
\bibfield{author}{\bibinfo{person}{Pengfei Li}, \bibinfo{person}{Yu Hua},
  \bibinfo{person}{Pengfei Zuo}, \bibinfo{person}{Zhangyu Chen}, {and}
  \bibinfo{person}{Jiajie Sheng}.} \bibinfo{year}{2023}\natexlab{b}.
\newblock \showarticletitle{{ROLEX}: A Scalable {RDMA-oriented} Learned
  {Key-Value} Store for Disaggregated Memory Systems}. In
  \bibinfo{booktitle}{\emph{USENIX FAST}}. \bibinfo{publisher}{USENIX
  Association}, \bibinfo{address}{Santa Clara, CA}, \bibinfo{pages}{99--114}.
\newblock
\showISBNx{978-1-939133-32-8}
\urldef\tempurl%
\url{https://www.usenix.org/conference/fast23/presentation/li-pengfei}
\showURL{%
\tempurl}


\bibitem[Lockerman et~al\mbox{.}(2020)]%
        {lockerman2020livia}
\bibfield{author}{\bibinfo{person}{Elliot Lockerman}, \bibinfo{person}{Axel
  Feldmann}, \bibinfo{person}{Mohammad Bakhshalipour},
  \bibinfo{person}{Alexandru Stanescu}, \bibinfo{person}{Shashwat Gupta},
  \bibinfo{person}{Daniel Sanchez}, {and} \bibinfo{person}{Nathan Beckmann}.}
  \bibinfo{year}{2020}\natexlab{}.
\newblock \showarticletitle{Livia: Data-centric computing throughout the memory
  hierarchy}. In \bibinfo{booktitle}{\emph{ACM ASPLOS}}.
\newblock


\bibitem[{Microsoft Corporation}(2024)]%
        {hash2}
\bibfield{author}{\bibinfo{person}{{Microsoft Corporation}}.}
  \bibinfo{year}{2024}\natexlab{}.
\newblock \bibinfo{booktitle}{\emph{SQL Server and Azure SQL Index Architecture
  and Design Guide}}.
\newblock
\newblock
\shownote{\url{https://learn.microsoft.com/en-us/sql/relational-databases/sql-server-index-design-guide?view=sql-server-ver16##hash_index}}.


\bibitem[Min et~al\mbox{.}(2024)]%
        {sephash}
\bibfield{author}{\bibinfo{person}{Xinhao Min}, \bibinfo{person}{Kai Lu},
  \bibinfo{person}{Pengyu Liu}, \bibinfo{person}{Jiguang Wan},
  \bibinfo{person}{Changsheng Xie}, \bibinfo{person}{Daohui Wang},
  \bibinfo{person}{Ting Yao}, {and} \bibinfo{person}{Huatao Wu}.}
  \bibinfo{year}{2024}\natexlab{}.
\newblock \showarticletitle{SepHash: {A} Write-Optimized Hash Index On
  Disaggregated Memory via Separate Segment Structure}.
\newblock \bibinfo{journal}{\emph{Proc. {VLDB} Endow.}} \bibinfo{volume}{17},
  \bibinfo{number}{5} (\bibinfo{year}{2024}), \bibinfo{pages}{1091--1104}.
\newblock
\urldef\tempurl%
\url{https://www.vldb.org/pvldb/vol17/p1091-lu.pdf}
\showURL{%
\tempurl}


\bibitem[MongoDB(2024)]%
        {mongodb}
\bibfield{author}{\bibinfo{person}{Inc. MongoDB}.}
  \bibinfo{year}{2024}\natexlab{}.
\newblock \bibinfo{title}{{WiredTiger storage engine.}}
\newblock
  \bibinfo{howpublished}{\url{https://docs.mongodb.com/manual/core/wiredtiger/}}.
\newblock


\bibitem[Morrison(1968)]%
        {trie3}
\bibfield{author}{\bibinfo{person}{Donald~R. Morrison}.}
  \bibinfo{year}{1968}\natexlab{}.
\newblock \showarticletitle{{PATRICIA - Practical Algorithm To Retrieve
  Information Coded in Alphanumeric}}.
\newblock \bibinfo{journal}{\emph{JACM}} (\bibinfo{year}{1968}).
\newblock


\bibitem[Mutlu et~al\mbox{.}(2019)]%
        {mutlu2019processing}
\bibfield{author}{\bibinfo{person}{Onur Mutlu}, \bibinfo{person}{Saugata
  Ghose}, \bibinfo{person}{Juan G{\'o}mez-Luna}, {and} \bibinfo{person}{Rachata
  Ausavarungnirun}.} \bibinfo{year}{2019}\natexlab{}.
\newblock \showarticletitle{Processing data where it makes sense: Enabling
  in-memory computation}.
\newblock \bibinfo{journal}{\emph{Microprocessors and Microsystems}}
  \bibinfo{volume}{67} (\bibinfo{year}{2019}), \bibinfo{pages}{28--41}.
\newblock


\bibitem[Mutlu et~al\mbox{.}(2022)]%
        {mutlu2022modern}
\bibfield{author}{\bibinfo{person}{Onur Mutlu}, \bibinfo{person}{Saugata
  Ghose}, \bibinfo{person}{Juan G{\'o}mez-Luna}, {and} \bibinfo{person}{Rachata
  Ausavarungnirun}.} \bibinfo{year}{2022}\natexlab{}.
\newblock \showarticletitle{A modern primer on processing in memory}.
\newblock In \bibinfo{booktitle}{\emph{Emerging Computing: From Devices to
  Systems: Looking Beyond Moore and Von Neumann}}.
  \bibinfo{publisher}{Springer}, \bibinfo{pages}{171--243}.
\newblock


\bibitem[Nishtala et~al\mbox{.}(2013)]%
        {memcache}
\bibfield{author}{\bibinfo{person}{Rajesh Nishtala}, \bibinfo{person}{Hans
  Fugal}, \bibinfo{person}{Steven Grimm}, \bibinfo{person}{Marc Kwiatkowski},
  \bibinfo{person}{Herman Lee}, \bibinfo{person}{Harry~C Li},
  \bibinfo{person}{Ryan McElroy}, \bibinfo{person}{Mike Paleczny},
  \bibinfo{person}{Daniel Peek}, \bibinfo{person}{Paul Saab}, {et~al\mbox{.}}}
  \bibinfo{year}{2013}\natexlab{}.
\newblock \showarticletitle{Scaling memcache at facebook}. In
  \bibinfo{booktitle}{\emph{{{USENIX NSDI}}}}.
\newblock


\bibitem[Novakovic et~al\mbox{.}(2019)]%
        {storm_systor_19}
\bibfield{author}{\bibinfo{person}{Stanko Novakovic}, \bibinfo{person}{Yizhou
  Shan}, \bibinfo{person}{Aasheesh Kolli}, \bibinfo{person}{Michael Cui},
  \bibinfo{person}{Yiying Zhang}, \bibinfo{person}{Haggai Eran},
  \bibinfo{person}{Boris Pismenny}, \bibinfo{person}{Liran Liss},
  \bibinfo{person}{Michael Wei}, \bibinfo{person}{Dan Tsafrir}, {and}
  \bibinfo{person}{Marcos Aguilera}.} \bibinfo{year}{2019}\natexlab{}.
\newblock \showarticletitle{{Storm: A Fast Transactional Dataplane for Remote
  Data Structures}}. In \bibinfo{booktitle}{\emph{SYSTOR}}.
  \bibinfo{pages}{97–108}.
\newblock


\bibitem[Oh et~al\mbox{.}(2021)]%
        {fpgatrusted1}
\bibfield{author}{\bibinfo{person}{Hyunyoung Oh}, \bibinfo{person}{Kevin Nam},
  \bibinfo{person}{Seongil Jeon}, \bibinfo{person}{Yeongpil Cho}, {and}
  \bibinfo{person}{Yunheung Paek}.} \bibinfo{year}{2021}\natexlab{}.
\newblock \showarticletitle{MeetGo: A trusted execution environment for remote
  applications on FPGA}.
\newblock \bibinfo{journal}{\emph{IEEE Access}}  \bibinfo{volume}{9}
  (\bibinfo{year}{2021}), \bibinfo{pages}{51313--51324}.
\newblock


\bibitem[Olgun et~al\mbox{.}(2022)]%
        {olgun2022pidram}
\bibfield{author}{\bibinfo{person}{Ataberk Olgun},
  \bibinfo{person}{Juan~G{\'o}mez Luna}, \bibinfo{person}{Konstantinos
  Kanellopoulos}, \bibinfo{person}{Behzad Salami}, \bibinfo{person}{Hasan
  Hassan}, \bibinfo{person}{Oguz Ergin}, {and} \bibinfo{person}{Onur Mutlu}.}
  \bibinfo{year}{2022}\natexlab{}.
\newblock \showarticletitle{PiDRAM: A Holistic End-to-end FPGA-based Framework
  for Processing-in-DRAM}.
\newblock \bibinfo{journal}{\emph{ACM Transactions on Architecture and Code
  Optimization}} \bibinfo{volume}{20}, \bibinfo{number}{1}
  (\bibinfo{year}{2022}), \bibinfo{pages}{1--31}.
\newblock


\bibitem[Oliveira et~al\mbox{.}(2022)]%
        {oliveira2022accelerating}
\bibfield{author}{\bibinfo{person}{Geraldo~F Oliveira}, \bibinfo{person}{Juan
  G{\'o}mez-Luna}, \bibinfo{person}{Saugata Ghose}, \bibinfo{person}{Amirali
  Boroumand}, {and} \bibinfo{person}{Onur Mutlu}.}
  \bibinfo{year}{2022}\natexlab{}.
\newblock \showarticletitle{Accelerating neural network inference with
  processing-in-DRAM: from the edge to the cloud}.
\newblock \bibinfo{journal}{\emph{IEEE Micro}} \bibinfo{volume}{42},
  \bibinfo{number}{6} (\bibinfo{year}{2022}), \bibinfo{pages}{25--38}.
\newblock


\bibitem[Ousterhout et~al\mbox{.}(2019)]%
        {ousterhout_shenango_19_nsdi}
\bibfield{author}{\bibinfo{person}{Amy Ousterhout}, \bibinfo{person}{Joshua
  Fried}, \bibinfo{person}{Jonathan Behrens}, \bibinfo{person}{Adam Belay},
  {and} \bibinfo{person}{Hari Balakrishnan}.} \bibinfo{year}{2019}\natexlab{}.
\newblock \showarticletitle{Shenango: Achieving High {CPU} Efficiency for
  Latency-sensitive Datacenter Workloads}. In \bibinfo{booktitle}{\emph{USENIX
  NSDI}}.
\newblock


\bibitem[Page et~al\mbox{.}(1999)]%
        {pagerank}
\bibfield{author}{\bibinfo{person}{Lawrence Page}, \bibinfo{person}{Sergey
  Brin}, \bibinfo{person}{Rajeev Motwani}, {and} \bibinfo{person}{Terry
  Winograd}.} \bibinfo{year}{1999}\natexlab{}.
\newblock \bibinfo{booktitle}{\emph{{The PageRank Citation Ranking: Bringing
  Order to the Web}}}.
\newblock \bibinfo{type}{Technical Report}.
\newblock


\bibitem[Pereira et~al\mbox{.}(2021)]%
        {fpgatrusted5}
\bibfield{author}{\bibinfo{person}{S{\'e}rgio Pereira}, \bibinfo{person}{David
  Cerdeira}, \bibinfo{person}{Cristiano Rodrigues}, {and}
  \bibinfo{person}{Sandro Pinto}.} \bibinfo{year}{2021}\natexlab{}.
\newblock \showarticletitle{Towards a trusted execution environment via
  reconfigurable FPGA}.
\newblock \bibinfo{journal}{\emph{arXiv preprint arXiv:2107.03781}}
  (\bibinfo{year}{2021}).
\newblock


\bibitem[Reda et~al\mbox{.}(2022)]%
        {redn}
\bibfield{author}{\bibinfo{person}{Waleed Reda}, \bibinfo{person}{Marco
  Canini}, \bibinfo{person}{Dejan Kosti{\'c}}, {and} \bibinfo{person}{Simon
  Peter}.} \bibinfo{year}{2022}\natexlab{}.
\newblock \showarticletitle{{RDMA} is Turing complete, we just did not know it
  yet!}. In \bibinfo{booktitle}{\emph{{{USENIX NSDI}}}}.
\newblock


\bibitem[{Redis}(2024)]%
        {redis}
\bibfield{author}{\bibinfo{person}{{Redis}}.} \bibinfo{year}{2024}\natexlab{}.
\newblock \bibinfo{title}{Redis - The Real-time Data Platform}.
\newblock \bibinfo{howpublished}{\url{https://redis.io/}}.
\newblock


\bibitem[Reiss(2016)]%
        {memoverprovisioning}
\bibfield{author}{\bibinfo{person}{Charles Reiss}.}
  \bibinfo{year}{2016}\natexlab{}.
\newblock \emph{\bibinfo{title}{Understanding Memory Configurations for
  In-Memory Analytics}}.
\newblock \bibinfo{thesistype}{Ph.\,D. Dissertation}. \bibinfo{school}{EECS
  Department, University of California, Berkeley}.
\newblock
\urldef\tempurl%
\url{http://www2.eecs.berkeley.edu/Pubs/TechRpts/2016/EECS-2016-136.html}
\showURL{%
\tempurl}


\bibitem[Reynolds(1993)]%
        {continuation}
\bibfield{author}{\bibinfo{person}{John~C Reynolds}.}
  \bibinfo{year}{1993}\natexlab{}.
\newblock \showarticletitle{The discoveries of continuations}.
\newblock \bibinfo{journal}{\emph{Lisp and symbolic computation}}
  \bibinfo{volume}{6} (\bibinfo{year}{1993}), \bibinfo{pages}{233--247}.
\newblock


\bibitem[Rivitti et~al\mbox{.}(2023)]%
        {ebpfjump}
\bibfield{author}{\bibinfo{person}{Alessandro Rivitti},
  \bibinfo{person}{Roberto Bifulco}, \bibinfo{person}{Angelo Tulumello},
  \bibinfo{person}{Marco Bonola}, {and} \bibinfo{person}{Salvatore
  Pontarelli}.} \bibinfo{year}{2023}\natexlab{}.
\newblock \showarticletitle{eHDL: Turning eBPF/XDP Programs into Hardware
  Designs for the NIC}. In \bibinfo{booktitle}{\emph{ACM ASPLOS}}.
\newblock


\bibitem[Robinson(2021)]%
        {samsung_pim}
\bibfield{author}{\bibinfo{person}{Daniel Robinson}.}
  \bibinfo{year}{2021}\natexlab{}.
\newblock \bibinfo{title}{Samsung to Bring In-Memory Processing to Standard
  {{DIMMs}} and Mobile Memory}.
\newblock
  \bibinfo{howpublished}{\url{https://blocksandfiles.com/2021/08/24/samsung-to-bring-in-memory-processing-to-standard-dimms-and-mobile-memory/}}.
\newblock


\bibitem[Rosero-Montalvo et~al\mbox{.}(2023)]%
        {fpgatrusted3}
\bibfield{author}{\bibinfo{person}{Paul~D. Rosero-Montalvo},
  \bibinfo{person}{Zsolt István}, {and} \bibinfo{person}{Wilmar Hernandez}.}
  \bibinfo{year}{2023}\natexlab{}.
\newblock \showarticletitle{A Survey of Trusted Computing Solutions Using
  FPGAs}.
\newblock \bibinfo{journal}{\emph{IEEE Access}}  \bibinfo{volume}{11}
  (\bibinfo{year}{2023}), \bibinfo{pages}{31583--31593}.
\newblock
\urldef\tempurl%
\url{https://doi.org/10.1109/ACCESS.2023.3261802}
\showDOI{\tempurl}


\bibitem[Ruan et~al\mbox{.}(2020)]%
        {aifm}
\bibfield{author}{\bibinfo{person}{Zhenyuan Ruan}, \bibinfo{person}{Malte
  Schwarzkopf}, \bibinfo{person}{Marcos~K. Aguilera}, {and}
  \bibinfo{person}{Adam Belay}.} \bibinfo{year}{2020}\natexlab{}.
\newblock \showarticletitle{{AIFM}: {High-Performance},
  {Application-Integrated} Far Memory}. In \bibinfo{booktitle}{\emph{{{USENIX
  OSDI}}}}.
\newblock


\bibitem[Schuiki et~al\mbox{.}(2018)]%
        {schuiki2018scalable}
\bibfield{author}{\bibinfo{person}{Fabian Schuiki}, \bibinfo{person}{Michael
  Schaffner}, \bibinfo{person}{Frank~K G{\"u}rkaynak}, {and}
  \bibinfo{person}{Luca Benini}.} \bibinfo{year}{2018}\natexlab{}.
\newblock \showarticletitle{A scalable near-memory architecture for training
  deep neural networks on large in-memory datasets}.
\newblock \bibinfo{journal}{\emph{IEEE Trans. Comput.}} \bibinfo{volume}{68},
  \bibinfo{number}{4} (\bibinfo{year}{2018}), \bibinfo{pages}{484--497}.
\newblock


\bibitem[Seshadri and Mutlu(2017)]%
        {seshadri2017simple}
\bibfield{author}{\bibinfo{person}{Vivek Seshadri} {and} \bibinfo{person}{Onur
  Mutlu}.} \bibinfo{year}{2017}\natexlab{}.
\newblock \showarticletitle{Simple operations in memory to reduce data
  movement}.
\newblock In \bibinfo{booktitle}{\emph{Advances in Computers}}.
  Vol.~\bibinfo{volume}{106}. \bibinfo{publisher}{Elsevier},
  \bibinfo{pages}{107--166}.
\newblock


\bibitem[Shan et~al\mbox{.}(2018)]%
        {legoos}
\bibfield{author}{\bibinfo{person}{Yizhou Shan}, \bibinfo{person}{Yutong
  Huang}, \bibinfo{person}{Yilun Chen}, {and} \bibinfo{person}{Yiying Zhang}.}
  \bibinfo{year}{2018}\natexlab{}.
\newblock \showarticletitle{{LegoOS: A Disseminated, Distributed OS for
  Hardware Resource Disaggregation}}. In \bibinfo{booktitle}{\emph{USENIX
  OSDI}}.
\newblock


\bibitem[Shen et~al\mbox{.}(2023a)]%
        {ditto}
\bibfield{author}{\bibinfo{person}{Jiacheng Shen}, \bibinfo{person}{Pengfei
  Zuo}, \bibinfo{person}{Xuchuan Luo}, \bibinfo{person}{Yuxin Su},
  \bibinfo{person}{Jiazhen Gu}, \bibinfo{person}{Hao Feng},
  \bibinfo{person}{Yangfan Zhou}, {and} \bibinfo{person}{Michael~R. Lyu}.}
  \bibinfo{year}{2023}\natexlab{a}.
\newblock \showarticletitle{Ditto: An Elastic and Adaptive Memory-Disaggregated
  Caching System}. In \bibinfo{booktitle}{\emph{SOSP}}.
\newblock


\bibitem[Shen et~al\mbox{.}(2023b)]%
        {fusee}
\bibfield{author}{\bibinfo{person}{Jiacheng Shen}, \bibinfo{person}{Pengfei
  Zuo}, \bibinfo{person}{Xuchuan Luo}, \bibinfo{person}{Tianyi Yang},
  \bibinfo{person}{Yuxin Su}, \bibinfo{person}{Yangfan Zhou}, {and}
  \bibinfo{person}{Michael~R. Lyu}.} \bibinfo{year}{2023}\natexlab{b}.
\newblock \showarticletitle{{FUSEE}: A Fully {Memory-Disaggregated} {Key-Value}
  Store}. In \bibinfo{booktitle}{\emph{USENIX FAST}}.
\newblock


\bibitem[Shi et~al\mbox{.}(2020)]%
        {flighttracker}
\bibfield{author}{\bibinfo{person}{Xiao Shi}, \bibinfo{person}{Scott Pruett},
  \bibinfo{person}{Kevin Doherty}, \bibinfo{person}{Jinyu Han},
  \bibinfo{person}{Dmitri Petrov}, \bibinfo{person}{Jim Carrig},
  \bibinfo{person}{John Hugg}, {and} \bibinfo{person}{Nathan Bronson}.}
  \bibinfo{year}{2020}\natexlab{}.
\newblock \showarticletitle{{FlightTracker}: Consistency across
  {Read-Optimized} Online Stores at Facebook}. In
  \bibinfo{booktitle}{\emph{{{USENIX OSDI}}}}.
\newblock


\bibitem[Shiratake(2020)]%
        {memscaling1}
\bibfield{author}{\bibinfo{person}{Shigeru Shiratake}.}
  \bibinfo{year}{2020}\natexlab{}.
\newblock \showarticletitle{Scaling and Performance Challenges of Future DRAM}.
  In \bibinfo{booktitle}{\emph{International Memory Workshop (IMW)}}.
\newblock


\bibitem[Sidler et~al\mbox{.}(2020)]%
        {strom}
\bibfield{author}{\bibinfo{person}{David Sidler}, \bibinfo{person}{Zeke Wang},
  \bibinfo{person}{Monica Chiosa}, \bibinfo{person}{Amit Kulkarni}, {and}
  \bibinfo{person}{Gustavo Alonso}.} \bibinfo{year}{2020}\natexlab{}.
\newblock \showarticletitle{StRoM: Smart Remote Memory}. In
  \bibinfo{booktitle}{\emph{{{EuroSys}}}}.
\newblock


\bibitem[Singh et~al\mbox{.}(2021)]%
        {singh2021fpga}
\bibfield{author}{\bibinfo{person}{Gagandeep Singh}, \bibinfo{person}{Mohammed
  Alser}, \bibinfo{person}{Damla~Senol Cali}, \bibinfo{person}{Dionysios
  Diamantopoulos}, \bibinfo{person}{Juan G{\'o}mez-Luna}, \bibinfo{person}{Henk
  Corporaal}, {and} \bibinfo{person}{Onur Mutlu}.}
  \bibinfo{year}{2021}\natexlab{}.
\newblock \showarticletitle{FPGA-based near-memory acceleration of modern
  data-intensive applications}.
\newblock \bibinfo{journal}{\emph{IEEE Micro}} \bibinfo{volume}{41},
  \bibinfo{number}{4} (\bibinfo{year}{2021}), \bibinfo{pages}{39--48}.
\newblock


\bibitem[Stewart et~al\mbox{.}(2016)]%
        {upmu}
\bibfield{author}{\bibinfo{person}{Emma~M. Stewart}, \bibinfo{person}{Anna
  Liao}, {and} \bibinfo{person}{Ciaran Roberts}.}
  \bibinfo{year}{2016}\natexlab{}.
\newblock \showarticletitle{Open $\mu$PMU: A Real World Reference Distribution
  Micro-phasor Measurement Unit Data Set for Research and Application
  Development}.
\newblock  (\bibinfo{year}{2016}).
\newblock


\bibitem[Sun et~al\mbox{.}(2023)]%
        {sun2023demystifying}
\bibfield{author}{\bibinfo{person}{Yan Sun}, \bibinfo{person}{Yifan Yuan},
  \bibinfo{person}{Zeduo Yu}, \bibinfo{person}{Reese Kuper},
  \bibinfo{person}{Ipoom Jeong}, \bibinfo{person}{Ren Wang}, {and}
  \bibinfo{person}{Nam~Sung Kim}.} \bibinfo{year}{2023}\natexlab{}.
\newblock \bibinfo{title}{Demystifying CXL Memory with Genuine CXL-Ready
  Systems and Devices}.
\newblock
\newblock


\bibitem[Tang et~al\mbox{.}(2024)]%
        {cxlperformance}
\bibfield{author}{\bibinfo{person}{Yupeng Tang}, \bibinfo{person}{Ping Zhou},
  \bibinfo{person}{Wenhui Zhang}, \bibinfo{person}{Henry Hu},
  \bibinfo{person}{Qirui Yang}, \bibinfo{person}{Hao Xiang},
  \bibinfo{person}{Tongping Liu}, \bibinfo{person}{Jiaxin Shan},
  \bibinfo{person}{Ruoyun Huang}, \bibinfo{person}{Cheng Zhao},
  \bibinfo{person}{Cheng Chen}, \bibinfo{person}{Hui Zhang},
  \bibinfo{person}{Fei Liu}, \bibinfo{person}{Shuai Zhang},
  \bibinfo{person}{Xiaoning Ding}, {and} \bibinfo{person}{Jianjun Chen}.}
  \bibinfo{year}{2024}\natexlab{}.
\newblock \showarticletitle{Exploring Performance and Cost Optimization with
  ASIC-Based CXL Memory}. In \bibinfo{booktitle}{\emph{EuroSys}}.
\newblock


\bibitem[Tsai et~al\mbox{.}(2020)]%
        {clover}
\bibfield{author}{\bibinfo{person}{Shin-Yeh Tsai}, \bibinfo{person}{Yizhou
  Shan}, {and} \bibinfo{person}{Yiying Zhang}.}
  \bibinfo{year}{2020}\natexlab{}.
\newblock \showarticletitle{Disaggregating Persistent Memory and Controlling
  Them Remotely: An Exploration of Passive Disaggregated Key-Value Stores}. In
  \bibinfo{booktitle}{\emph{USENIX ATC}}.
\newblock


\bibitem[Tu et~al\mbox{.}(2022)]%
        {tu2022redcim}
\bibfield{author}{\bibinfo{person}{Fengbin Tu}, \bibinfo{person}{Yiqi Wang},
  \bibinfo{person}{Zihan Wu}, \bibinfo{person}{Ling Liang},
  \bibinfo{person}{Yufei Ding}, \bibinfo{person}{Bongjin Kim},
  \bibinfo{person}{Leibo Liu}, \bibinfo{person}{Shaojun Wei},
  \bibinfo{person}{Yuan Xie}, {and} \bibinfo{person}{Shouyi Yin}.}
  \bibinfo{year}{2022}\natexlab{}.
\newblock \showarticletitle{ReDCIM: Reconfigurable digital computing-in-memory
  processor with unified FP/INT pipeline for cloud AI acceleration}.
\newblock \bibinfo{journal}{\emph{IEEE Journal of Solid-State Circuits}}
  \bibinfo{volume}{58}, \bibinfo{number}{1} (\bibinfo{year}{2022}),
  \bibinfo{pages}{243--255}.
\newblock


\bibitem[Vuppalapati et~al\mbox{.}(2022)]%
        {shortstack}
\bibfield{author}{\bibinfo{person}{Midhul Vuppalapati}, \bibinfo{person}{Kushal
  Babel}, \bibinfo{person}{Anurag Khandelwal}, {and} \bibinfo{person}{Rachit
  Agarwal}.} \bibinfo{year}{2022}\natexlab{}.
\newblock \showarticletitle{{SHORTSTACK}: Distributed, Fault-tolerant,
  Oblivious Data Access}. In \bibinfo{booktitle}{\emph{{USENIX} OSDI}}.
\newblock


\bibitem[Wang et~al\mbox{.}(2020)]%
        {wang2020_semeru}
\bibfield{author}{\bibinfo{person}{Chenxi Wang}, \bibinfo{person}{Haoran Ma},
  \bibinfo{person}{Shi Liu}, \bibinfo{person}{Yuanqi Li},
  \bibinfo{person}{Zhenyuan Ruan}, \bibinfo{person}{Khanh Nguyen},
  \bibinfo{person}{Michael~D. Bond}, \bibinfo{person}{Ravi Netravali},
  \bibinfo{person}{Miryung Kim}, {and} \bibinfo{person}{Guoqing~Harry Xu}.}
  \bibinfo{year}{2020}\natexlab{}.
\newblock \showarticletitle{Semeru: {{A Memory-Disaggregated Managed
  Runtime}}}. In \bibinfo{booktitle}{\emph{USENIX OSDI}}.
\newblock


\bibitem[Wang et~al\mbox{.}(2022b)]%
        {wang2022_memLiner}
\bibfield{author}{\bibinfo{person}{Chenxi Wang}, \bibinfo{person}{Haoran Ma},
  \bibinfo{person}{Shi Liu}, \bibinfo{person}{Yifan Qiao},
  \bibinfo{person}{Jonathan Eyolfson}, \bibinfo{person}{Christian Navasca},
  \bibinfo{person}{Shan Lu}, {and} \bibinfo{person}{Guoqing~Harry Xu}.}
  \bibinfo{year}{2022}\natexlab{b}.
\newblock \showarticletitle{{{MemLiner}}: {{Lining}} up Tracing and Application
  for a {{Far-Memory-Friendly}} Runtime}. In \bibinfo{booktitle}{\emph{USENIX
  OSDI}}.
\newblock


\bibitem[Wang et~al\mbox{.}(2022a)]%
        {sherman}
\bibfield{author}{\bibinfo{person}{Qing Wang}, \bibinfo{person}{Youyou Lu},
  {and} \bibinfo{person}{Jiwu Shu}.} \bibinfo{year}{2022}\natexlab{a}.
\newblock \showarticletitle{Sherman: A Write-Optimized Distributed B+Tree Index
  on Disaggregated Memory}. In \bibinfo{booktitle}{\emph{SIGMOD}}.
\newblock


\bibitem[Wang et~al\mbox{.}(2024)]%
        {cxl2}
\bibfield{author}{\bibinfo{person}{Zhonghua Wang}, \bibinfo{person}{Yixing
  Guo}, \bibinfo{person}{Kai Lu}, \bibinfo{person}{Jiguang Wan},
  \bibinfo{person}{Daohui Wang}, \bibinfo{person}{Ting Yao}, {and}
  \bibinfo{person}{Huatao Wu}.} \bibinfo{year}{2024}\natexlab{}.
\newblock \showarticletitle{Rcmp: Reconstructing RDMA-Based Memory
  Disaggregation via CXL}.
\newblock \bibinfo{journal}{\emph{ACM Transactions on Architecture and Code
  Optimization}} \bibinfo{volume}{21}, \bibinfo{number}{1}
  (\bibinfo{year}{2024}), \bibinfo{pages}{1--26}.
\newblock


\bibitem[Wang et~al\mbox{.}(2022c)]%
        {wang2022_Nearstream}
\bibfield{author}{\bibinfo{person}{Zhengrong Wang}, \bibinfo{person}{Jian
  Weng}, \bibinfo{person}{Sihao Liu}, {and} \bibinfo{person}{Tony Nowatzki}.}
  \bibinfo{year}{2022}\natexlab{c}.
\newblock \showarticletitle{Near-Stream Computing: {{General}} and Transparent
  near-Cache Acceleration}. In \bibinfo{booktitle}{\emph{HPCA}}.
  \bibinfo{pages}{331--345}.
\newblock


\bibitem[Wang et~al\mbox{.}(2021)]%
        {wang2021stream}
\bibfield{author}{\bibinfo{person}{Zhengrong Wang}, \bibinfo{person}{Jian
  Weng}, \bibinfo{person}{Jason Lowe-Power}, \bibinfo{person}{Jayesh Gaur},
  {and} \bibinfo{person}{Tony Nowatzki}.} \bibinfo{year}{2021}\natexlab{}.
\newblock \showarticletitle{Stream floating: Enabling proactive and
  decentralized cache optimizations}. In \bibinfo{booktitle}{\emph{2021 IEEE
  International Symposium on High-Performance Computer Architecture (HPCA)}}.
  IEEE, \bibinfo{pages}{640--653}.
\newblock


\bibitem[Xia et~al\mbox{.}(2021)]%
        {fpgatrusted4}
\bibfield{author}{\bibinfo{person}{Ke Xia}, \bibinfo{person}{Yukui Luo},
  \bibinfo{person}{Xiaolin Xu}, {and} \bibinfo{person}{Sheng Wei}.}
  \bibinfo{year}{2021}\natexlab{}.
\newblock \showarticletitle{Sgx-fpga: Trusted execution environment for
  cpu-fpga heterogeneous architecture}. In \bibinfo{booktitle}{\emph{2021 58th
  ACM/IEEE Design Automation Conference (DAC)}}. IEEE,
  \bibinfo{pages}{301--306}.
\newblock


\bibitem[Xie et~al\mbox{.}(2023)]%
        {xie2023mpu}
\bibfield{author}{\bibinfo{person}{Xinfeng Xie}, \bibinfo{person}{Peng Gu},
  \bibinfo{person}{Yufei Ding}, \bibinfo{person}{Dimin Niu},
  \bibinfo{person}{Hongzhong Zheng}, {and} \bibinfo{person}{Yuan Xie}.}
  \bibinfo{year}{2023}\natexlab{}.
\newblock \showarticletitle{MPU: Memory-centric SIMT Processor via In-DRAM
  Near-bank Computing}.
\newblock \bibinfo{journal}{\emph{ACM Transactions on Architecture and Code
  Optimization}} \bibinfo{volume}{20}, \bibinfo{number}{3}
  (\bibinfo{year}{2023}), \bibinfo{pages}{1--26}.
\newblock


\bibitem[Xie et~al\mbox{.}(2021)]%
        {xie2021spacea}
\bibfield{author}{\bibinfo{person}{Xinfeng Xie}, \bibinfo{person}{Zheng Liang},
  \bibinfo{person}{Peng Gu}, \bibinfo{person}{Abanti Basak},
  \bibinfo{person}{Lei Deng}, \bibinfo{person}{Ling Liang},
  \bibinfo{person}{Xing Hu}, {and} \bibinfo{person}{Yuan Xie}.}
  \bibinfo{year}{2021}\natexlab{}.
\newblock \showarticletitle{SpaceA: Sparse matrix vector multiplication on
  processing-in-memory accelerator}. In \bibinfo{booktitle}{\emph{2021 IEEE
  International Symposium on High-Performance Computer Architecture (HPCA)}}.
  IEEE, \bibinfo{pages}{570--583}.
\newblock


\bibitem[Xilinx(2022)]%
        {xilinx_network}
\bibfield{author}{\bibinfo{person}{Xilinx}.} \bibinfo{year}{2022}\natexlab{}.
\newblock \bibinfo{title}{{XUP Vitis Network Example (VNx)}}.
\newblock
  \bibinfo{howpublished}{\url{https://github.com/Xilinx/xup_vitis_network_example}}.
\newblock


\bibitem[Yang et~al\mbox{.}(2020)]%
        {twittercache}
\bibfield{author}{\bibinfo{person}{Juncheng Yang}, \bibinfo{person}{Yao Yue},
  {and} \bibinfo{person}{K.~V. Rashmi}.} \bibinfo{year}{2020}\natexlab{}.
\newblock \showarticletitle{A large scale analysis of hundreds of in-memory
  cache clusters at Twitter}. In \bibinfo{booktitle}{\emph{USENIX OSDI}}.
\newblock


\bibitem[Yoon et~al\mbox{.}(2023)]%
        {prefetch3}
\bibfield{author}{\bibinfo{person}{Wonsup Yoon}, \bibinfo{person}{Jisu Ok},
  \bibinfo{person}{Jinyoung Oh}, \bibinfo{person}{Sue Moon}, {and}
  \bibinfo{person}{Youngjin Kwon}.} \bibinfo{year}{2023}\natexlab{}.
\newblock \showarticletitle{DiLOS: Do Not Trade Compatibility for Performance
  in Memory Disaggregation}. In \bibinfo{booktitle}{\emph{EuroSys}}.
\newblock


\bibitem[You et~al\mbox{.}(2021)]%
        {kayak_nsdi_21}
\bibfield{author}{\bibinfo{person}{Jie You}, \bibinfo{person}{Jingfeng Wu},
  \bibinfo{person}{Xin Jin}, {and} \bibinfo{person}{Mosharaf Chowdhury}.}
  \bibinfo{year}{2021}\natexlab{}.
\newblock \showarticletitle{{Ship Compute or Ship Data? Why Not Both?}}. In
  \bibinfo{booktitle}{\emph{USENIX NSDI}}. \bibinfo{pages}{633--651}.
\newblock


\bibitem[Young et~al\mbox{.}(2018)]%
        {young2018accord}
\bibfield{author}{\bibinfo{person}{Vinson Young}, \bibinfo{person}{Chiachen
  Chou}, \bibinfo{person}{Aamer Jaleel}, {and} \bibinfo{person}{Moinuddin
  Qureshi}.} \bibinfo{year}{2018}\natexlab{}.
\newblock \showarticletitle{Accord: Enabling associativity for gigascale dram
  caches by coordinating way-install and way-prediction}. In
  \bibinfo{booktitle}{\emph{ISCA}}.
\newblock


\bibitem[Yu et~al\mbox{.}(2014)]%
        {db1000}
\bibfield{author}{\bibinfo{person}{Xiangyao Yu}, \bibinfo{person}{George
  Bezerra}, \bibinfo{person}{Andrew Pavlo}, \bibinfo{person}{Sahana Devadas},
  {and} \bibinfo{person}{Michael Stonebraker}.}
  \bibinfo{year}{2014}\natexlab{}.
\newblock \showarticletitle{Staring into the abyss: An evaluation of
  concurrency control with one thousand cores}.
\newblock \bibinfo{journal}{\emph{Proceedings of the VLDB Endowment}}
  \bibinfo{volume}{8} (\bibinfo{date}{11} \bibinfo{year}{2014}).
\newblock
\urldef\tempurl%
\url{https://doi.org/10.14778/2735508.2735511}
\showDOI{\tempurl}


\bibitem[Yu et~al\mbox{.}(2009)]%
        {netlock}
\bibfield{author}{\bibinfo{person}{Zhuolong Yu}, \bibinfo{person}{Yiwen Zhang},
  \bibinfo{person}{Vladimir Bravermann}, \bibinfo{person}{Mosharaf Chowdhury},
  {and} \bibinfo{person}{Xin Jin}.} \bibinfo{year}{2009}\natexlab{}.
\newblock \showarticletitle{{NetLock: Fast, Centralized Lock Management Using
  Programmable Switches}}. In \bibinfo{booktitle}{\emph{SIGCOMM}}.
\newblock


\bibitem[Zaharia et~al\mbox{.}(2012)]%
        {spark}
\bibfield{author}{\bibinfo{person}{Matei Zaharia}, \bibinfo{person}{Mosharaf
  Chowdhury}, \bibinfo{person}{Tathagata Das}, \bibinfo{person}{Ankur Dave},
  \bibinfo{person}{Justin Ma}, \bibinfo{person}{Murphy McCauly},
  \bibinfo{person}{Michael~J. Franklin}, \bibinfo{person}{Scott Shenker}, {and}
  \bibinfo{person}{Ion Stoica}.} \bibinfo{year}{2012}\natexlab{}.
\newblock \showarticletitle{Resilient Distributed Datasets: A {Fault-Tolerant}
  Abstraction for {In-Memory} Cluster Computing}. In
  \bibinfo{booktitle}{\emph{{{USENIX NSDI}}}}.
\newblock


\bibitem[Zhang et~al\mbox{.}(2018)]%
        {surf}
\bibfield{author}{\bibinfo{person}{Huanchen Zhang}, \bibinfo{person}{Hyeontaek
  Lim}, \bibinfo{person}{Viktor Leis}, \bibinfo{person}{David~G. Andersen},
  \bibinfo{person}{Michael Kaminsky}, \bibinfo{person}{Kimberly Keeton}, {and}
  \bibinfo{person}{Andrew Pavlo}.} \bibinfo{year}{2018}\natexlab{}.
\newblock \showarticletitle{SuRF: Practical Range Query Filtering with Fast
  Succinct Tries}. In \bibinfo{booktitle}{\emph{SIGMOD}}.
\newblock


\bibitem[Zhang et~al\mbox{.}(2022)]%
        {zhang2022_teleport}
\bibfield{author}{\bibinfo{person}{Qizhen Zhang}, \bibinfo{person}{Xinyi Chen},
  \bibinfo{person}{Sidharth Sankhe}, \bibinfo{person}{Zhilei Zheng},
  \bibinfo{person}{Ke Zhong}, \bibinfo{person}{Sebastian Angel},
  \bibinfo{person}{Ang Chen}, \bibinfo{person}{Vincent Liu}, {and}
  \bibinfo{person}{Boon~Thau Loo}.} \bibinfo{year}{2022}\natexlab{}.
\newblock \showarticletitle{Optimizing Data-Intensive Systems in Disaggregated
  Data Centers with {{TELEPORT}}}. In \bibinfo{booktitle}{\emph{SIGMOD}}.
\newblock


\bibitem[Zhao et~al\mbox{.}(2022)]%
        {fpgatrusted2}
\bibfield{author}{\bibinfo{person}{Mark Zhao}, \bibinfo{person}{Mingyu Gao},
  {and} \bibinfo{person}{Christos Kozyrakis}.} \bibinfo{year}{2022}\natexlab{}.
\newblock \showarticletitle{ShEF: shielded enclaves for cloud FPGAs}. In
  \bibinfo{booktitle}{\emph{ACM ASPLOS}}.
\newblock


\bibitem[Zhong et~al\mbox{.}(2022)]%
        {xrp}
\bibfield{author}{\bibinfo{person}{Yuhong Zhong}, \bibinfo{person}{Haoyu Li},
  \bibinfo{person}{Yu~Jian Wu}, \bibinfo{person}{Ioannis Zarkadas},
  \bibinfo{person}{Jeffrey Tao}, \bibinfo{person}{Evan Mesterhazy},
  \bibinfo{person}{Michael Makris}, \bibinfo{person}{Junfeng Yang},
  \bibinfo{person}{Amy Tai}, \bibinfo{person}{Ryan Stutsman}, {and}
  \bibinfo{person}{Asaf Cidon}.} \bibinfo{year}{2022}\natexlab{}.
\newblock \showarticletitle{{XRP}: {In-Kernel} Storage Functions with {eBPF}}.
  In \bibinfo{booktitle}{\emph{USENIX OSDI}}.
\newblock


\end{thebibliography}
}
\end{sloppypar}

\newpage
\appendix

\section*{Appendix}
\label{sec:appendix}

\renewcommand\thesection{\Alph{section}.}
\renewcommand\thesubsection{\Alph{section}.\arabic{subsection}}
\setcounter {subfigure} {0}
\setcounter {figure} {0}
\setcounter {section} {0}
\setcounter {page} {1}

\section{Multiplexing $M+N$ Iterator Executions for Maximizing Pipeline Utilization}

We claimed in \S\ref{ssec:architecture} that if $t_c = \eta \cdot t_d$ for all offloaded iterator executions, it is always possible to multiplex $m + n$ concurrent iterator executions and fully utilize all memory and logic pipelines. We prove our claim by providing a staggered scheduling algorithm (Algorithm~\ref{alg:scheduling}) that ensures such multiplexing across $m+n$ iterator executions. The scheduler processes $m+n$ iterator execution requests, assigning each a memory pipeline, a logic pipeline, and staggered start times. These requests are then executed in the respective memory pipelines. Through this staggered scheduling approach, \name fully utilizes the $n$ memory pipelines and $m$ logic pipelines, ensuring no resources are wasted. Note that this algorithm is a simplified version to illustrate the potential for full pipeline saturation under the given condition. \name's scheduler implements a real-time algorithm to multiplex incoming requests on the fly.

\begin{algorithm}
\caption{Staggered-Scheduling}
\label{alg:scheduling}
\begin{algorithmic}[1]
\State $m, n \gets$ number of logic, memory pipelines
\State $L_i, M_j \gets$ $i^{th}$ logic pipeline, $j^{th}$ memory pipeline
\State $t_d \gets$ data fetch time per pointer traversal iteration

\While{true}
    \State Dequeue $n + m$ requests from network stack
    \For{$i \gets 1$ \textbf{to} $m + n$}
        \State Assign request $R_i$ to ($M_{i \bmod n}$,  $L_{i \bmod m}$)

        \State Schedule $R_i$ to start at time $(i-1) \cdot \frac{t_d}{n}$

    \EndFor
    
    \State Start requests as scheduled at memory pipelines

\EndWhile
\end{algorithmic}
\end{algorithm}

\subsection{\name Empirical Analysis}
Prior studies have shown that real-world data-centric cloud applications spend a significant fraction of time traversing pointers, as summarized in Fig.~\ref{fig:sup_motivation}.

\begin{figure}[b]
    \centering
    \footnotesize
     \subfigure[Survey from prior studies]{
                \begin{tabular}[b]{c|c} 
                    \textbf{Application} & \specialcell{\textbf{\% of time spent}\\\textbf{in pointer traversal}} \\ \hline
                    GraphChi~\cite{graphchi} & $\sim 93\%$ \\ \hline
                    MonetDB\cite{monetdb} & $70\%-97\%$ \\ \hline
                    GC in Spark~\cite{spark} & $\sim 72$\% \\ \hline
                    VoltDB~\cite{voltdb} & Up to $49.55$\% \\ \hline
                    MemC3~\cite{memc3} & Up to $21.15$\% \\ \hline
                    DBx1000 \cite{db1000} & $\sim 9$\% \\\hline
                    Memcached\cite{memcached} & $\sim 7$\% \\ \hline\hline
                \end{tabular} 
    }
            \label{tab:sup_motivation}
    \caption{\textbf{Time cloud applications spend in pointer traversals} based on prior studies}
    \label{fig:sup_motivation}
\end{figure}

\begin{table*}[ht]
  \centering
  \captionof{table}{\small Additional data structure supported by \name.}
  \scriptsize
  \begin{tabular}{L{.12\textwidth}|L{.06\textwidth}|L{.05\textwidth}|L{.22\textwidth}|L{.22\textwidth}|L{.09\textwidth}|L{.07\textwidth}}
    \hline
    {\bf Data Structure} & {\bf Category} & {\bf Library} & {\bf Data structure API} & {\bf Internal function} & {\bf Original code} & {\bf \name code}\\\hline
    \hline
    List & \multirow{5}{*}{List} & \multirow{2}{*}{STL} & \multirow{2}{*}{\texttt{std::find(start, end, value)}} & \multirow{2}{*}{\texttt{std::find(start, end, value)}} &\multirow{2}{*}{Listing~\ref{lst:list}} & \multirow{2}{*}{Listing~\ref{lst:list_mod}}  \\\cline{1-1}
    Forward list & & &  & & & \\\cline{1-1}\cline{3-7}
    Bimap &  & \multirow{3}{*}{Boost} &  \multirow{3}{*}{\texttt{find(key, hash)}} & \multirow{3}{*}{\texttt{find(key, hash)}} &\multirow{3}{*}{Listing \ref{lst:bimap}} & \multirow{3}{*}{Listing \ref{lst:bimap_mod}}\\\cline{1-1}
    Unordered map  &  &  &  &  & & \\\cline{1-1}
    Unordered set  &  &  &  &  & & \\\hline
    Btree & \multirow{8}{*}{Tree} & Google & \multirow{8}{*}{\texttt{find(\&key)}} & \makecell[l]{\texttt{internal\_locate\_plain}\\\texttt{\_compare(key, iter)}} & Listing \ref{lst:btree} & Listing \ref{lst:btree_mod}\\\cline{1-1}\cline{3-3}\cline{5-7}
    Map & & \multirow{4}{*}{STL} & & \multirow{4}{*}{\texttt{\_M\_lower\_bound(x, y, key)}} &\multirow{4}{*}{Listing \ref{lst:map}} & \multirow{4}{*}{Listing \ref{lst:map_mod}}\\\cline{1-1}
    Set & & & & &  & \\\cline{1-1}
    Multimap & & & & & & \\\cline{1-1}
    Multiset & & & & & & \\\cline{1-1}\cline{3-3}\cline{5-7}
    AVL tree & & \multirow{3}{*}{Boost} & & \multirow{3}{*}{\texttt{lower\_bound\_loop(x, y, key)}} &\multirow{3}{*}{Listing \ref{lst:avltree}} & \multirow{3}{*}{Listing~\ref{lst:avltree_mod}}\\\cline{1-1}
    Splay tree & & & & & & \\\cline{1-1}
    Scapegoat tree & & & & & & \\
    \hline
  \end{tabular}
  \label{table:extra-apps-2}
\end{table*}

\section{\name Supported Data Structures}\label{sec:appendix_data_structures}

We adapt $13$ data structures across $4$ popular open-sourced libraries to \name's iterator abstraction (\S\ref{sec:interface}). In particular, we outline how the data structure implementations for certain operations can be expressed using \code{init()}, \code{next()}, and \code{end()}. For simplicity and readability, (i) we assume that the data structure developer defines a macro, \code{SP\_PTR\_(variable\_name)}, as the address of the variable resides on the \code{scratch\_pad}, and (ii) we omit obvious type conversions for de-referenced pointers. 

We analyze two widely used categories of data structures: lists and trees. In our analysis, we find that the top-level data structure APIs (\ie, the APIs used by applications) use the same base function under the hood. For instance, list and forward list in the STL library share the same internal function, \code{std::find()}. We summarize our findings in Table \ref{table:extra-apps-2}, including the data structure libraries, their category, the top-level data structure APIs, and the internal base function.

\paragraphb{List structures} Our surveyed list structures already follow the execution flow of \name iterator: \code{init()}, \code{next()}, and \code{end()}.

These data structures generally have compute-intensive \code{end()} functions to check multiple termination conditions, while their \code{next()} function simply dereferences a single pointer to the next node. Listing \ref{lst:list} and Listing \ref{lst:list_mod} demonstrate a linked list with two termination conditions: (i) value is found or (ii) search reaches the end. To indicate which condition is met, a special flag (\eg, \code{KEY\_NOT\_FOUND}) is written on the \code{scratch\_pad}. Listing~\ref{lst:bimap} and Listing~\ref{lst:bimap_mod} describe a bitmap that uses a hashtable internally, where colliding entries are stored in linked lists within the same bucket. As such, the \name iterator interface resembles that of \code{std::list} quite closely.

\paragraphb{Tree-like data structures}
Compared to list structures, tree data structures require more computation in the \code{next()} function, as the next pointer is determined based on the value in the child node. For instance, in \code{Btree} (Listing \ref{lst:btree}, \ref{lst:btree_mod}), the next function iterates through internal node keys, comparing them to the search key. Interestingly, \code{std::map} (Listing \ref{lst:map}, \ref{lst:map_mod}) and Boost AVL trees (Listing \ref{lst:avltree}, \ref{lst:avltree_mod}) share the same offload function structure, with only minor implementation and naming differences.

\lstset{frame=tb,
  xleftmargin=0cm,
  linewidth=0.95\textwidth
}
\captionsetup[lstlisting]{style=centered_lstlisting}


\subsection{List data structure in STL library}

\begin{minipage}{0.47\textwidth}
\centering
\begin{lstlisting}[caption={C++ STL realization for \code{std::find()}},label={lst:list}, captionpos=t]
struct node {
    value_type value;
    struct node* next;
};

node* find(node* first, node* last, const value_type& value)
{
    for (; first != last; first=first->next)
        if (first->value == value)
            return first;
    return last;
}
\end{lstlisting}

\begin{lstlisting}[caption={\name realization for \code{std::find()}},label={lst:list_mod}, captionpos=t]
class list_find : chase_iterator {

    init(void *value, void* first) {
        *SP_PTR_VALUE = value;
        cur_ptr = first;
    }
  
    void* next() {
        return cur_ptr->next;
    }
  
    bool end() {
        if (*SP_PTR_VALUE == cur_ptr->value) {
            *SP_PTR_RETURN = cur_ptr;  
            return true;
        }
        if (cur_ptr->next == NULL) {
            *SP_PTR_RETURN = KEY_NOT_FOUND;  
            return true;
        }
        return false;
    }
}
\end{lstlisting}
\end{minipage}

\subsection{List data structure in Boost library}

\begin{minipage}{0.47\textwidth}
\centering
\begin{lstlisting}[caption={Boost realization for \code{bimap::find()}},label={lst:bimap}, captionpos=t]
struct node {
    key_type key;
    struct node* next;
    value_type value;
};
void* find(const key_type& key, const hash_type& hash) const
{
    // The bucket start pointer can be pre-computed before offloading
    std::size_t buc = buckets.position(hash(key));
    node_ptr start = buckets.at(buc)
    for(node_ptr x = start; x != NULL; x = x->next){
        if(key == x->key){
            return x;
        }
    }
    return NULL;
}
\end{lstlisting}

\begin{lstlisting}[caption={\name realization for \code{bimap::find()}},label={lst:bimap_mod}, captionpos=t]
class bimap_find : chase_iterator {
public:
    key_type key;
  
    init(void *key, void* start) {
        *SP_PTR_KEY = key;
        cur_ptr = start;
    }
  
    void* next() {
        return cur_ptr->next;
    }
  
    bool end() {
        if (*SP_PTR_KEY == cur_ptr->key) {
            *SP_PTR_RETURN = cur_ptr;
            return true;
        }
        if (cur_ptr->next == NULL) {
            *SP_PTR_RETURN = NULL;  
            return true;
        }
        return false;
    }
}
\end{lstlisting}
\end{minipage}

\subsection{Tree data structure in Google library}
\begin{minipage}{0.47\textwidth}
\centering
\begin{lstlisting}[caption={Original code for \code{btree::internal\_locate\_plain\_compare()}},label={lst:btree}, captionpos=t]
#define kNodeValues 8
struct btree_node {
    bool is_leaf;    
    int num_keys;
    key_type keys[kNodeValues];
    btree_node* child[kNodeValues + 1];
};
IterType btree::internal_locate_plain_compare(const key_type &key, IterType iter) const {
    for (;;) { 
        int i;
        for(int i = 0; i < iter->num_keys; i++) {
            if(key <= iter->keys[i]) break;
        }
        if (iter.node->is_leaf) break;
        iter.node = iter.node->child(i);
    }
    return iter;
}
\end{lstlisting}

\begin{lstlisting}[caption={\name code for \code{btree::internal\_locate\_plain\_compare()}},label={lst:btree_mod}, captionpos=t]
class btree_find_unique : chase_iterator {
    init(void *key, void* iter) {
        *SP_PTR_KEY = key;
        cur_ptr = iter;
    }
  
    void* next() {
        *SP_PTR_I = 0;
        for(; *SP_PTR_I < cur_ptr->num_keys; *SP_PTR_I++) {
            if(*SP_PTR_KEY <= cur_ptr->keys[*SP_PTR_I]) 
                break;
        }
        cur_ptr = cur_ptr->child[*SP_PTR_I];
    }
  
    bool end() {
        if(cur_ptr->is_leaf) {
            *SP_PTR_RETURN = cur_ptr;
            return true;
        } else return false;
    }
}
\end{lstlisting}
\end{minipage}

\subsection{Tree data structure in STL library}
\begin{minipage}{0.47\textwidth}
\centering
\begin{lstlisting}[caption={C++ STL realization for \code{map::find()}},label={lst:map}, captionpos=t]
struct node {
    key_type key;
    node* left;
    node* right;
};

_M_lower_bound(node* x, node* y, const key_type& key)
{
    while (x != 0) {
        if (x->key <= key) {
            y = x; 
            x = x->left;
        } else {
            x = x->right;
        }
    }
    return y;
}
\end{lstlisting}

\begin{lstlisting}[caption={\name realization for \code{map::find()}},label={lst:map_mod}, captionpos=t]
class map_find : chase_iterator {
    init(void *key, void* x, void* y) {
        *SP_PTR_KEY = key;
        *SP_PTR_Y = y;
        cur_ptr = x;
    }
  
    void* next() {
        if (cur_ptr->key <= *SP_PTR_KEY) {
            *SP_PTR_Y = cur_ptr;  
            cur_ptr = cur_ptr->left;
        } else {
            cur_ptr= cur_ptr->right;
        }
        return cur_ptr->left;
    }
  
    bool end() {
        if (cur_ptr == NULL) {
            *SP_PTR_RETURN = *SP_PTR_Y;  
            return true;
        } else {
            return false;
        }
    }
}
\end{lstlisting}
\end{minipage}

\begin{minipage}{0.47\textwidth}
\subsection{Tree data structure in Boost library}
\centering
\begin{lstlisting}[caption={Boost realization for \code{avltree::find()}},label={lst:avltree}, captionpos=t]
static node_ptr lower_bound_loop
(node_ptr x, node_ptr y, const KeyType &key)
{
    while(x){
        if(x->key >= key)) {
            x = x->right;
        }
        else{
            y = x;
            x = x->left;
        }
    }
    return y;
}
\end{lstlisting}

\begin{lstlisting}[caption={\name realization for \code{avltree::find()}},label={lst:avltree_mod}, captionpos=t]
class avltree_find : chase_iterator {
public:
    key_type key;
    void* y;
  
    init(void *key, void* x, void* y) {
        *SP_PTR_KEY = key;
        *SP_PTR_Y = y;
        cur_ptr = x;
    }
  
    void* next() {
        if(cur_ptr->key >= *SP_PTR_KEY) {
            cur_ptr = cur_ptr->right;
        }
        else{
            *SP_PTR_Y = cur_ptr;
            cur_ptr = cur_ptr->left;
        }
    }
  
    bool end() {
        // The result is already stored at SP_PTR_Y
        if(cur_ptr == NULL) {
            return true;
        } else {
            return false;
        }
    }
}
\end{lstlisting}
\end{minipage}

\newpage
\section{\name Additional Evaluation Results}

In this section, we provide additional evaluation results for \name.

\begin{figure*}[t]
\centering
  \includegraphics[width=\textwidth]{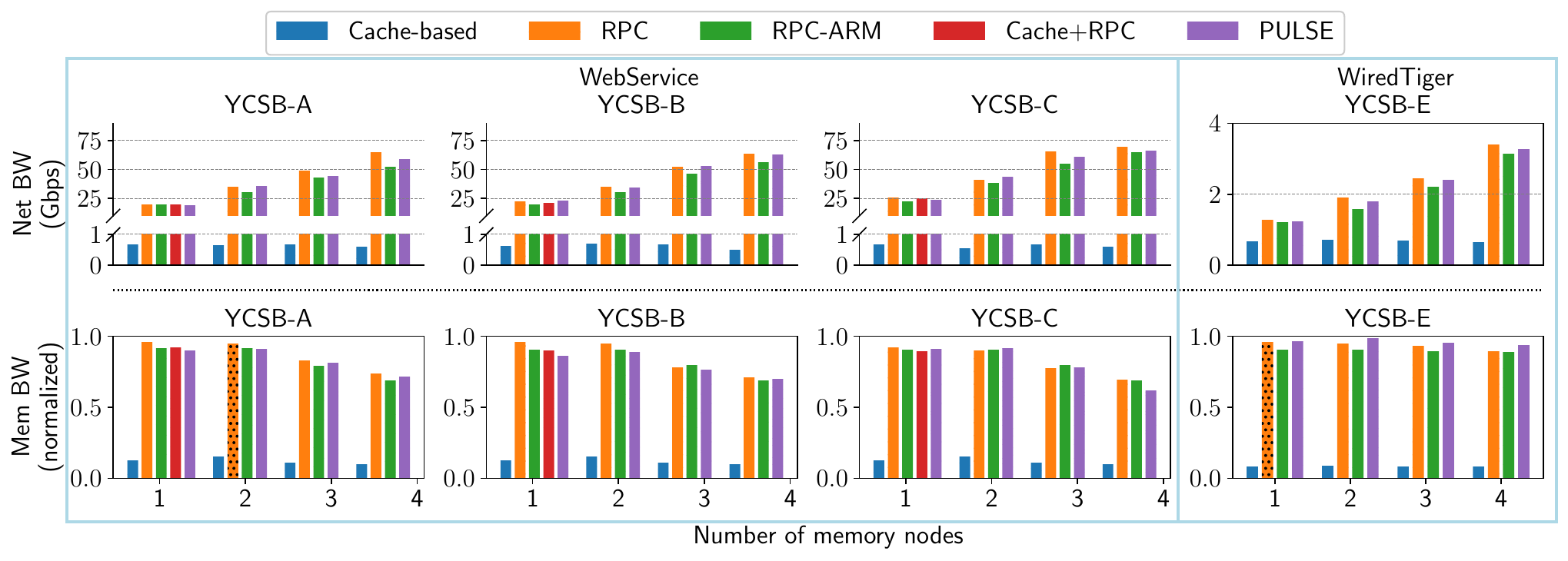}
  \caption{\textbf{Network and memory bandwidth utilization.} \name and RPC utilize over 90\% of the available memory bandwidth, while the cache-based approach suffers from swap system overhead. In Webservice, the network bandwidth becomes the bottleneck due to large 8 KB data transfers.}
  %
\label{fig:sup_eval_perf_e2e_utilization}
\end{figure*}

\begin{figure}[t]
\centering
\subfigure[Impact of access pattern]{
  \includegraphics[width=0.45\columnwidth]{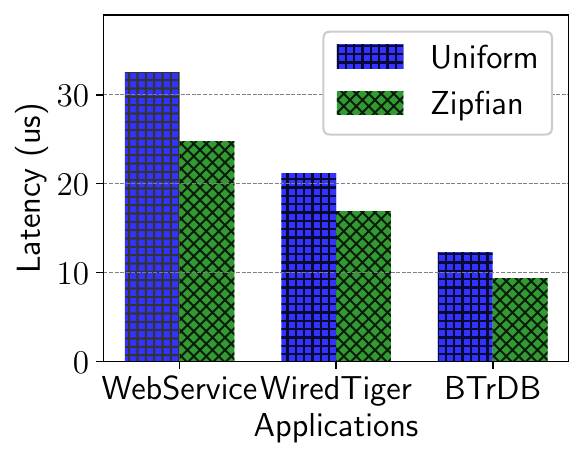}
  \label{fig:sup_cache_friendly}}
\subfigure[Impact of modifications]{
  \includegraphics[width=0.45\columnwidth]{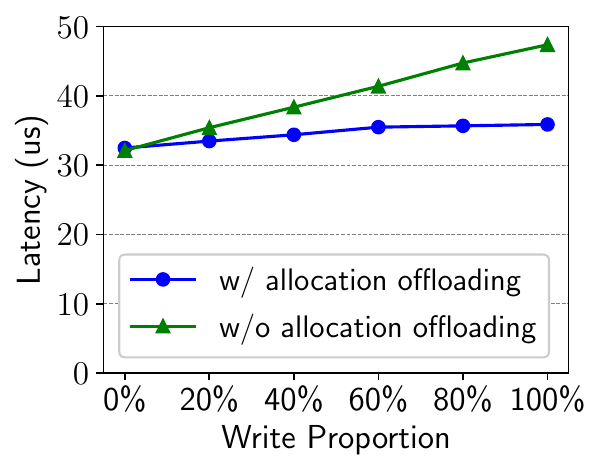}
  \label{fig:sup_write}}
\caption{(a) \name latency is up to $1.3\times$ lower for skewed than uniform access patterns due to caching. (b) Offloaded allocations in \name improve the WebService request latencies as the proportion of writes increases by reducing the number of round trips per allocation.}
\label{fig:sup_eval_cache_friendly}
\end{figure}

\begin{figure}[!ht]
    \centering
    \subfigure[Traversal length]{
        \includegraphics[width=0.45\columnwidth]{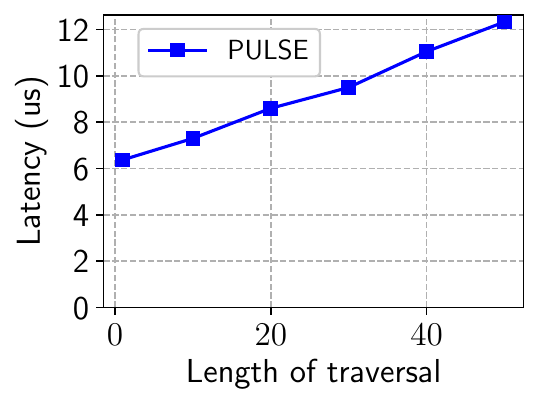}
        \label{fig:sup_eval_sensitivity_length}    
    }%
    \subfigure[Number of memory pipelines]{
        \includegraphics[width=0.45\columnwidth]{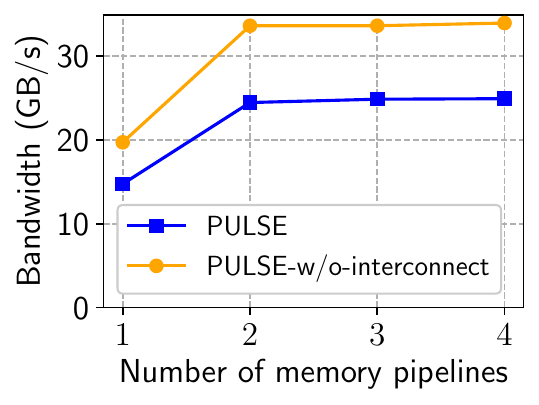}
        \label{fig:sup_eval_sensitivity_core}
    }
    \caption{\textbf{Sensitivity to traversal length and the number of memory pipelines.} (a) \name latency scales linearly with the length of traversal. (b) \name accelerator can saturate memory bandwidth with just two \name memory pipelines.}
\end{figure}

\begin{figure}[b]
\centering
  \subfigure[Latency]{
    \includegraphics[width=0.45\columnwidth]{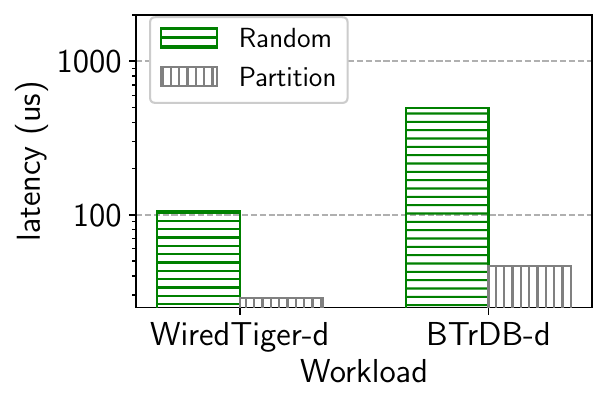}
    \label{fig:sup_eval_sensitivity_allocation_latency}
  }%
  \subfigure[Throughput]{
    \includegraphics[width=0.45\columnwidth]{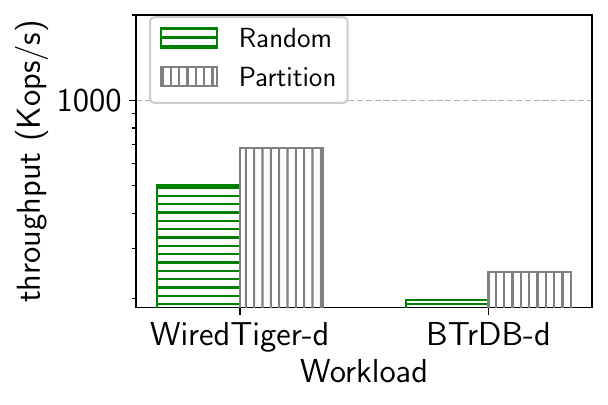}
    \label{fig:sup_eval_sensitivity_allocation_throughput}
  }%
  \caption{\textbf{Allocation policy.} \name performs better with the partitioned allocation since it minimizes cross-node traversals.}
\end{figure}

\subsection{Network and Memory Bandwidth Utilization} 
We evaluate the network and memory bandwidth utilization of the three applications in Fig. \ref{fig:sup_eval_perf_e2e_utilization}. For WiredTiger, \name and RPC utilize over 90\% of the available memory bandwidth, while the Cache-based approach suffers from low network bandwidth and memory utilization due to swap system overhead. For WebService, the large 8 KB data transfers saturate the maximum bandwidth that the DPDK stack can sustain~\cite{erpc}. As a result, network bandwidth becomes the bottleneck, reducing \name and RPC memory bandwidth utilization under 3 and 4 memory nodes. The memory bandwidth is normalized, where $1.0$ corresponds to $25$ GB/s per node.

\begin{figure*}[t]
\centering
  \includegraphics[width=0.9\textwidth]{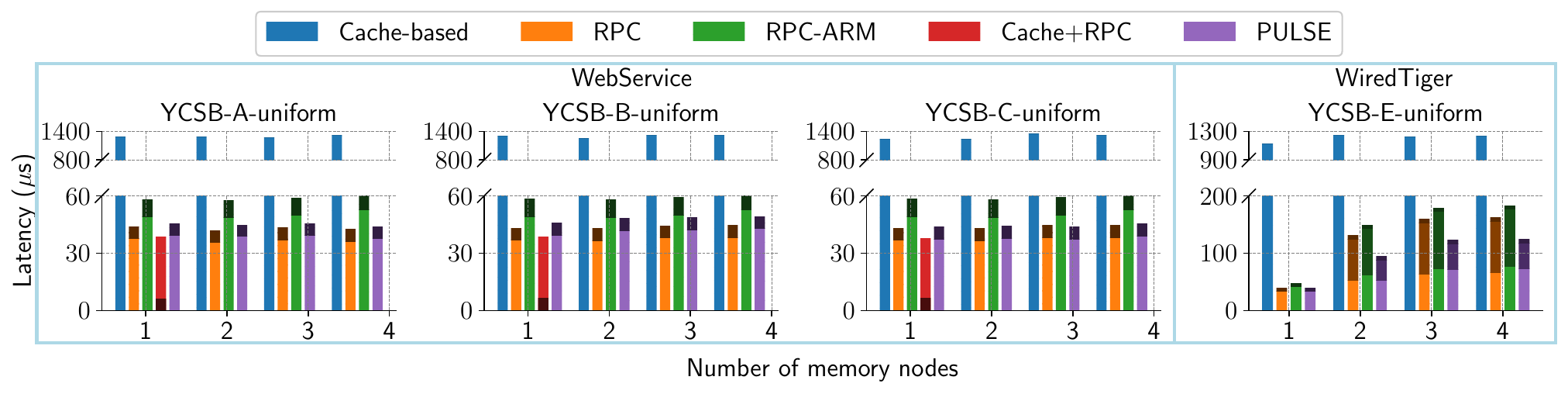}
  \label{fig:sup_eval_perf_e2e_latency_uniform}
  \includegraphics[width=0.9\textwidth]{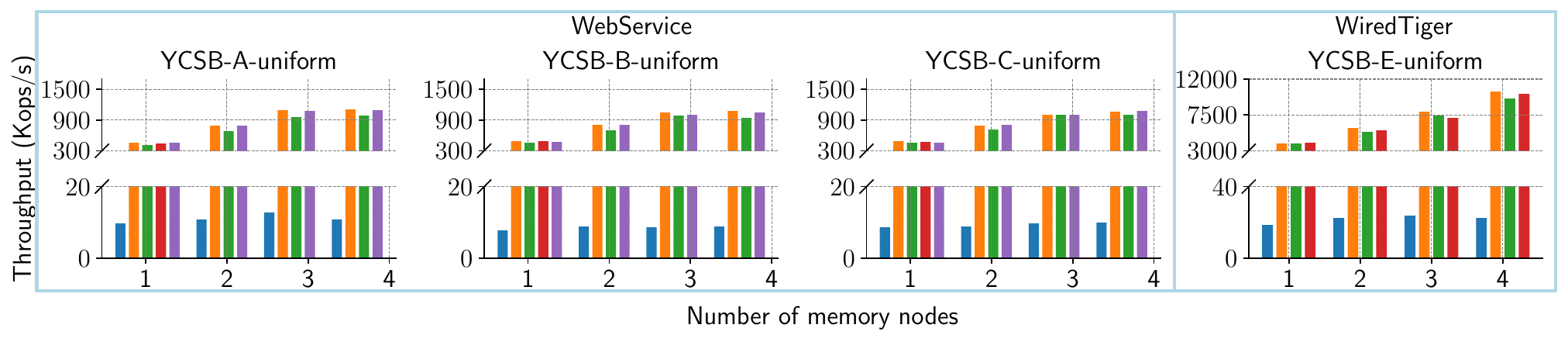}
  \label{fig:sup_eval_perf_e2e_throughput_uniform}
  \includegraphics[width=0.9\textwidth]{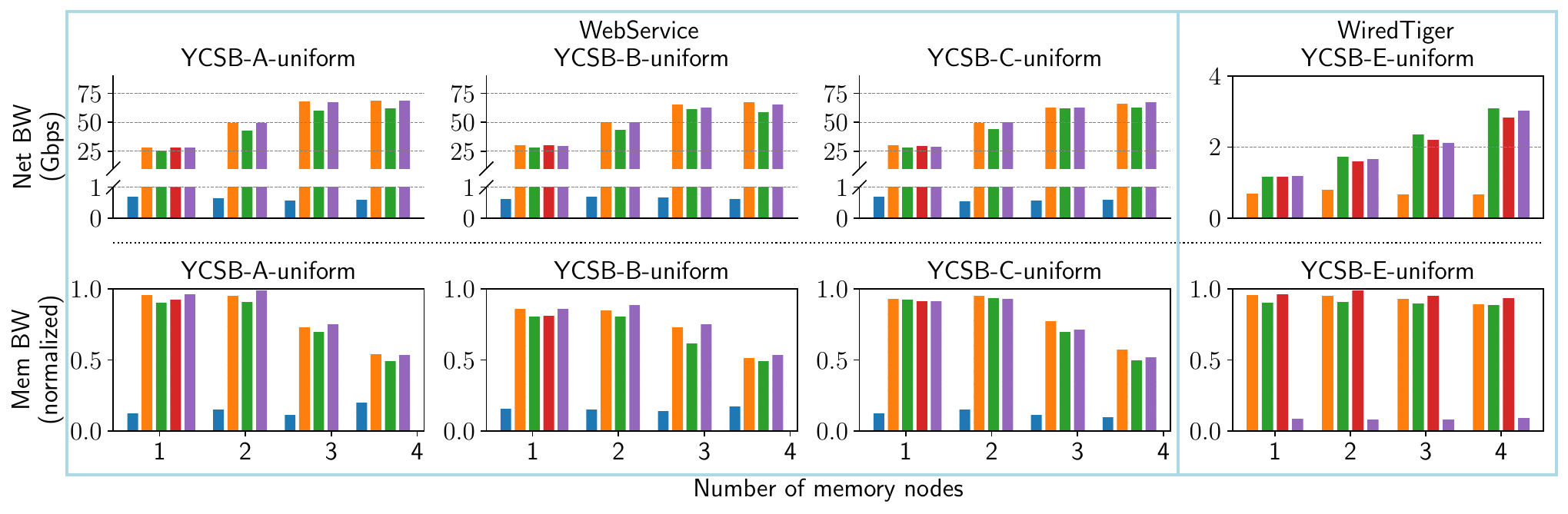}
  \label{fig:sup_eval_perf_e2e_utilization_uniform}
  \caption{\textbf{Application performance using workload with uniform distribution.}}
  \label{fig:sup_eval_uniform}
  %

\end{figure*}

\subsection{\name Sensitivity Analysis}
\label{ssec:sensitivity}

We evaluate \name's sensitivity to workload characteristics and system parameters: access pattern, data structure modifications, traversal length, allocation policy, and the number of \name memory pipelines.

\paragraphb{Impact of access pattern}
While our evaluation so far has been confined to Zipfian workloads, we evaluate the impact of skewed access patterns on \name performance for all three applications. Our setup comprises a single $32$GB memory node with a $2$GB CPU node cache. Figure~\ref{fig:sup_cache_friendly} shows caching at the CPU node reduces the number of iterator requests offloaded to the \name accelerator for the skewed (Zipfian) workload, improving \name performance for such workloads by up to $1.33\times$ relative to uniform ones.

\paragraphb{Impact of data structure modifications} Operations that modify data structures can require new memory allocations during traversal. Instead of returning control to the CPU node for allocations, \name populates the scratchpad for every request with a fixed number of pre-allocated memory regions. When a new allocation is initiated at the \name accelerator, it uses a pre-allocated memory region on the scratchpad. If all such regions ($16$ in our implementation) are used up in a single request, the traversal interrupts and returns to the CPU node. \name periodically replenishes pre-allocated entries, ensuring that allocation-triggered traversal interruptions are rare.

We evaluate the impact of data structure modifications in \name (\S\ref{sec:impl}) by increasing the proportion of writes for the WebService application on a single memory node. Figure~\ref{fig:sup_write} shows that as the proportion of writes increases, \name without offloaded allocations experiences higher latencies (up to $1.4\times$) since each new node allocation requires two additional round trips; offloaded allocations reduce the allocation overhead to $<1.1\%$.

\paragraphb{Length of traversal} For simplicity, we evaluate traversal queries on a single linked list with varying numbers of nodes traversed per query. As expected, Fig.~\ref{fig:sup_eval_sensitivity_length} shows that the end-to-end execution latency for a linked list traversal scales linearly with the number of nodes traversed.

\paragraphb{Allocation policy} We find that the allocation policy used for a data structure has a significant impact on application performance specifically for distributed traversals (Figs.~\ref{fig:sup_eval_sensitivity_allocation_latency} and~\ref{fig:sup_eval_sensitivity_allocation_throughput}). We evaluated the WiredTiger and BTrDB workloads (that employ B+-Tree as their underlying data structure) with two allocation policies: one that partitions allocations in a way that ensures all nodes in half the subtree are placed on one memory node and the other half on another, and another that allocates memory uniformly across the two nodes (as in \code{glibc} allocator). The average latency for random allocations is $3.7-10.8\times$ higher than partitioned allocation since it incurs significantly more cross-node traversals. This shows that while uniformly distributed allocations can enable better system-wide resource utilization, it may be preferable to exploit application-specific partitioned allocations for workloads where performance is the primary concern. 

\paragraphb{Number of \name memory pipelines} We evaluate the number of \name memory access pipelines required to saturate \name's memory bandwidth on a single memory node. We used the same linked list as our traversal-length experiment due to its relatively low $\eta$ value ($\sim$$0.06$), which allows us to stress the memory access pipeline without saturating the logic pipeline. Fig.~\ref{fig:sup_eval_sensitivity_core} shows that just $2$ memory pipelines can saturate \name's the per-node memory bandwidth of $25$ GB/s. We note that our $25$ GB/s limit does not match the hardware-specified memory channel bandwidths; this is primarily due to our use of the vendor-supplied memory interconnect IP, required to connect all memory pipelines to all memory channels. Indeed, if we remove the IP and measure memory bandwidth when each memory pipeline is connected to a dedicated memory channel, \name can achieve a memory bandwidth up to $34$ GB/s (shown as \name w/o Interconnect in Fig.~\ref{fig:sup_eval_sensitivity_core}). 

\paragraphb{\name performance with uniform workload} As illustrated in Fig.~\ref{fig:sup_eval_uniform}, while sharing a similar trend as Zipfian distribution, all approaches experience higher latency compared to Zipfian distribution due to the ineffectiveness of caching. \name provides lower (vs. Cache-based, RPC-ARM, and Cache$+$RPC) or comparable (vs. RPC) latency for a single memory node and $2.2$--$29\%$ lower latency (vs. RPC) for multi-memory nodes.

%

\end{document}